\documentclass[
superscriptaddress,
nofootinbib,
 amsmath,amssymb,
 aps,
prd,
twocolumn
]{revtex4-2}

\usepackage{graphicx}
\usepackage{dcolumn}
\usepackage{bm}

\usepackage{comment}

\usepackage[caption=false]{subfig}
\usepackage{xcolor}

\begin{document}

\title{Low-$T/|W|$ instabilities in differentially rotating neutron stars resembling merger remnants} 

\author{Georgios Lioutas}
    \email{georgios.lioutas@h-its.org}
    \affiliation{Heidelberger Institut für Theoretische Studien (HITS), Schloss-Wolfsbrunnenweg 35, 69118 Heidelberg, Germany}
    \affiliation{GSI Helmholtzzentrum für Schwerionenforschung, Planckstraße 1, D-64291 Darmstadt, Germany}
\author{Panagiotis Iosif}%
 \email{panagiotis.iosif@units.it}
\affiliation{Department of Physics, University of Trieste, I-34127 Trieste, Italy}%
\affiliation{INFN, Sezione di Trieste, I-34127 Trieste, Italy}%
\author{Andreas Bauswein}
\affiliation{GSI Helmholtzzentrum für Schwerionenforschung, Planckstraße 1, D-64291 Darmstadt, Germany}%
\affiliation{Helmholtz Research Academy Hesse for FAIR (HFHF), GSI Helmholtz Center for Heavy Ion Research, Campus Darmstadt, Germany}%
\author{Nikolaos Stergioulas}
\affiliation{Department of Physics, Aristotle University of Thessaloniki, 54124 Thessaloniki, Greece}%

\date{\today}

\begin{abstract}
We construct constant rest-mass sequences of equilibrium models of differentially rotating neutron stars which resemble binary neutron star post-merger remnants. For a more realistic description of the post-merger remnant, we impose that each model carries approximately $95\%$ of the angular momentum that a binary system with the same total rest mass has at the moment of merging, based on an empirical relation informed from neutron star merger simulations. We account for equation of state effects by employing two distinct microphysical descriptions for high density matter. We dynamically evolve the equilibrium models with a three-dimensional general relativistic hydrodynamics code that employs the conformal flatness approximation. We investigate the connection between the occurrence of the instability and the existence of corotation radii within the stellar configurations and determine the instability window for both equation of state sequences. The occurrence of low-$T/|W|$ instabilities leads to pronounced gravitational wave emission in the range $0.13 \lessapprox\beta \lessapprox 0.2$, while models outside this range exhibit less pronounced features in the gravitational wave spectrum. The prominence of gravitational wave emission is primarily determined by $\beta$, while the equation of state seems to have a more minor effect. We present correlations between the strength of the gravitational wave emission associated with the instability and properties of the equilibrium models. Stellar configurations modelled by different equations of state display differences in the timescales over which the various dynamical features develop, as well as whether they exhibit a pronounced $m=1$ deformation. Potential relations between the instability growth timescales and properties of the stellar models are studied.
\end{abstract}

\maketitle


\section{Introduction}
\label{sec:intro}

A binary neutron star (BNS) merger with a total binary mass below an equation-of-state dependent threshold mass does not undergo a prompt gravitational collapse to a black hole, but forms a massive rotating neutron star (NS) remnant \cite{2011PhRvD..83l4008H,2013PhRvL.111m1101B}. Rotation can stabilize the system against black hole formation, even if the remnant's mass exceeds the maximum mass of nonrotating or uniformly rotating NSs \cite{PhysRevLett.94.201101,2006PhRvD..73f4027S,2008PhRvD..78h4033B}. The merger results in a complex velocity distribution during the first milliseconds: two irrotational high-density cores orbit around each other forming a shear layer in-between them and a differentially rotating envelope surrounding the central parts of the remnant. Violent oscillations of the remnant produce gravitational-wave emission with a rich spectrum. Energy and angular momentum losses by gravitational waves and neutrinos as well as angular momentum redistribution govern the evolution of the remnant in the early post-merger stage (see e.g.\ \cite{2017RPPh...80i6901B, 2019JPhG...46k3002B, 2019ARNPS..69...41S, 2020GReGr..52..108B, 2021GReGr..53...27D,2021GReGr..53...59S} for reviews). Over several dynamical timescales the system approaches a quasi-equilibrium state, which to a certain extent can be resembled by a stationary differentially rotating NS and which can be seen as background configuration of distinct oscillation modes~\cite{2011MNRAS.418..427S}.

Self-consistent studies of BNS mergers, as well as NS dynamics in general, typically require three dimensional (3D) numerical simulations, which include a high level of sophistication and can be computationally expensive. A complementary approach for the study of neutron star dynamics lies in constructing equilibrium models of rotating relativistic stars (see e.g.\ \cite{2017LRR....20....7P} for a review). Such equilibrium models provide a testbed, where certain aspects of the dynamics can be studied in a simplified, but controlled setup.

Early efforts to construct models of differentially rotating relativistic stars employed a simple, monotonic rotation law, the so-called $j$-constant law \cite{1989MNRAS.237..355K, 1989MNRAS.239..153K}. This rotation law can adequately describe the rotation profiles of proto-neutron stars from core-collapse supernovae (see e.g. \cite{1998A&A...330.1005G, 2006ApJS..164..130O}). However, it can only roughly approximate the rotation profiles of merger remnants, where the peak of the angular velocity appears off-center (e.g.\ \cite{2005PhRvD..71h4021S,2016PhRvD..94d4060K,2017PhRvD..96d3004H}). This motivated studies to develop novel differential rotation laws with different level of complexity and number of parameters \cite{2012A&A...541A.156G, Bauswein_Stergioulas_2017, 2016PhRvD..93d4056U, Uryu_etal_2017, 2021PhRvD.103f3014C, 2024MNRAS.532..945C}. Specifically, Ref.~\cite{Uryu_etal_2017} was among the first studies to model differential rotation systematically and introduced 3- and 4-parameter rotation laws that could qualitatively capture the behaviour of a remnant's rotation profile. A number of studies investigate aspects of the dynamics of the post-merger like rotation profiles introduced in \cite{Uryu_etal_2017} (see e.g.\ \cite{2019PhRvD.100d3015Z, 2020MNRAS.498.5904P, 2020PhRvD.102d4040X, 2021MNRAS.503..850I, 2022MNRAS.510.2948I, 2022Univ....8..172F, 2023PhRvD.108b4058S, 2024PhRvD.110l4063M,2024PhRvD.110d3015C, 2025PhRvD.111j4030L}).

Particularly interesting is the study of non-axisymmetric instabilities, which can potentially be excited in the BNS remnant and leave an observable imprint on the GW emission. In this context, the ratio of the kinetic energy $T$, to the gravitational potential energy $W$ (hereafter $\beta = T/|W|$) provides a measure of the available rotational kinetic energy that can fuel the instability. The well-known bar-mode instability appears for high values of $\beta\ge0.24$ \cite{2000ApJ...542..453S,2007PhRvD..75d4023B}. In later works, additional non-axisymmetric instabilities were found, even for low values of $\beta$ assuming a high degree of differential rotation \cite{2003MNRAS.343..619S,2001ApJ...550L.193C,2002MNRAS.334L..27S,2003ApJ...595..352S}. Owing to the fact that these dynamical instabilities appear also for low values of $\beta$, they are generally referred to as low-$T/|W|$ instabilities (see \cite{2006MNRAS.368.1429S} for a distinction between the classical high $\beta$ bar instability and the low $\beta$ instability).

The exact nature and the criterion for the onset of low-$T/|W|$ instabilities is not yet fully understood. It has been proposed that these instabilities belong to the more generic class of shear instabilities in differentially rotating stars, and that their occurrence requires the existence of corotation points (i.e.\ points where the mode pattern speed matches the local angular velocity) of the unstable modes \cite{Watts_2005}. Furthermore, previous analyses, primarily in the context of self-gravitating discs, have highlighted the importance of the vortensity profile in the stability of these configurations \cite{2000ApJ...533.1023L,1999ApJ...513..805L,1989ApJ...344..645P,1978ApJ...221...51L}. The discussion on the connection of the vortensity profile and unstable modes was extended to neutron stars in \cite{2006ApJ...651.1068O}, where the authors suggest that the local minimum of the vortensity can act as a resonance cavity and amplify the amplitude of the mode. Further studies have focused on low-$T/|W|$ instabilities \cite{2015MNRAS.446..555P,2010CQGra..27k4104C,2007CoPhC.177..288C,2007CoPhC.177..288C,2017MNRAS.466..600Y,2016PhRvD..94h4032S}, also in the context of rotation laws that exhibit two corotation points \cite{2020PhRvD.102d4040X,2020MNRAS.498.5904P}. These studies identify, for the considered rotation laws, a potential relation between the growth timescale of an unstable mode and the distance of the respective corotation point from the local minimum of the vortensity, as well as the possibility that each corotation point might drive an instability. In the context of 3D BNS merger simulations, recent works also discuss the potential for low-$T/|W|$ instabilities to leave an imprint on the GW signal \cite{2020PhRvD.101f4052D,2022PhRvD.105d3020S}.

In this work, we study how the low-$T/|W|$ instability develops across a range of $\beta$. To this end, we study a rotation law, which resembles early post-merger BNS remnants \cite{Uryu_etal_2017}. We construct sequences of constant rest mass covering a broad range of $\beta$ values for two microphysical equations of state (EOS), namely SFHo \cite{2013ApJ...774...17S} and DD2 \cite{2010PhRvC..81a5803T,2010NuPhA.837..210H,1997ADNDT..66..131M}. We note that SFHo and DD2 are relatively soft/stiff, respectively, thus leading to different stellar properties e.g.\ radii and maximum supported mass. For the construction of the initial data, we employ a slice of each EOS at zero temperature and in $\beta$-equilibrium. We evolve a subset of these models for roughly $20$~ms employing a new 3D Eulerian general relativistic hydrodynamics (GRHD) code which evolves the space time dynamically adopting the conformal flatness condition (CFC) \cite{1980grg1.conf...23I,2008IJMPD..17..265I, 1996PhRvD..54.1317W}. We employ the full temperature- and composition-dependent EOS tables in the simulations. We report the occurrence of a low-$T/|W|$ instability and investigate how its development depends on the properties of the equilibrium model, e.g.\ the ratio $\beta$, the rotation profile and the Newtonian vortensity profile. We also comment on the observability of the instability via GW emission. Towards a more quantitative resembling of BNS merger remnants, as compared to previous studies, we construct sequences based on the 4-parameter rotation law from \cite{Uryu_etal_2017} with bulk properties (mass and angular momentum) similar to those found in merger simulations, while we also employ microphysical EOS tables to capture the EOS influence.

The paper is organized as follows: Section~\ref{sec:label2} describes the 3D GRHD code, which we employ for our evolutions (see also Appendix~\ref{app:Codeimplementation} and \ref{app:Codetests} for details on the implementation and code tests). Section~\ref{sec:RNS} focuses on the properties of the equilibrium models, which we construct. In Sec.~\ref{sec:IV}, we provide a detailed discussion for two reference models, one for each EOS employed in this study. In Sec.~\ref{sec:V} we apply a quadrupolar perturbation to the reference models, as a way to emulate up to some extent late time dynamics. Section~\ref{sec:VI} extends the discussion to the rest of the models which we evolve, focusing on the characteristics of the instability with respect to $\beta$. In Sec.~\ref{sec:VII}, we evaluate the extent to which the connections reported in the literature between the development/growth of the instability and the rotation and vortensity profiles apply to the rotation law under consideration. Finally, we provide a resolution study, as well as additional plots in Appendix~\ref{app:ResStudy} and \ref{app:Diffbeta}, respectively. Unless otherwise stated, we employ $c=G=1$ throughout the paper.

\section{Theoretical and numerical framework}\label{sec:label2}

We construct rotating relativistic neutron star configurations with post-merger rotation profiles using the {\tt rns} code \cite{1995ApJ...444..306S}, employing the more recent version of the code introduced in \cite{2021MNRAS.503..850I}. We evolve these equilibrium models with a 3D code, which solves the GRHD equations in a finite volume fashion on a uniform Cartesian grid. The 3D code adopts the ADM $3+1$ formalism \cite{2008GReGr..40.1997A} to foliate the spacetime into a set of non-intersecting spacelike hypersurfaces with constant coordinate time $t$. In the ADM formalism, the line element reads
\begin{eqnarray}
    ds^2 &=& g_{\mu\nu} dx^\mu dx^\nu \nonumber \\
         &=& \left( -\alpha^2 + \beta_i \beta^i \right) dt^2 + 2 \beta_i dx^i dt + \gamma_{ij} dx^i dx^j,
\end{eqnarray}
where $g_{\mu\nu}$ denotes the $4$-metric, $\alpha$ is the lapse function, $\beta^i$ is the shift vector and $\gamma_{ij}$ the $3$-metric. The code solves the Einstein equations adopting the CFC approximation~\cite{1980grg1.conf...23I,2008IJMPD..17..265I,1996PhRvD..54.1317W}. We devote the current section to the description of the 3D hydrodynamics code (see Appendix~\ref{app:Codeimplementation} for details on the numerical implementation and Appendix~\ref{app:Codetests} for code tests), while Sec.~\ref{sec:RNS} provides more details on the equilibrium models considered in this work.

\subsection{Hydrodynamics equations}\label{sec:HydroEq}
The GRHD evolution equations for a perfect fluid with energy-momentum tensor $T^{\mu\nu} = \rho h u^\mu u^\nu + p g^{\mu\nu}$ and rest-mass density current $J^\mu = \rho u^\mu$, follow from a set of local conservation laws, which read
\begin{eqnarray}
\nabla_\mu J^\mu &=& 0, \\
\nabla_\mu T^{\mu\nu} &=& 0, 
\end{eqnarray}
where $\rho$ is the rest-mass density, $h=1+\epsilon+p/\rho$ is the specific enthalpy, $p$ is the pressure, $\epsilon$ is the specific internal energy and $u^\mu$ is the fluid $4$-velocity. We denote the covariant derivative with respect to the $4$-metric $g_{\mu\nu}$ by $\nabla_\mu$.

The GRHD equations can be cast into a first-order flux-conservative hyperbolic system by introducing a set of conserved variables $(D, S_i, \tau)$ \cite{1997ApJ...476..221B}, which depend on the primitive variables $(\rho, \upsilon^i, \epsilon)$ as
\begin{eqnarray}
D &=& \rho W_\mathrm{L}, \\
S_i &=& \rho h W_\mathrm{L}^2 \upsilon_i, \\
\tau &=& \rho h W_\mathrm{L}^2 -p -D.
\end{eqnarray}
Here, $W_\mathrm{L}=\alpha u^0 = \left(1 - \gamma_{ij} \upsilon^i \upsilon^j \right)^{-1/2}$ is the Lorentz factor and $\upsilon^i = u^i/W_\mathrm{L} + \beta^i/\alpha$ is the $3$-velocity of the fluid as measured by an Eulerian observer. The hyperbolic system, as it is implemented in our code, then reads
\begin{equation}\label{eq:DEhydro}
    \partial_t  \bm{U} + \partial_i \bm{F}^i = \bm{S},
\end{equation}
where $\bm{U} = \sqrt{\gamma} (D, S_i, \tau)$ is the evolved state vector \cite{1997ApJ...476..221B,2008LRR....11....7F}. Furthermore, $\bm{F}^i$ and $\bm{S}$ refer to the flux vector
\begin{equation} 
\bm{F}^i = \alpha \sqrt{\gamma} \left(
\begin{array}{c}
D \left( \upsilon^i -\frac{\beta^i}{\alpha}\right) \\
S_j \left( \upsilon^i -\frac{\beta^i}{\alpha}\right) + p \delta^i_j \\
\tau \left( \upsilon^i -\frac{\beta^i}{\alpha}\right) + p \upsilon^i
\end{array} 
\right),
\end{equation} 
and the source vector
\begin{equation} 
\bm{S} = \alpha \sqrt{\gamma} \left(
\begin{array}{c}
0 \\
T^{\mu\nu} \left( \partial_\mu g_{\nu j} - \Gamma^\lambda_{\nu\mu} g_{\lambda j}\right) \\
\alpha \left( T^{\mu 0} \partial_\mu \ln{\alpha} - T^{\mu\nu} \Gamma^0_{\nu\mu} \right)
\end{array} 
\right),
\end{equation}
respectively. Here, $\gamma=\det{\left(\gamma_{ij}\right)}$ is the determinant of the $3$-metric and $\Gamma^\lambda_{\nu\mu}$ are the Christoffel symbols.

\subsection{Field equations}
The spacetime dynamics is determined by solving Einstein's field equations $G_{\mu\nu}=8\pi T_{\mu\nu}$, where $G_{\mu\nu}$ is the Einstein tensor. The ADM formalism \cite{2008GReGr..40.1997A} casts the Einstein equations into a set of evolution equations for the $3$-metric $\gamma_{ij}$ and the extrinsic curvature $K_{ij}$, which read
\begin{eqnarray}
    \partial_t \gamma_{ij} = &-& 2\alpha K_{ij} + \nabla_i \beta_j + \nabla_j \beta_i, \label{eq:evol3met} \\
    \partial_t K_{ij} = &-&\nabla_i \nabla_j \alpha + \alpha \left( R_{ij} + K K_{ij} - 2 K_{ik} K^k{}_j \right) \nonumber \\
    &+& \beta^k \nabla_k K_{ij} + K_{ik} \nabla_j \beta^k + K_{jk} \nabla_i \beta^k \nonumber \\
    &-& 4 \pi \alpha \left[ 2 S_{ij} - \gamma_{ij} \left(S - E \right) \right], \label{eq:evolextcurv}
\end{eqnarray}
and a set of constraint equations, which must be satisfied at all times,
\begin{eqnarray}
    R + K^2 - K_{ij} K^{ij} &=& 16 \pi E, \label{eq:constHam} \\
    \nabla_i \left( K^{ij} - \gamma^{ij} K \right) &=& 8 \pi S^i. \label{eq:constmom}
\end{eqnarray}
In Eqs.~\eqref{eq:evol3met}-\eqref{eq:constmom}, $\nabla_i$ is the covariant derivative associated to the $3$-metric, $R_{ij}$ is the Ricci tensor, $R$ denotes the corresponding Ricci scalar and $K=K_i{}^i$ refers to the trace of the extrinsic curvature. The source terms include matter contributions, which are given by
\begin{eqnarray}
    E   &=& \rho h W_\mathrm{L}^2-p, \\
    S^i &=& \rho h W_\mathrm{L} u^\mu \gamma^i_\mu,
\end{eqnarray}
while $S=S_i{}^i$ is the trace of $S^{ij}=\gamma^i_\mu \gamma^i_\nu T^{\mu\nu}$.

In this work, we adopt the conformal flatness condition \cite{1980grg1.conf...23I,2008IJMPD..17..265I,1996PhRvD..54.1317W}, namely we approximate the spatial $3$-metric as
\begin{equation}
    \gamma_{ij} = \psi^4 \hat{\gamma}_{ij},
\end{equation}
where $\hat{\gamma}_{ij}$ is the flat metric and $\psi$ is the conformal factor, which depends on the spacetime coordinates. In Cartesian isotropic coordinates, which our metric solver employs, $\hat{\gamma}_{ij}=\delta_{ij}$.

Under the assumption of the maximal slicing condition $K=0$, the field equations \eqref{eq:constHam}, \eqref{eq:evolextcurv} and \eqref{eq:constmom} become a set of five coupled elliptic non-linear equations for $\psi$, $\alpha\psi$ and $\beta^i$ respectively (see e.g.\ \cite{1996PhRvD..54.1317W,1998PhRvD..57.7299B}), which read
\begin{eqnarray}
 \Delta\psi          &=& - 2\pi\psi^5 E - \frac{1}{8}\psi^5K_{ij}K^{ij}, \label{eq:DEcfcpsi}\\
 \Delta(\alpha\psi)  &=& 2 \pi\alpha\psi^5 (E+2S) + \frac{7}{8}\alpha\psi^5 K_{ij}K^{ij}, \label{eq:DEcfcalppsi}\\
 \Delta\beta^i       &=& - \frac{1}{3}\partial^i\partial_j\beta^j 
                       + 2 \psi^{10} K^{ij} \partial_j \left(\frac{\alpha}{\psi^6}\right) \nonumber \\
                       && + 16\pi\alpha\psi^4 S^i \label{eq:DEcfcbeta},
\end{eqnarray}
where $\Delta$ is the Laplace operator with respect to the flat metric. Moreover, the trace of Eq.~\eqref{eq:evol3met} yields the relation
\begin{equation}
    \partial_t \psi = \frac{\psi}{6} \nabla_k \beta^k. \label{eq:evolpsi}
\end{equation}
By introducing the definition $\beta^i = B^i - \frac{1}{4}\partial_i\chi$, Eq.~\eqref{eq:DEcfcbeta} can be replaced by two simpler Poisson-like differential equations for the fields $B^i$ and $\chi$ \cite{1998PhRvD..57.7299B},
\begin{eqnarray}
 \Delta B^i   &=& 2 \psi^{10} K^{ij} \partial_j \left(\frac{\alpha}{\psi^6}\right) 
                +16\pi\alpha\psi^4 S^i, \label{DEcfc:Beta} \\
 \Delta \chi  &=& \partial_i B^i \label{eq:DEcfcchi},
\end{eqnarray}
which can be solved iteratively.

Note that by requiring that the $3$-metric remains conformally flat and that the maximal slicing condition is preserved throughout the whole evolution (i.e. Eq.~\eqref{eq:evolpsi} holds), the extrinsic curvature becomes time-independent and is given by the expression
\begin{equation}
 K_{ij} = \frac{\psi^4}{2\alpha} \left( \delta_{ik}\partial_j\beta^k + \delta_{jk}\partial_i\beta^k - \frac{2}{3} \delta_{ij} \partial_k\beta^k \right).
\end{equation}

\subsection{Equation of state}\label{sec:EOS}
We implement modules both for analytic, as well as microphysical EOSs. In particular, the code can perform evolutions with:
\begin{enumerate}
    \item The (isentropic) polytropic EOS
    \begin{equation}\label{eq:EOSpolyP}
        p = K \rho^\Gamma,
    \end{equation}
    where $p$ is pressure, $\rho$ is rest-mass density, $K$ denotes the polytropic constant and $\Gamma$ is the polytropic index. We can employ the polytropic EOS for barotropic evolutions, namely we do not solve the equation for $\tau$ and instead fix the specific internal energy based on
    \begin{equation}\label{eq:EOSpolyE}
        \epsilon = \frac{K \rho^{\Gamma-1}}{\Gamma-1}.
    \end{equation}
    
    \item The ideal gas EOS
    \begin{equation}
        p = (\Gamma-1) \rho \epsilon,
    \end{equation}
    where the pressure depends on both the rest-mass density and the specific internal energy, i.e.\ it requires solving the full system of Eqs.~\eqref{eq:DEhydro}. In contrast to the polytropic EOS, the ideal gas EOS permits the conversion of kinetic energy to thermal energy, which allows for a more realistic description of certain dynamical processes (e.g.\ shock heating).
    
    \item Hybrid EOSs, comprised of a zero-temperature tabulated microphysical EOS, supplemented with a thermal ideal gas component \cite{1993A&A...268..360J}. If we denote the contributions from the microphysical barotropic EOS to the pressure and specific internal energy as $p_\mathrm{cold}(\rho)$ and $\epsilon_\mathrm{cold}(\rho)$, respectively, the total pressure is given by
    \begin{equation}
        p = p_\mathrm{cold}(\rho) + p_\mathrm{th},
    \end{equation}
    where $p_\mathrm{th}$ is the thermal ideal gas contribution
    \begin{equation}\label{eq:EOSpth}
        p_\mathrm{th} = (\Gamma_\mathrm{th}-1) \rho \epsilon_\mathrm{th}.
    \end{equation}
    In Eq.~\eqref{eq:EOSpth}, the thermal specific energy density follows from
    \begin{equation}
        \epsilon_\mathrm{th} = \epsilon - \epsilon_\mathrm{cold}(\rho),
    \end{equation}
    where $\epsilon$ is obtained from the numerical evolution of the GRHD equations and, in a similar fashion to the pressure, is decomposed into a cold contribution from the barotropic EOS and a thermal part. The value of $\Gamma_\mathrm{th}$ is constant and lies typically between $1.5$ and $2$ for studies on NSs \cite{2010PhRvD..82h4043B}.
    
    \item Microphysical EOSs of the form $p=p(\rho, Y_\mathrm{e}, \epsilon)$, which provide the full temperature and composition dependence (via the electron fraction $Y_\mathrm{e}$). These EOSs are available in tabulated form. To account for the fact that the composition is required as input to these EOSs, we advect the initial $Y_\mathrm{e}$ by solving the equation
    \begin{equation}
        \partial_t \left( \sqrt{\gamma} D Y_\mathrm{e} \right) + \partial_i \left( \sqrt{\gamma} D (\alpha \upsilon^i - \beta^i) Y_\mathrm{e} \right) = 0,
    \end{equation}
    in addition to the GRHD system of equations (i.e.\ Eqs.~\eqref{eq:DEhydro}).
\end{enumerate}

\section{Equilibrium models}\label{sec:RNS}

\subsection{Approximating properties of post-merger remnants}\label{sec:IIIA}

We construct quasi-spherical relativistic equilibrium models, which are stable against gravitational collapse and reproduce selected properties of BNS post-merger remnants. We assume a baryonic mass corresponding to that of a binary system composed of two neutron stars with gravitational masses of $1.35 M_{\odot}$ each, measured at infinite separation. The resulting equilibrium models are massive in comparison to the maximum mass of nonrotating NSs. 
To account for the fact that BNS merger remnants have a significant angular momentum content (see e.g. \cite{2018MNRAS.481.3670R}), the constructed models have large angular momentum guided by an empirical relation between the angular momentum at the time of merger, $J_\text{merger}$, and neutron star properties \cite{Bauswein_Stergioulas_2017}, where we set $J \sim 95 \% \, J_\text{merger}$. This empirical relation is based on simulations that employ equal-mass, irrotational neutron stars as initial data. We note that for equal-mass binary systems with two $1.35 M_{\odot}$ NSs, we find $J_\text{merger} \left[G {M_\odot}^2 / c\right]\approx 6.1 $ and $6.4$ for SFHo and DD2, respectively (see e.g.\ Fig.\ 4 in \cite{Bauswein_Stergioulas_2017}, Table I in \cite{2016PhRvD..94b4023B} and Fig.\ 2 in \cite{2018MNRAS.481.3670R}).
Finally, we use a differential rotation law that allows for post-merger-like rotation profiles where the maximum of the angular velocity $\Omega_\text{max}$ is off-center, namely the 4-parameter law \cite{Uryu_etal_2017}:

\begin{equation}
\Omega = \Omega_c \frac{1 + \left( \dfrac{F}{B^2 \Omega_c} \right)^p}{1 + \left( \dfrac{F}{A^2 \Omega_c} \right)^{q+p}} \, . \label{eq:Uryuetal_rotlaw8}
\end{equation}

In the above, $F \equiv u^t u_{\phi} $ is the gravitationally redshifted angular momentum per unit rest mass and enthalpy and $ \Omega_c $ is the angular velocity at the center of the star. The parameter $p$ controls the growth of the rotation curve near the rotation axis, while $q$ controls the asymptotic behaviour of $\Omega(r)$ (setting $q = 3$ recovers the Keplerian rotation law in the Newtonian limit). Following \cite{Uryu_etal_2017, 2021MNRAS.503..850I}, we set $p=1$ and $q=3$ and determine parameters $A$ and $B$ by fixing the ratios $\lambda_1 = \Omega_\text{max} / \Omega_c$ and $\lambda_2 = \Omega_e / \Omega_c$, where $\Omega_e$ is the angular velocity at the equator. Setting $\lambda_1= 1.6$ and $\lambda_2=1.0$ has been shown to yield massive, quasi-spherical models \cite{2021MNRAS.503..850I, 2022MNRAS.510.2948I}.

\subsection{Framework and assumptions}\label{sec:IIIB}
Our framework follows standard choices within the equilibrium modeling literature. We construct axisymmetric, and asymptotically flat configurations and approximate the matter as a perfect fluid (see Sec.~\ref{sec:HydroEq} for the expression of the stress-energy tensor). In quasi-isotropic coordinates, the line element is written as
\begin{eqnarray}
ds^2 = &-e^{\gamma + \rho} dt^2 + e^{\gamma - \rho} r^2 \sin^2 \theta (d\phi - \omega dt)^2  \nonumber \\
& + e^{2\mu} (dr^2 + r^2 d\theta^2) \label{eq:stationary_axisym_metric} \, , 
\end{eqnarray}
with metric functions $ \gamma $, $ \rho $, $ \omega $ and $ \mu $ depending only on the coordinates $r$ and $\theta$. We note that $\omega$ is the angular velocity of a zero-angular-momentum observer (ZAMO) and describes the relativistic dragging of inertial frames due to rotation. The reader can find more details of the basic concepts in \cite[Chapter~1]{Friedman_Stergioulas_2013}.

To construct such equilibrium models we employ a recent version of the {\tt rns} code. At its core, {\tt rns} implements the Komatsu, Eriguchi and Hachisu (KEH) numerical scheme \cite{1989MNRAS.237..355K} with modifications by Cook, Shapiro and Teukolsky (CST) \cite{1992ApJ...398..203C}. The KEH/CST scheme is presented in detail in \cite[Section 4.1]{Friedman_Stergioulas_2013}.

Since our evolution code adopts the CFC approximation for the spacetime, we need to ensure that our initial data are also consistent with this choice. Therefore, we impose the condition $\mu = (\gamma-\rho)/2$ between these three metric functions in the KEH/CST iterative scheme, essentially enforcing conformal flatness to the full general relativistic solutions provided by {\tt rns} \cite{1996PhRvD..53.5533C,2014GReGr..46.1800I}. We note that the CFC approximation has been shown to be accurate and robust for constructing differentially rotating equilibrium models, with errors up to a few percent in physical quantities, even for the most relativistic and rapidly rotating configurations \cite{2014GReGr..46.1800I, 2021MNRAS.503..850I}.

We adopt two tabulated, hadronic candidate EOS\footnote{We construct the rotating models employing EOS slices at zero temperature and in $\beta$-equilibrium. In our time evolutions, we employ the full temperature- and composition-dependent tables.}, namely SFHo~\cite{2013ApJ...774...17S} and DD2~\cite {2010PhRvC..81a5803T,2010NuPhA.837..210H,1997ADNDT..66..131M} that yield typical non-rotating neutron star radii between 11 and 13 km. These EOS models are compatible with different observational constraints posing limits on the maximum mass \cite{2010Natur.467.1081D,2013Sci...340..448A, 2020NatAs...4...72C, 2018ApJ...859...47A, 2021ApJ...915L..12F}, tidal deformability \cite{2017PhRvL.119p1101A,PhysRevLett.121.161101} and radii of NSs \cite{2021ApJ...918L..28M,2021ApJ...918L..27R,2021ApJ...918L..29R}.

\begin{figure}
    \includegraphics[width=\columnwidth]{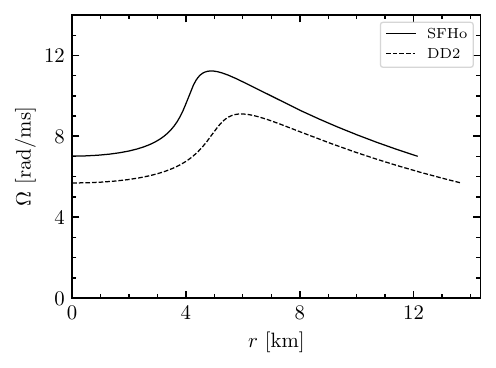}
    \caption{Rotation profiles for our reference models versus the coordinate radius $r$, calculated via the {\tt rns} code. The angular velocity $\Omega$ follows rotation law \eqref{eq:Uryuetal_rotlaw8}.}
    \label{fig:rns_rot_profiles}
\end{figure}

\begin{table*}
    \centering
    \caption{Properties of selected models belonging to a $M_0 = 2.964 M_\odot$ constant rest-mass sequence for SFHo and a $M_0 = 2.939 M_\odot$ constant rest-mass sequence for DD2. From left to right the columns show the polar to equatorial axis ratio, the central energy density, the gravitational mass, the angular momentum, the ratio of the rotational kinetic energy to the absolute value of the gravitational binding energy, the angular velocity at the rotation axis, the frame dragging metric potential at the rotation axis, the difference of the two previous quantities, the maximum value of the angular velocity, the angular velocity at the equator, the angular velocity of a free particle in circular orbit at the equator, the equatorial circumferential radius, and the equatorial coordinate radius. We mark the subset of models which we dynamically evolve with an `$\checkmark$' in the last column. The reference models, which we extensively discuss in Sec.~\ref{sec:IV}, are also marked in the last column.}
    \label{tab:SFHO_DD2_M0_const_sequence}
    \resizebox{\textwidth}{!}{
    \begin{tabular}{ccccccccccccccc}
        \hline
        EOS & $r_p / r_e$ & $\epsilon_c \, (\times 10^{15})$ & $M$ & $J$ & $T/|W|$ & $\Omega_c$ & $\omega_c$ & $(\Omega-\omega)_c$ & $\Omega_\text{max}$ & $\Omega_e$ & $\Omega_K$ & $R_e$ & $r_e$ & Evolved \\ 
        & & $[\text{g}/\text{cm}^3]$ & $[M_\odot]$ & $\left[\frac{G {M_\odot}^2}{c}\right]$ &   & [rad/ms] & [rad/ms] & [rad/ms] & [rad/ms] & [rad/ms] & [rad/ms] & [km] & [km] & Models \\
        \hline
        SFHo & 0.605 & 1.700 & 2.543 & 4.855 & 0.161 & 8.187 & 8.044 & 0.142 & 13.098 & 8.187 & 12.002 & 13.194 & 8.687 & $\checkmark$ \\
          & 0.600 & 1.650 & 2.544 & 4.892 & 0.163 & 8.107 & 7.925 & 0.183 & 12.972 & 8.107 & 11.833 & 13.334 & 8.828 &  \\
          & 0.593 & 1.600 & 2.548 & 4.955 & 0.166 & 8.039 & 7.815 & 0.224 & 12.862 & 8.039 & 11.648 & 13.496 & 8.985 &  \\
          & 0.586 & 1.550 & 2.551 & 5.011 & 0.169 & 7.961 & 7.691 & 0.271 & 12.738 & 7.961 & 11.454 & 13.667 & 9.155 &  \\
          & 0.578 & 1.500 & 2.554 & 5.076 & 0.172 & 7.882 & 7.562 & 0.321 & 12.612 & 7.882 & 11.246 & 13.856 & 9.342 &  \\
          & 0.568 & 1.450 & 2.560 & 5.168 & 0.176 & 7.809 & 7.438 & 0.371 & 12.494 & 7.809 & 11.018 & 14.076 & 9.553 & $\checkmark$  \\
          & 0.558 & 1.400 & 2.563 & 5.250 & 0.180 & 7.721 & 7.294 & 0.427 & 12.354 & 7.721 & 10.777 & 14.311 & 9.784 &  \\
          & 0.546 & 1.350 & 2.569 & 5.358 & 0.185 & 7.633 & 7.148 & 0.485 & 12.213 & 7.633 & 10.511 & 14.584 & 10.050 &  \\
          & 0.533 & 1.300 & 2.575 & 5.471 & 0.190 & 7.533 & 6.985 & 0.548 & 12.052 & 7.533 & 10.221 & 14.893 & 10.352 & $\checkmark$  \\
          & 0.518 & 1.250 & 2.582 & 5.607 & 0.196 & 7.423 & 6.809 & 0.613 & 11.876 & 7.423 & 9.896 & 15.257 & 10.707 &  \\
          & 0.501 & 1.200 & 2.590 & 5.761 & 0.203 & 7.295 & 6.611 & 0.684 & 11.672 & 7.295 & 9.527 & 15.691 & 11.134 & $\checkmark$  \\
          & 0.481 & 1.150 & 2.599 & 5.940 & 0.211 & 7.140 & 6.379 & 0.761 & 11.425 & 7.140 & 9.085 & 16.241 & 11.678 &  \\
          & 0.466 & 1.119 & 2.605 & 6.066 & 0.216 & 7.020 & 6.204 & 0.816 & 11.232 & 7.020 & 8.743 & 16.690 & 12.127 & $\checkmark$ (Ref. Model) \\
          & 0.454 & 1.100 & 2.610 & 6.163 & 0.220 & 6.925 & 6.070 & 0.855 & 11.080 & 6.925 & 8.463 & 17.073 & 12.512 &  \\
          & 0.445 & 1.090 & 2.613 & 6.227 & 0.222 & 6.855 & 5.973 & 0.882 & 10.969 & 6.855 & 8.246 & 17.378 & 12.822 & $\checkmark$  \\
        \hline   
        DD2 & 0.827 & 1.200 & 2.500 & 2.978 & 0.064 & 4.718 & 4.314 & 0.404 & 7.549 & 4.718 & 11.281 & 13.590 & 9.397 & $\checkmark$  \\
          & 0.805 & 1.150 & 2.507 & 3.191 & 0.073 & 4.914 & 4.458 & 0.457 & 7.863 & 4.914 & 11.057 & 13.806 & 9.590 &  \\
          & 0.781 & 1.100 & 2.514 & 3.416 & 0.083 & 5.100 & 4.586 & 0.514 & 8.160 & 5.100 & 10.816 & 14.050 & 9.810 & $\checkmark$  \\
          & 0.755 & 1.050 & 2.521 & 3.653 & 0.093 & 5.269 & 4.693 & 0.576 & 8.430 & 5.269 & 10.554 & 14.327 & 10.061 &  \\
          & 0.727 & 1.000 & 2.528 & 3.902 & 0.105 & 5.420 & 4.778 & 0.642 & 8.672 & 5.420 & 10.269 & 14.641 & 10.349 & $\checkmark$  \\
          & 0.695 & 0.950 & 2.540 & 4.193 & 0.118 & 5.563 & 4.852 & 0.711 & 8.902 & 5.563 & 9.950 & 15.020 & 10.692 &  \\
          & 0.661 & 0.900 & 2.551 & 4.498 & 0.132 & 5.672 & 4.888 & 0.785 & 9.076 & 5.672 & 9.596 & 15.460 & 11.098 & $\checkmark$  \\
          & 0.624 & 0.850 & 2.564 & 4.835 & 0.148 & 5.747 & 4.885 & 0.862 & 9.196 & 5.747 & 9.197 & 15.989 & 11.590 &  \\
          & 0.584 & 0.800 & 2.577 & 5.202 & 0.165 & 5.776 & 4.832 & 0.944 & 9.242 & 5.776 & 8.739 & 16.638 & 12.202 & $\checkmark$  \\
          & 0.539 & 0.750 & 2.596 & 5.641 & 0.185 & 5.752 & 4.724 & 1.028 & 9.203 & 5.752 & 8.192 & 17.489 & 13.008 &  \\
          & 0.508 & 0.718 & 2.607 & 5.939 & 0.198 & 5.691 & 4.604 & 1.086 & 9.105 & 5.691 & 7.773 & 18.185 & 13.683 & $\checkmark$ (Ref. Model) \\
          & 0.488 & 0.700 & 2.617 & 6.145 & 0.207 & 5.638 & 4.519 & 1.119 & 9.021 & 5.638 & 7.489 & 18.691 & 14.174 &  \\
          & 0.463 & 0.680 & 2.626 & 6.389 & 0.217 & 5.549 & 4.390 & 1.160 & 8.879 & 5.549 & 7.098 & 19.421 & 14.894 &  \\
          & 0.447 & 0.670 & 2.631 & 6.528 & 0.222 & 5.480 & 4.295 & 1.185 & 8.767 & 5.480 & 6.818 & 19.965 & 15.440 & $\checkmark$  \\
        \hline
    \end{tabular}}
\end{table*}

\begin{figure}
    \includegraphics[width=\columnwidth]{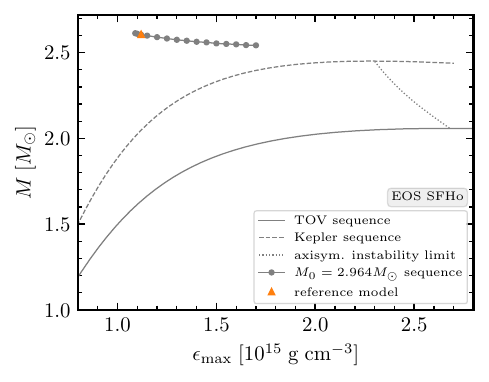}
    \caption{Gravitational mass $M$ versus maximum energy density $\epsilon_\text{max}$ for the SFHo EOS. The solid line denotes the nonrotating sequence, the dashed line the Kepler sequence, and the dotted line the axisymmetric instability limit for uniform rotation. The  filled circle line denotes a constant rest mass sequence with $M_0$ equal to our model's rest mass (see Table~\ref{tab:SFHO_DD2_M0_const_sequence}). The reference model is shown with an orange triangle.}
    \label{Mass_density_SFHO}
\end{figure}

\begin{figure}
    \includegraphics[width=\columnwidth]{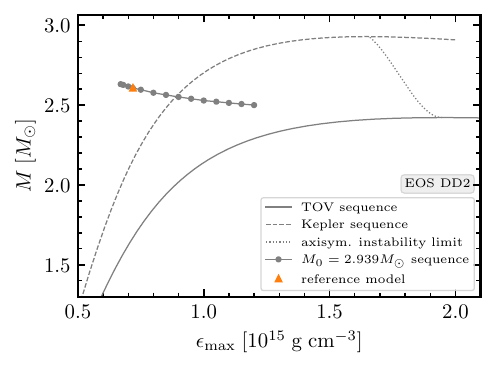}
    \caption{Same as Figure \ref{Mass_density_SFHO}, but for the DD2 EOS.}
    \label{Mass_density_DD2}
\end{figure}

We define two reference models, the properties of which are listed in Table \ref{tab:SFHO_DD2_M0_const_sequence}. 
Figure~\ref{fig:rns_rot_profiles} shows their rotation profiles, which qualitatively resemble that of a BNS merger remnant (see e.g.\ \cite{2005PhRvD..71h4021S,2017PhRvD..96d3004H}). To better illustrate the location of our models relative to the nonrotating Tolman-Oppenheimer-Volkoff (TOV) sequence and the mass-shedding (Kepler) limit for uniformly rotating stars, we plot the gravitational mass $M$ versus the maximum energy density $\epsilon_\text{max}$ in Figures \ref{Mass_density_SFHO} and \ref{Mass_density_DD2}.\footnote{For quasi-spherical models as the ones considered in this work, $\epsilon_\text{max}$ coincides with the central energy density $\epsilon_c$.} It is worth noting that the rest mass of our SFHo model exceeds the maximum rest mass $M_\text{0 max,rot}$ of a cold, uniformly rotating neutron star ($M_\text{0 max,rot}\big|_\text{SFHo} = 2.88 M_\odot$) making it a \textit{hypermassive} configuration \cite{2000ApJ...528L..29B}. This is not the case for our DD2 model ($M_\text{0 max,rot}\big|_\text{DD2} = 3.48 M_\odot$). 

Additionally, we construct constant rest-mass sequences with a mass value equal to that of our reference models ($M_0 = 2.964 M_\odot$ for SFHo and $M_0 = 2.939 M_\odot$ for DD2). Table~\ref{tab:SFHO_DD2_M0_const_sequence} also reports the physical properties of the extra models. We include these configurations in Figs.~\ref{Mass_density_SFHO} and \ref{Mass_density_DD2} and dynamically evolve a subset of them, marked in Table~\ref{tab:SFHO_DD2_M0_const_sequence} (see Secs.~\ref{sec:VI} and \ref{sec:VII}), in addition to the reference models (see Secs.~\ref{sec:IV} and \ref{sec:V}). 

\begin{figure}
    \includegraphics[width=\columnwidth]{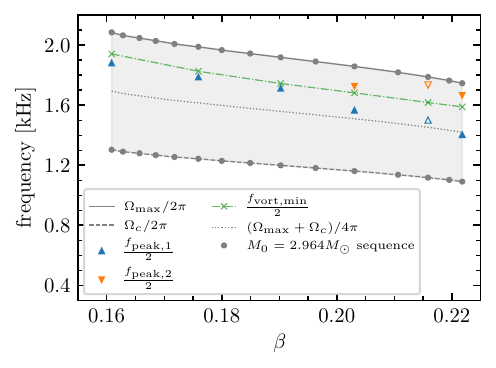}
    \caption{Corotation band for SFHo: The filled circles represent $\Omega/2\pi$ values of a constant rest-mass sequence with $M_0$ equal to our reference model's rest mass (see Table~\ref{tab:SFHO_DD2_M0_const_sequence}). The solid and dashed lines represent the edges $\Omega_\textrm{max}/2\pi$ and $\Omega_c/2\pi$ of the corotation region (grey shaded area) respectively. The pattern frequencies of the selected evolved models are shown as blue up triangles, whereas the orange down triangles represent secondary frequencies identified for some of the models (see Table~\ref{tab:frequencies} and Secs.~\ref{sec:IV}, \ref{sec:VI}). The reference model's frequencies are shown as empty up and down triangles. The grey dotted line represents the center of the corotation band, while the green crosses depict estimated frequencies based on the assumption that the local vortensity minimum is a corotation radius (see discussion in Sec.~\ref{sec:VII}). The green dash-dotted line connects the aforementioned frequencies.}
    \label{fig:corot_sfho}
\end{figure}

\begin{figure}
    \includegraphics[width=\columnwidth]{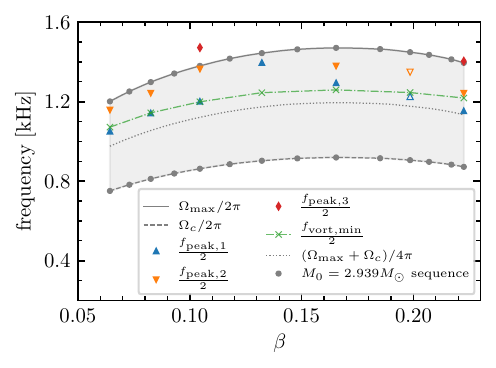}
    \caption{Same as Figure \ref{fig:corot_sfho} for the DD2 EOS. For two models, we identify additional tertiary frequencies shown as red diamonds.}
    \label{fig:corot_dd2}
\end{figure}

Figures \ref{fig:corot_sfho} and \ref{fig:corot_dd2} display the corotation band (grey shaded area; defined by the limits $\Omega_\textrm{max}/2\pi$ and $\Omega_\textrm{min}/2 \pi = \Omega_c/ 2 \pi$; note $\lambda_2=1$, see Sec.~\ref{sec:IIIA}) as a function of the rotational parameter $\beta$, for the two considered EOS models. We present the corotation band based on the equilibrium properties of all the models reported in Table~\ref{tab:SFHO_DD2_M0_const_sequence}. For the models which we evolve, we also include the pattern frequencies\footnote{We note that we extract the GW strain amplitude based on the quadrupole formula for an observer along the polar direction. The quadrupole formula extracts the $l=2$ component. In addition, along the polar axis only the $m=2$ index survives. Hence, in all GW strain plots, we show the $(l,m)=(2,2)$ contribution. For an $(l,m)$-mode with frequency $f$, the pattern speed is given by $f/m$, which explains the division of the GW frequency by a factor $2$ in the relevant plots.}, which follow from the GW frequencies identified in our time evolutions (see Secs.~\ref{sec:IV} and \ref{sec:VI}). The presence of corotation points, i.e.\ radii where the pattern speed of a mode matches the local angular velocity inside the star, relates to the rise of shear instabilities \cite{Watts_2005}. We discuss our findings in Sec.~\ref{sec:IV} and \ref{sec:VI}.

\subsection{Domain of quasi-spherical solutions}\label{terminal_models}
From Table~\ref{tab:SFHO_DD2_M0_const_sequence} we note that the SFHo sequence (i.e., the softer EOS) terminates at a value of $\beta=0.161$. This means that beyond this terminal model, it is not possible to construct configurations with slower rotation and higher density in the region of the parameter space defined by the specific EOS, differential rotation law, and rest-mass constraints. In contrast, for the DD2 sequence (i.e., the stiffer EOS), we do construct models with a slower rotation, reaching a value of $\beta=0.064$.\footnote{For DD2, we show only part of the full sequence allowed by our chosen parameters, as a thorough study of equilibrium sequences is beyond the scope of our study. However, we verified that models can be constructed up to $\epsilon_\text{max}=1.9 \times 10^{15} \text{g}/\text{cm}^3$, with $\beta \sim 0.016$. In that density region, a \textit{minimum} also appears along the $M_0=\text{const}$ sequence.} 

A possible explanation for this behaviour is the existence of different types of solutions for models of differentially rotating neutron stars (denoted Type A, B, C and D in the literature) \cite{2009MNRAS.396.2359A}. 
The types of models that can be constructed depend on the stiffness of the EOS employed, the degree of differential rotation, and the maximum density \cite{2009MNRAS.396.2359A, 2016MNRAS.463.2667S, 2017ApJ...837...58G, 2019ApJ...879...44S, 2019PhRvD..99h3017E, 2021MNRAS.503..850I, 2022MNRAS.510.2948I}.
In this work, we focus on constructing quasi-spherical (Type A) solutions, assuming a specific degree of differential rotation.
As a result, for each EOS employed and varying the rotation rate, the domain of quasi-spherical solutions is constrained in a specific density range that is different for each EOS. Beyond this range, only other types of solutions exist, such as quasi-toroidal (Type C) models where the maximum density is off-center, or Type B and D models.\footnote{While quasi-toroidal remnants have been observed in eccentric BNS merger simulations \cite{2015PhRvD..92l1502P,2016PhRvD..93b4011E}, Type B and D configurations seem less likely to occur in nature.}

\section{Results for reference models}\label{sec:IV}

We evolve the models described in Sec.~\ref{sec:RNS} for roughly $20~\mathrm{ms}$. We employ a resolution of $\mathrm{dx}=0.15$ with $257$ hydro cells/metric grid points per direction, $3$rd-order piecewise parabolic reconstruction and $3$rd-order time integration (see Appendix~\ref{app:Codeimplementation}). We devote the current section to the reference models for SFHo and DD2, while we discuss additional models with different $\beta$ values in Sec.~\ref{sec:VI}. We consider additional runs with different resolutions to evaluate the impact of numerics on our results in Appendix~\ref{app:ResStudy}. Our simulations use the fully temperature and composition dependent EOS models for both the SFHo and DD2 configurations.

\subsection{SFHo} \label{sec:res_sfho}

\subsubsection{Density modes}

To understand the dynamics of the equilibrium models, when evolved in time, we compute the density modal decomposition in the equatorial plane as (see e.g.\ \cite{2015PhRvD..92l1502P,2016PhRvD..94d3003L,2016PhRvD..93b4011E,2016PhRvD..94f4011R})
\begin{equation} \label{eq:ModeDec}
    P_m = \int_{\mathbb{R}^2} \rho W_\mathrm{L} e^{-i m \phi} \sqrt{\gamma} dx dy.
\end{equation}

\begin{figure}
    \includegraphics[width=\columnwidth]{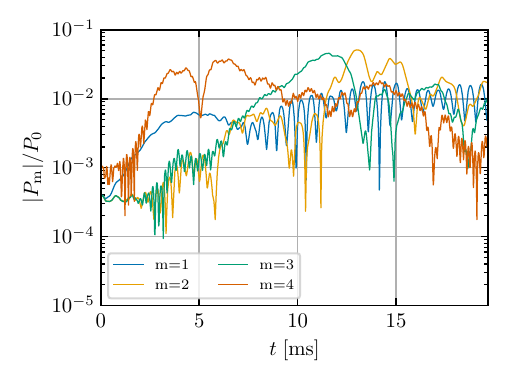}
    \caption{Amplitudes of the $m=1,2,3,4$ density modes on the equatorial plane normalized to $P_
    \mathrm{0}$ for the SFHo model.}
    \label{SFHO_ModalDecomp_plot}
\end{figure}

Figure~\ref{SFHO_ModalDecomp_plot} shows the normalized amplitude of the $m=1,2,3,4$ modes. Numerical discretization errors excite the $m=4$ mode more strongly from the beginning of the simulation, due to adopting a Cartesian grid (see e.g.\ \cite{2020PhRvD.102d4040X,2007PhRvD..75d4023B,2010CQGra..27k4104C}). As soon as the evolution starts, the amplitude of all shown modes increases. The $m=4$ mode dominates for the first roughly $9~\mathrm{ms}$ and is then surpassed by the $m=3$ mode for a few milliseconds. Around $t=13~\mathrm{ms}$, the $m=2$ mode becomes dominant, while at the later stages of the simulation the amplitude of the $m=1,2,3$ modes saturates and they all have comparable amplitudes. The growth of the modal amplitudes is typical of an instability (e.g. \cite{2007PhRvD..75d4023B,2010CQGra..27k4104C,2020PhRvD.102d4040X,2015PhRvD..92l1502P,2016PhRvD..94d3003L,2016PhRvD..93b4011E,2016PhRvD..94f4011R}).

\begin{figure*}
    \includegraphics[width=\textwidth]{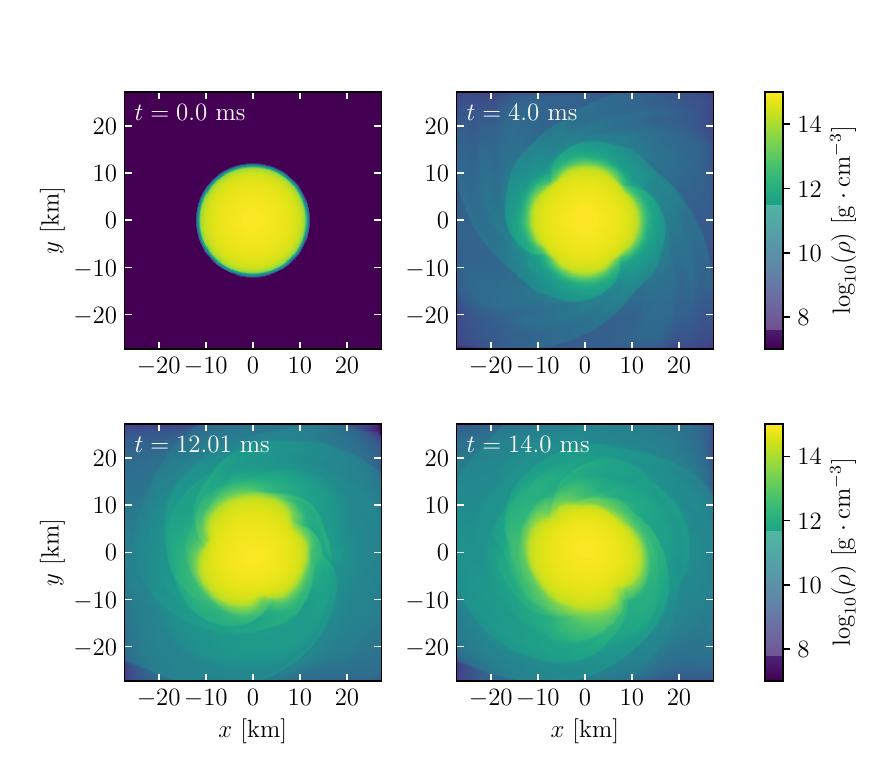}
    \caption{Rest-mass density in the equatorial plane for the SFHo configuration. Different panels refer to different times throughout the simulation. The structure of the configuration highlights that the $m=4$ (upper right), $m=3$ (lower left) and $m=2$ (lower right) modes, respectively, are more dominant at different times (see also Fig.~\ref{SFHO_ModalDecomp_plot}).}
    \label{sfho_rhosnap_plot}
\end{figure*}

To better visualize the evolution of the discussed modes, we present snapshots of the rest-mass density in the equatorial plane at different times in Fig.~\ref{sfho_rhosnap_plot}. The structure of the configuration varies considerably throughout the simulated time. In agreement with Fig.~\ref{SFHO_ModalDecomp_plot}, we identify a very pronounced $m=4$ structure at $t=4~\mathrm{ms}$ (upper right panel), which evolves initially into a $m=3$ ($t=12~\mathrm{ms}$; lower left panel) and on longer timescales a $m=2$ deformation ($t=14~\mathrm{ms}$; lower right panel). In addition, we can identify the presence of density spiral arms originating from the configuration, which push material away from the stellar model. As an estimate, we consider matter with density $\rho<10^{12}~\mathrm{g\cdot cm^{-3}}$ and find that $\approx 2.04 \times 10^{-2} ~ M_\odot$ of matter surrounds the stellar model at the end of the simulation, compared to $1.1\times 10^{-4} ~ M_\odot$ in the beginning of the simulation. Notably, the structure of the rotating model around $t=14~\mathrm{ms}$ resembles binary neutron star merger remnants during the early post-merger phase. We note though that our models have a single high-density core, while they also do not capture the thermal properties of early post-merger remnants.

\subsubsection{Gravitational waves}

\begin{figure}
    \includegraphics[width=\columnwidth]{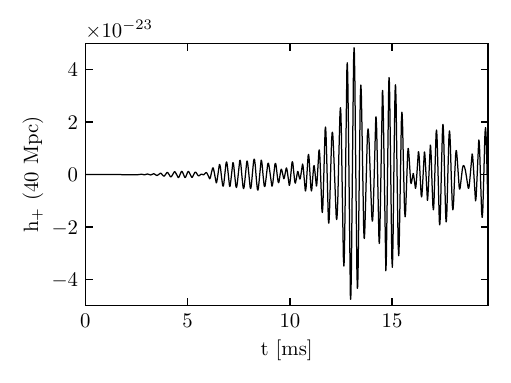}
    \caption{Plus polarization of the GW strain at a distance of $40~\mathrm{Mpc}$ along the polar direction for the SFHo model. The amplitude of the GW signal increases as the instability grows.}
    \label{SFHO_hplus_plot}
\end{figure}

The excited non-axisymmetric modes in the rotating star lead to GW emission. We extract the plus polarization $h_+$ of the GW strain amplitude as seen at a distance of $40~\mathrm{Mpc}$ along the polar direction based on the quadrupole formula (see Fig.~\ref{SFHO_hplus_plot}). The amplitude of the GW signal is very low during the initial stages of the simulation. Over the first $3.5~\mathrm{ms}$, the amplitude remains below $1\%$ of the maximum amplitude reached within the $\approx20~\mathrm{ms}$ of the evolution (i.e.\ at $t\approx13~\mathrm{ms}$). The GW emission becomes stronger, as the instability grows. In addition, we observe strong modulation in the GW signal. Overall, the amplitude of the GW signal agrees well with the evolution of the $m=2$ mode (see Fig.~\ref{SFHO_ModalDecomp_plot}).

We compare the amplitude of the GW emission in Fig.~\ref{SFHO_hplus_plot} to the amplitude of the GW re-excitations in some of the BNS merger simulations discussed in \cite{2022PhRvD.105d3020S}. We find that the amplitude of the GW emission in our simulation is lower by only some $10\%$. We note that (i) in the current work we employ the quadrupole formula to extract the GW emission, which typically underestimates the GW signal amplitude by some $10\%$ compared to the full GR extraction performed in \cite{2022PhRvD.105d3020S} (see e.g. \cite{2005PhRvD..71h4021S,2022EPJA...58...74D}) and (ii) we compare an SFHo differentially rotating model with a gravitational mass of $M=2.7~M_\mathrm{\odot}$ to BNS mergers modelled by the MPA1 EOS with gravitational masses of $M=3,3.1~M_\mathrm{\odot}$. We find similar agreement between the GW re-excitations in the BNS port-merger remnants and additional differentially rotating configurations with prominent GW emission that we evolve in this work (see Secs.~\ref{sec:res_dd2}, \ref{sec:V} and \ref{sec:VI}). This comparison adds support that the GW re-excitations in the post-merger GW signal from BNS mergers might be due to the excitation of low-$T/|W|$ instabilities.

\begin{figure}
    \includegraphics[width=\columnwidth]{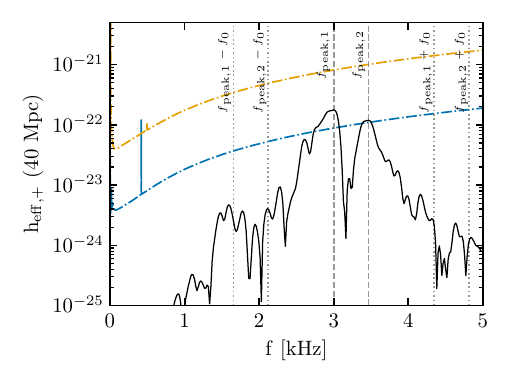}
    \caption{Spectrum of the GW signal for the SFHo stellar configuration (see Fig.~\ref{SFHO_hplus_plot}). The vertical dashed lines identify the main peaks in the spectrum. The vertical dotted lines mark the frequencies where the couplings between the main frequencies and the radial mode lie. The dash-dotted lines indicate the design sensitivities of Advanced LIGO \cite{2015CQGra..32g4001L} (upper, orange line) and the Einstein Telescope \cite{2010CQGra..27s4002P,2020JCAP...03..050M,2026JCAP...03..081A} (lower, blue line). The relevant sensitivity curves can be found in \cite{LIGO-T1800044-v5} for Advanced LIGO and in \cite{ET-D_sensitivity_curve} denoted as ET-D \cite{2011CQGra..28i4013H} for the Einstein Telescope, respectively.}
    \label{SFHO_GWspectrum_plot}
\end{figure}

\begin{figure}
    \includegraphics[width=\columnwidth]{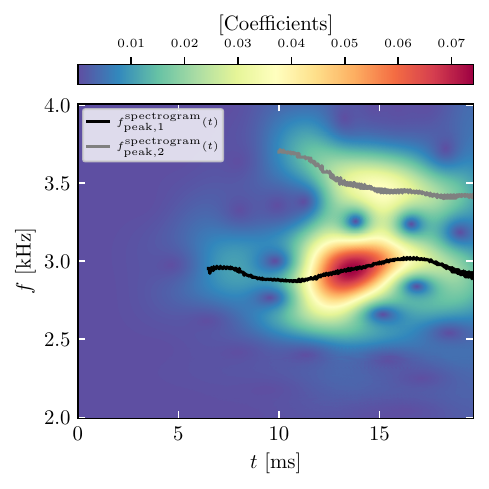}
    \caption{Spectrogram of the plus polarization of the GW strain shown in Fig.~\ref{SFHO_hplus_plot}. The black line shows the time evolution of the $f_\mathrm{peak,1}$ frequency (see Fig.~\ref{SFHO_GWspectrum_plot}) and corresponds to the maximum wavelet coefficient at any time $t$ considering only frequencies below $3.3~\mathrm{kHz}$. The grey line illustrates the evolution of the $f_\mathrm{peak,2}$ frequency and is defined similarly (restricting to $f>3.3~\mathrm{kHz}$). The black (grey) line starts at $t=6.5~\mathrm{ms}$ ($t=10~\mathrm{ms}$), when the feature can be clearly identified in the spectrogram.}
    \label{SFHO_GWspectrogram_plot}
\end{figure}

In Fig.~\ref{SFHO_GWspectrum_plot}, we compute $h_\mathrm{eff,+} = f \cdot \tilde{h}_+(f)$, where $\tilde{h}_+(f)$ is the Fourier transform of the plus polarization shown in Fig.~\ref{SFHO_hplus_plot}. We identify two pronounced peaks with frequencies $f_\mathrm{peak,1}=3.002~\mathrm{kHz}$ and $f_\mathrm{peak,2}=3.467~\mathrm{kHz}$. Notably, these frequencies are in good agreement with the dominant post-merger GW frequency from simulations of equal-mass $1.35 + 1.35~\mathrm{M}_\odot$ neutron star mergers with the SFHo EOS (e.g.\ \cite{2014PhRvD..90f2004C,2020PhRvD.101h4006K,2022arXiv220509979B}). We also remark that the superposition of two waves with frequencies $f_\mathrm{peak,1}$, $f_\mathrm{peak,2}$ should lead to an interference pattern where the total amplitude oscillates with a period of $1/|f_\mathrm{peak,1} - f_\mathrm{peak,2}|$. The observed modulation in the GW signal amplitude for $t>10~\mathrm{ms}$ exhibits a periodicity of roughly this value (see Fig.~\ref{SFHO_hplus_plot}; also Fig.~\ref{SFHO_GWspectrogram_plot} where both frequencies are present in the signal and how they evolve).

We also expect couplings between the main peaks in the GW spectrum and the quasi-radial mode. Such a coupling is also present in the post-merger GW signal of BNS mergers \cite{2011MNRAS.418..427S}. In Fig.~\ref{SFHO_GWspectrum_plot}, we indicate the frequencies $f_\mathrm{peak,1} \pm f_\mathrm{0}$, $f_\mathrm{peak,2} \pm f_\mathrm{0}$, where $f_\mathrm{0}=1.265~\mathrm{kHz}$ based on a Fourier analysis of the maximum density oscillation. In each case, we find that these frequencies match or are very close to secondary features of the GW spectrum.

\subsubsection{Corotation points}

The development of shear instabilities has been connected to the existence of corotation points within the star, i.e. points where the star's angular velocity matches the pattern speed of the unstable mode \cite{Watts_2005}. To better determine and track the evolution of the relevant pattern speeds during the simulation, we compute the spectrogram of the GW signal based on a wavelet analysis (see Fig.~\ref{SFHO_GWspectrogram_plot}). Similar to the GW spectrum, we can identify two distinct frequencies with an associated amplitude (i.e.\ wavelet coefficient) which increases as the instability grows. The feature related to $f_\mathrm{peak,1}$ exhibits a moderate amplitude already at $t\approx6.5~\mathrm{ms}$. For $t>10~\mathrm{ms}$, another component which can be associated to $f_\mathrm{peak,2}$ appears also pronounced in the spectrogram. 

We denote the time-dependent frequencies for these two features in the spectrogram as $f_\mathrm{peak,1}^\mathrm{spectrogram}(t)$ and $f_\mathrm{peak,2}^\mathrm{spectrogram}(t)$, respectively. To unambiguously define both frequencies, we impose that $f_\mathrm{peak,1}^\mathrm{spectrogram}(t)<3.3~\mathrm{kHz}$ and $f_\mathrm{peak,2}^\mathrm{spectrogram}(t)>3.3~\mathrm{kHz}$ for any given $t$. Under this condition, we extract both $f_\mathrm{peak,1}^\mathrm{spectrogram}(t)$ and $f_\mathrm{peak,2}^\mathrm{spectrogram}(t)$ as the frequency which corresponds to the maximum wavelet coefficient at any given time t. Figure~\ref{SFHO_GWspectrogram_plot} shows the evolution of both frequencies (see solid lines).

\begin{figure}
    \includegraphics[width=\columnwidth]{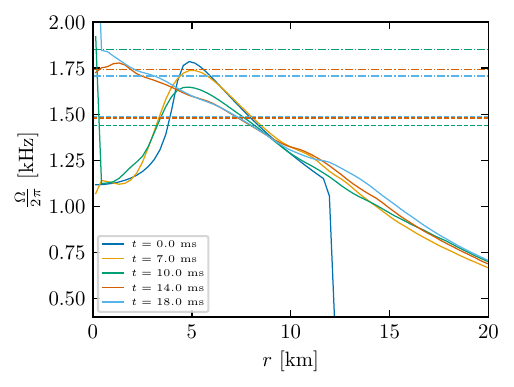}
    \caption{Rotation profile of the SFHo model as an azimuthal average in the equatorial plane. Different colors depict different times in the simulation (see legend). The horizontal dashed and dash-dotted lines depict the instantaneous value of the pattern velocities $f_\mathrm{peak,1}/2$ and $f_\mathrm{peak,2}/2$, respectively. At each time (i.e.\ comparing lines of the same color), the crossings of the horizontal lines with the angular velocity define the corotation points.}
    \label{SFHO_RotProf_plot}
\end{figure}

To identify the corotation points, we extract the instantaneous angular velocity at different times throughout the evolution. We define the angular velocity as
\begin{equation}
    \Omega = \frac{x \mathrm{v}^y - y \mathrm{v}^x}{x^2+y^2},
\end{equation}
where $\mathrm{v}^i=u^i/u^0$. We average along the azimuthal direction to obtain the rotation profile along the radial direction. Figure~\ref{SFHO_RotProf_plot} presents the rotation profile at different times. The horizontal lines depict the values of $f_\mathrm{peak,1}^\mathrm{spectrogram}(t)/2$ (dashed lines; only shown for $t>6.5~\mathrm{ms}$) and $f_\mathrm{peak,2}^\mathrm{spectrogram}(t)/2$ (dash-dotted lines; only shown for $t>10~\mathrm{ms}$)\footnote{As a reminder, the pattern frequency of an $(l,m)$-mode with frequency $f$ is given by $f/m$ and we consider the $(l,m)=(2,2)$ emission.}. Each color in Fig.~\ref{SFHO_RotProf_plot} refers to a specific time (see legend). The crossings between the angular velocity profile and the horizontal line(s) of the same color define the corotation point(s) at each shown time.

The shape of the angular velocity profile evolves throughout the simulation. The height of the off-center peak of the adopted rotation law gradually decreases, while at later stages in the simulation the central parts of the configuration exhibit high rotation rates (see also \cite{2020PhRvD.102d4040X} for a different rotation law). The evolution of the rotation profile affects the radii at which corotation points can be found. Focusing on the feature which corresponds to $f_\mathrm{peak,1}$, we find that for as long as the rotation profile admits an off-center maximum, two corotation points exist. The outer corotation radius for this mode is at roughly $7.4-8.2~\mathrm{km}$, while the inner corotation radius is at $\approx3.3~\mathrm{km}$. At later times, the outer corotation radius decreases slightly by about $0.1~\mathrm{km}$, while no inner corotation radius for this mode exists. The existence of at least one corotation point at all times sources the growth of the respective mode.

In the case of the $f_\mathrm{peak,2}$ feature, a corotation point is consistently present only at late times (see e.g. $t=14~\mathrm{ms}$) when the angular velocity near the center of the stellar model increases. Considering that the $f_\mathrm{peak,2}$ feature is present in Fig.~\ref{SFHO_GWspectrogram_plot} already at $t=10~\mathrm{ms}$, we carefully inspect times $t\le10~\mathrm{ms}$ for the existence of corotation points. We find that $f_\mathrm{peak,2}^\mathrm{spectrogram}(t)/2$ briefly intersects the rotation profile at $\approx 7.9-8~\mathrm{ms}$. We note that the rotation profile in Fig.~\ref{SFHO_RotProf_plot} is an azimuthal average, while Fig.~\ref{sfho_rhosnap_plot} clearly shows that the configuration deviates from axisymmetry. Hence, our analysis does not exclude the possibility that a local corotation point for $f_\mathrm{peak,2}$ might exist for longer than the brief time window that we identify. Overall, this observation is in agreement with the fact that the $f_\mathrm{peak,2}$ feature develops at later times than the $f_\mathrm{peak,1}$ mode and, marginally, supports the requirement of a corotation radius for the onset of the instability.

\subsubsection{Evolution of the rotation profile} \label{sec:sfho_rotprof_evol}

\begin{figure*}
    \includegraphics[width=\textwidth]{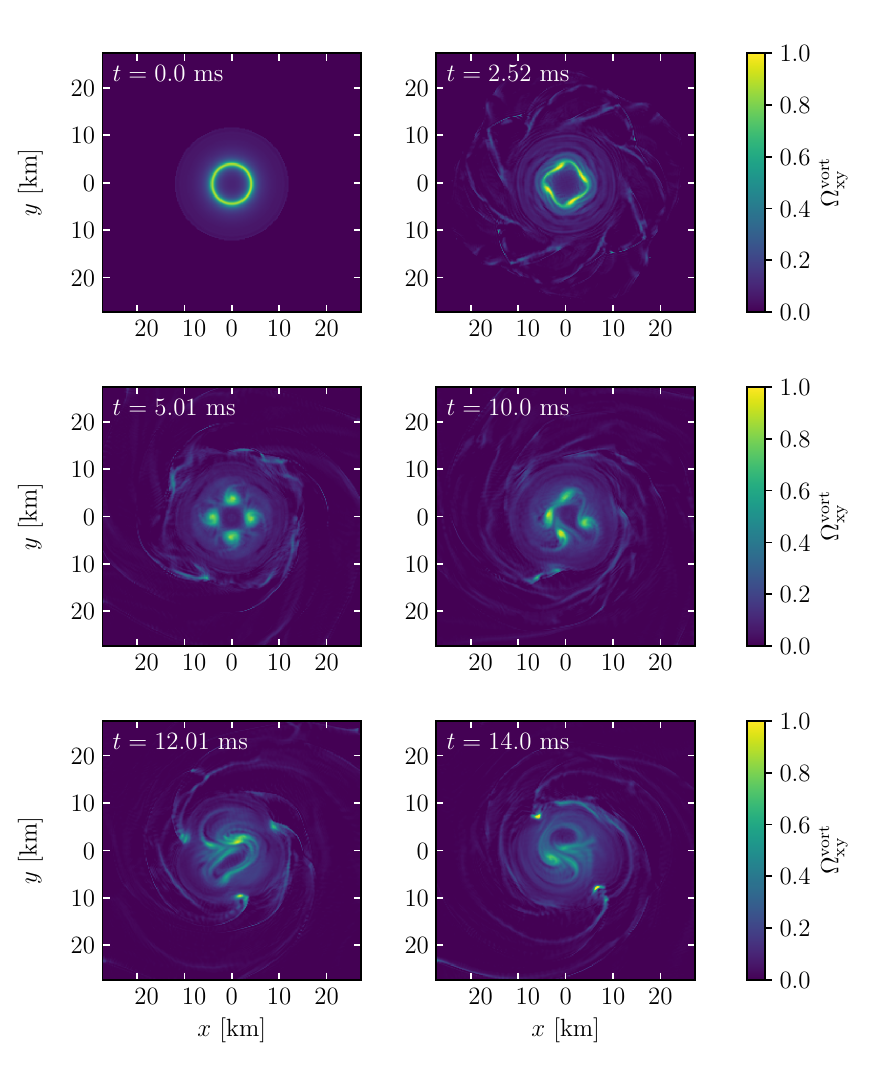}
    \caption{Vorticity snapshots in the equatorial plane for the SFHo reference model. Each panel corresponds to a different time.}
    \label{sfho_Vortsnap_plot}
\end{figure*}

Based on Fig.~\ref{SFHO_RotProf_plot}, the rotation profile exhibits an off-center peak during the first $t=10~\mathrm{ms}$ of the simulation and remains close to its initial morphology. At $t=14~\mathrm{ms}$, the shape of the angular velocity profile is drastically different. 
The maximum angular velocity moves towards the stellar center and the rotation profile evolves towards a monotonically decreasing form.
Here, we examine in closer detail the change in morphology of the rotation profile between $t=10~\mathrm{ms}$ and $t=14~\mathrm{ms}$. We define the vorticity 2-form
\begin{equation}
    \Omega_\mathrm{\mu\nu}^\mathrm{vort} = \nabla_\mu\left(h u_\nu\right) - \nabla_\nu\left(h u_\mu\right),
\end{equation}
and plot the xy-component of the vorticity at different simulation times in Fig.~\ref{sfho_Vortsnap_plot}. The vorticity distribution very quickly exhibits non-axisymmetric features (see $t=2.52~\mathrm{ms}$ panel), which become unstable and lead to the formation of 4 pronounced vortices. The early evolution of the vorticity, as well as the number of vortices that form, are primarily driven by the pronounced $m=4$ mode in the first few milliseconds (see Fig.~\ref{SFHO_ModalDecomp_plot})\footnote{
We also observe the formation of 4 vortices in the models discussed in Secs.~\ref{sec:res_dd2} and \ref{sec:VI}, where the $m=4$ mode dominates in the early stages. In simulations where we strongly excite the $m=2$ mode (Sec.~\ref{sec:V}), we observe the formation of 2 primary vortices during the first milliseconds.
}. 
The differential rotation distorts the vortices, which around $t=10~\mathrm{ms}$ start to interact. Between $t=10~\mathrm{ms}$ and $t=14~\mathrm{ms}$, the initial vortices merge and move towards the center of the stellar configuration. The vortex interaction facilitates pronounced angular momentum transport towards the inner parts of the star.

\begin{figure}
    \includegraphics[width=\columnwidth]{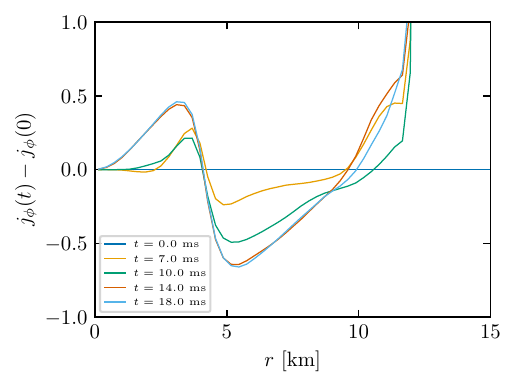}
    \caption{Azimuthal average of the specific angular momentum on the equatorial plane for the SFHo model. Different colors show different times in the simulation. The times match those in Fig~\ref{SFHO_RotProf_plot} and allow for a direct comparison.}
    \label{SFHO_specificAM}
\end{figure}

Fig.~\ref{SFHO_specificAM} shows the difference between the azimuthally averaged specific angular momentum profile at various times $j_\mathrm{\phi} (t)$ and the initial specific angular momentum profile $j_\mathrm{\phi} (0)$, where $j_\mathrm{\phi} = h u_\mathrm{\phi}$ is the specific angular momentum.
Over the first $7~\mathrm{ms}$, the specific angular momentum profile changes as angular momentum is transported from the intermediate stellar parts towards the central and outer layers.
The increase in specific angular momentum in the inner parts is mild within the first $10~\mathrm{ms}$, consistent with the overall small decrease in the height of the off-center peak and the small increase of the rotation profile in the central parts over this time window in Fig.~\ref{SFHO_RotProf_plot}.
As soon as the vortices start to interact and merge at $t=10~\mathrm{ms}$ (Fig.~\ref{sfho_Vortsnap_plot}), significant angular momentum transport towards the inner parts of the star takes place, as seen in Fig.~\ref{SFHO_specificAM} for $t>10~\mathrm{ms}$. Comparing Figs.~\ref{SFHO_RotProf_plot} and \ref{SFHO_specificAM}, the angular momentum transport explains the rise in the angular velocity profile that we observe at late times.

\subsubsection{Rest-mass density eigenfunction and vortensity}

\begin{figure}

    \subfloat[]{%
         \includegraphics[width=\columnwidth]{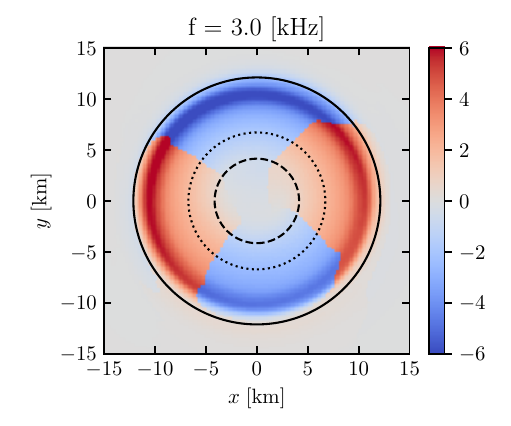}\label{fig:fft_equator_sfho_1}}
    
    \vspace*{8pt}%
    
    \subfloat[]{%
        \includegraphics[width=\columnwidth]{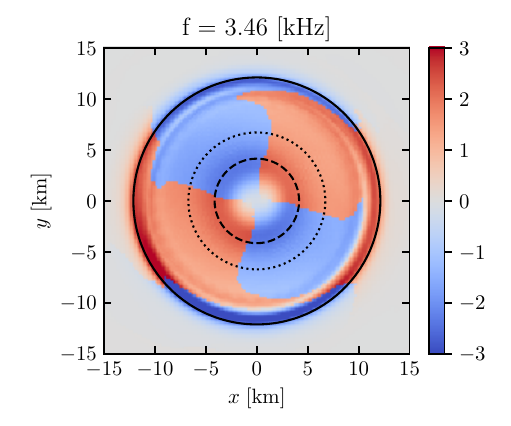}\label{fig:fft_equator_sfho_2}}

    \caption{Slice in the equatorial plane of the extracted m = 2 mode eigenfuctions for the density oscillations with a frequency of $3~\mathrm{kHz}$ (panel \protect\subref{fig:fft_equator_sfho_1}) and $3.46~\mathrm{kHz}$ (panel \protect\subref{fig:fft_equator_sfho_2}). Both panels refer to the SFHo model. The black circles depict the local maximum (dashed) and local minimum (dotted) of the Newtonian vortensity of the reference model in the equatorial plane, as well as the stellar surface (solid) of the reference model. The color scale has only a relative meaning, while the range is based on the values in the region $r<10$ km so that features within the star are easier to identify.}
    \label{fig:fft_equator_sfho}
\end{figure}

To better understand the occurrence of the two modes, we perform a Fourier analysis of the rest-mass density evolution on the equatorial plane (see e.g.\ \cite{2004MNRAS.352.1089S,2011MNRAS.418..427S}). During the evolution, we output the rest-mass density in each cell of the equatorial plane every $\approx 0.02~\mathrm{ms}$. For each cell, we record a time-series of the density evolution throughout the simulation and perform a Fourier analysis of each such time-series. We extract the amplitude of the Fourier spectrum at the (fixed) frequency of the mode at each cell. The amplitude is then colorcoded at each cell in the equatorial plane in Fig.~\ref{fig:fft_equator_sfho} as an estimate of the eigenfunction. We multiply the amplitude by the sign of the real part of the Fourier transform to correctly capture the nodal lines.

Figure~\ref{fig:fft_equator_sfho} shows the slice of the eigenfunction in the equatorial plane of the 2 modes with frequencies $3~\mathrm{kHz}$ (panel \ref{fig:fft_equator_sfho_1}) and $3.46~\mathrm{kHz}$ (panel \ref{fig:fft_equator_sfho_2}). In both panels, the black solid line depicts the surface of the equilibrium model. In addition, we plot the local maximum (black dashed line) and local minimum (black dotted line) of the Newtonian vortensity profile at the equatorial plane (Eq.~\eqref{eq:vor}). 

\begin{figure}
    \includegraphics[width=\columnwidth]{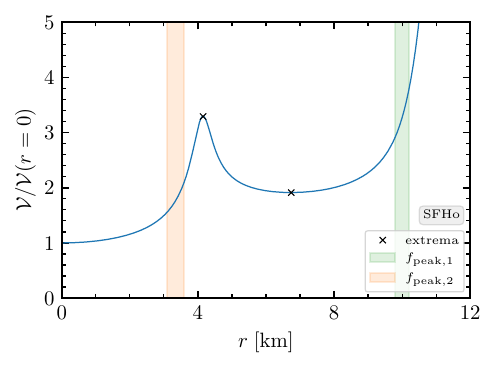}
    \caption{Newtonian vortensity at the equatorial plane versus the coordinate radius $r$ for the SFHo reference model. 
    The vertical shaded regions indicate the locations of the peaks of the eigenfunctions for each respective mode. The region widths follow the analysis discussed in the main text (i.e. the width at $99\%$ of the main peak height). The black crosses mark the minimum and maximum of the vortensity profile.}
    \label{fig:vortensity_sfho}
\end{figure}

The development of unstable modes has been connected to the vortensity profile of the stellar model. In \cite{2006ApJ...651.1068O}, the growth timescale of unstable modes has been associated with the distance of the corotation radius from the minimum of the vortensity well for the $j$-constant rotation law, which yields vortensity profiles with a single local minimum.
In \cite{2020PhRvD.102d4040X}, the authors find a vortensity profile that exhibits both a local minimum and maximum for a differential rotation law with an off-center angular velocity peak and discuss the connection between the vortensity local extrema, the angular velocity profile and the development of unstable modes. In Fig.~\ref{fig:vortensity_sfho}, we depict the Newtonian vortensity defined as the ratio between the radial vorticity and the rest-mass density 
\begin{equation}\label{eq:vor}
    \mathcal{V} = \frac{1}{\rho} \left( 2 \Omega + \varpi \partial_\varpi \Omega \right),
\end{equation}
where $\varpi$ is the radial cylindrical coordinate \cite{2010CQGra..27k4104C, 2020PhRvD.102d4040X}. We note that in the equatorial plane $\theta = \pi/2$, and therefore $\varpi = r \sin\theta = r$. Similar to \cite{2020PhRvD.102d4040X}, our vortensity profiles exhibit a local maximum and a local minimum away from the stellar center and surface.

In Fig.~\ref{fig:fft_equator_sfho}, we find that the modes corresponding to both frequencies exhibit the expected $m=2$ structure. Even though the peaks in the GW spectrum are relatively broad, they allow for the identification of the respective modes. Focusing on the $f_\mathrm{peak,1}$ mode (upper panel), we find that the mode is hardly present in the inner parts of the star, as its amplitude decreases noticeably at radial distances below the position of the local maximum of the vortensity profile. The amplitude of the mode seems to be most pronounced between the local minimum of the vortensity plot and the surface. Based on \ref{fig:fft_equator_sfho_2}, the $f_\mathrm{peak,2}$ mode is present throughout almost the whole star, excluding the central $\approx 1~\mathrm{km}$ region. We find that the mode is most pronounced right within the local maximum of the vortensity and at the stellar surface. We note that, by inspecting similar plots for frequencies that do not match any modes, we find that at(/right within) the surface the amplitude of the Fourier spectrum can be sizeable. We consider this to be noise, possibly related to the fact that the stellar surface changes significantly throughout the simulation, which makes it difficult to tie a single mode to this region. The analysis seems to suggest that the $f_\mathrm{peak,1}$ mode is primarily constrained within the vortensity well around the local minimum, while the $f_\mathrm{peak,2}$ mode eigenfunction is more pronounced in the parts of the star close and below the local vortensity maximum (considering the signal at the surface to be noise). We emphasize that the extrema refer to the Newtonian vortensity and that the vortensity profile is computed for the initial configuration. Thus, we do not consider how the dynamic evolution of the stellar configuration throughout the simulation affects the vortensity profile.

We also investigate where the respective eigenfunction peaks within the star. We azimuthally average the amplitude of each mode shown in Fig.~\ref{fig:fft_equator_sfho} to produce an averaged amplitude versus radius curve\footnote{We average the amplitude, which is a positive number. Namely, we do not include the sign shown in Fig.~\ref{fig:fft_equator_sfho}.}. As an attempt to eliminate the noise close to the surface, we perform the same type of analysis for frequencies in the range $[0.5, 10]~\mathrm{kHz}$ with a step of $100~\mathrm{Hz}$. We average the $95$ slices to obtain a frequency-independent background, which we subtract from the 2 eigenfunctions shown in Figs.~\ref{fig:fft_equator_sfho_1} and \ref{fig:fft_equator_sfho_2}. In both cases, we can identify a main peak after removing the background noise. Considering that the peaks are relatively broad, for each mode we compute the width at $99\%$ of the height of the main peak. We report the location of this area in Fig.~\ref{fig:vortensity_sfho} as a rough estimate of where the eigenfunction of each mode peaks. We emphasize that this procedure serves only as an estimate. Based on this analysis, the $f_\mathrm{peak,1}$ mode appears primarily outside the local minimum of the vortensity, while the $f_\mathrm{peak,2}$ mode lies within the local maximum (see also related discussion in Secs.~\ref{sec:res_dd2} and \ref{sec:V}).

\subsection{DD2} \label{sec:res_dd2}
\begin{figure}
    \includegraphics[width=\columnwidth]{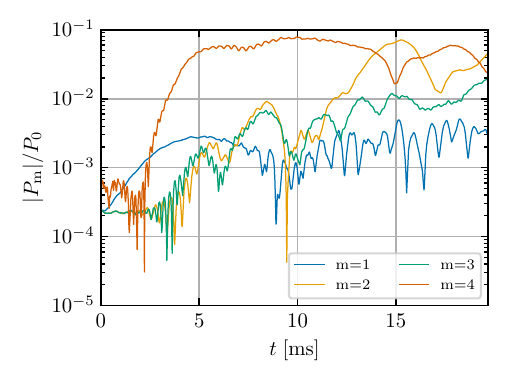}
    \caption{Density modal decomposition on the equatorial plane for the DD2 model. The normalized amplitudes for the $m=1,2,3,4$ modes are shown.}
    \label{DD2_ModalDecomp_plot}
\end{figure}

We perform an additional simulation for a stellar model described by the DD2 EOS and adopting the same rotation law to evaluate the dependence of our results on the EOS. Similar to the analysis in Sec.~\ref{sec:res_sfho}, we perform a modal decomposition of the density on the equatorial plane (see Eq.~\ref{eq:ModeDec}). Figure~\ref{DD2_ModalDecomp_plot} presents the normalized amplitudes of the $m=1,2,3,4$ modes. The evolution of the $m=1$ and $m=2$ modes exhibits a behaviour which is very similar to that of the SFHo model (see Fig.~\ref{SFHO_ModalDecomp_plot}). Namely, the $m=1$ shows an early increase and then roughly saturates for the duration of the simulation. At the same time, the $m=2$ gradually increases and around $t=15~\mathrm{ms}$ reaches its maximum amplitude. The $m=3$ and $m=4$ modes also progress rather similarly to the SFHo configuration over the first $\approx7~\mathrm{ms}$. The main differences arise at later times, where for the DD2 model the amplitude of the $m=4$ mode approximately saturates. Furthermore, the $m=3$ mode gradually grows, but becomes competitive with the even modes only at the latest stages of the simulation. As a result, the DD2 model displays an $m=4$ deformation throughout the whole simulation.

\begin{figure}
    \subfloat[]{%
        \includegraphics[width=\columnwidth]{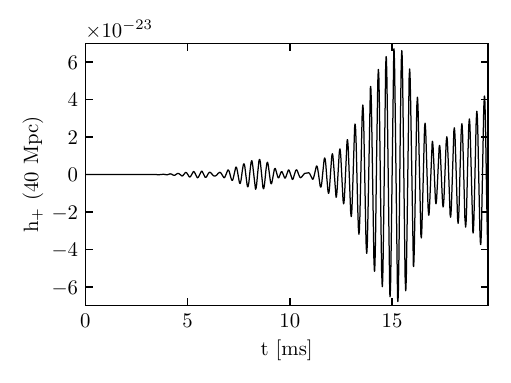}\label{DD2_hplus_plot}}

    \vspace*{8pt}%
    
    \subfloat[]{%
         \includegraphics[width=\columnwidth]{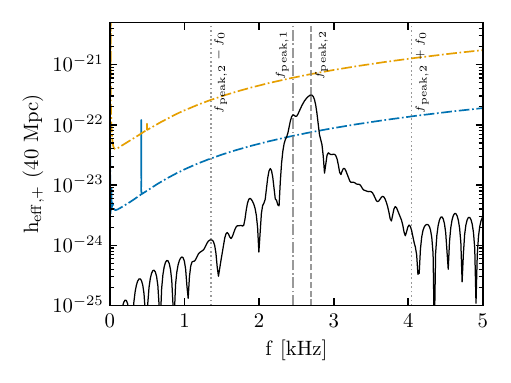}\label{DD2_GWspectrum_plot}}
    
    \caption{Panel \protect\subref{DD2_hplus_plot} shows the plus polarization of the GW signal at a distance of $40~\mathrm{Mpc}$. Panel \protect\subref{DD2_GWspectrum_plot} displays the respective GW spectrum. The dash-dotted lines illustrate the design sensitivity of the Advanced LIGO (upper line) and the Einstein Telescope (lower line). Both panels refer to the simulation with the DD2 EOS.}
    \label{dd2_GW_plots}
\end{figure}

Figure~\ref{DD2_hplus_plot} displays the plus polarization of the quadrupole emission. In a similar fashion to the SFHo model (see Fig.~\ref{SFHO_hplus_plot}), as the instability grows, the GW strain amplitude increases significantly and reaches its maximum values at $t \approx 15~\mathrm{ms}$. The maximum amplitude is about $40\%$ greater than the one extracted from the SFHo simulation. The signal features some modulation, particularly at the late stages of the simulation. The modulation in frequency is lower in comparison to the SFHo configuration.

\begin{figure}
    \includegraphics[width=\columnwidth]{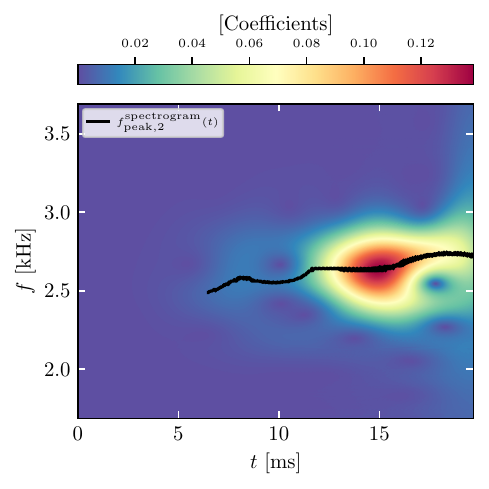}
    \caption{Spectrogram of the the GW strain shown in Fig.~\ref{DD2_hplus_plot}. The black line shows the time evolution of the main frequency (see Fig.~\ref{DD2_GWspectrum_plot}) defined as the maximum wavelet coefficient at any time. The black line starts at $t = 6.5~\mathrm{ms}$.}
    \label{DD2_GWspectrogram_plot}
\end{figure}

A Fourier analysis of the GW signal (see Fig.~\ref{DD2_GWspectrum_plot}) reveals a pronounced peak at $\approx 2.7~\mathrm{kHz}$ (denoted as $f_\mathrm{peak,2}$). Notably, this frequency is in excellent agreement with the dominant post-merger GW emission of equal-mass $1.35 + 1.35~\mathrm{M}_\odot$ neutron star mergers modelled with the DD2 EOS, see e.g.\ \cite{2015PhRvD..91l4056B,2017ApJ...842L..10R}. In addition, we mark the frequencies where we expect the couplings of the main frequency to the radial mode ($f_\mathrm{0} = 1.345~\mathrm{kHz}$) and find that they match small scale features of the spectrum. Contrary to the SFHo simulation, a pronounced second peak is not present in the spectrum. However, we note that the main peak seems to exhibit some substructure, particularly a second smaller peak is visible at roughly $2.45~\mathrm{kHz}$ (denoted as $f_\mathrm{peak,1}$). This secondary feature is more difficult to identify due to its overlap with the main peak, but we mark it because it is relevant for the discussion in Sec.~\ref{sec:V}. The frequency difference between $f_\mathrm{peak,2}$ and $f_\mathrm{peak,1}$ of about 150~Hz is compatible with the modulation of the signal in the time domain showing individual maxima at 8~ms and 15~ms. For completeness, we also include the respective spectrogram in Fig.~\ref{DD2_GWspectrogram_plot}.

\begin{figure}
    \includegraphics[width=\columnwidth]{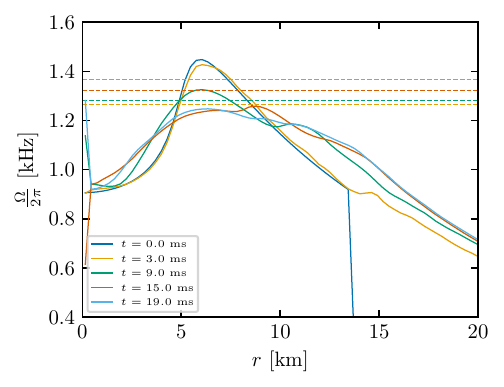}
    \caption{Azimuthally-averaged angular velocity of the DD2 model in the equatorial plane. The horizontal dashed lines display the instantaneous pattern frequency (extracted from a spectrogram of the GW signal for $f_\mathrm{peak,2}$). Intersection points between the rotation profile lines and the horizontal lines of the same color designate the corotation points.}
    \label{DD2_RotProf_plot}
\end{figure}

We present the evolution of the rotation profile in Fig.~\ref{DD2_RotProf_plot}. Throughout the simulation, the height of the off-center maximum gradually decreases. The inner parts of the star, as well as the regions close to the surface (further out than the original maximum of the angular velocity), rotate at a higher rate at later times. Contrary to the SFHo model, the rotation rate near the center remains close to the original rotation law and the maximum of the angular velocity is off-center for the whole duration of the simulation.

\begin{figure*}
    \includegraphics[width=\textwidth]{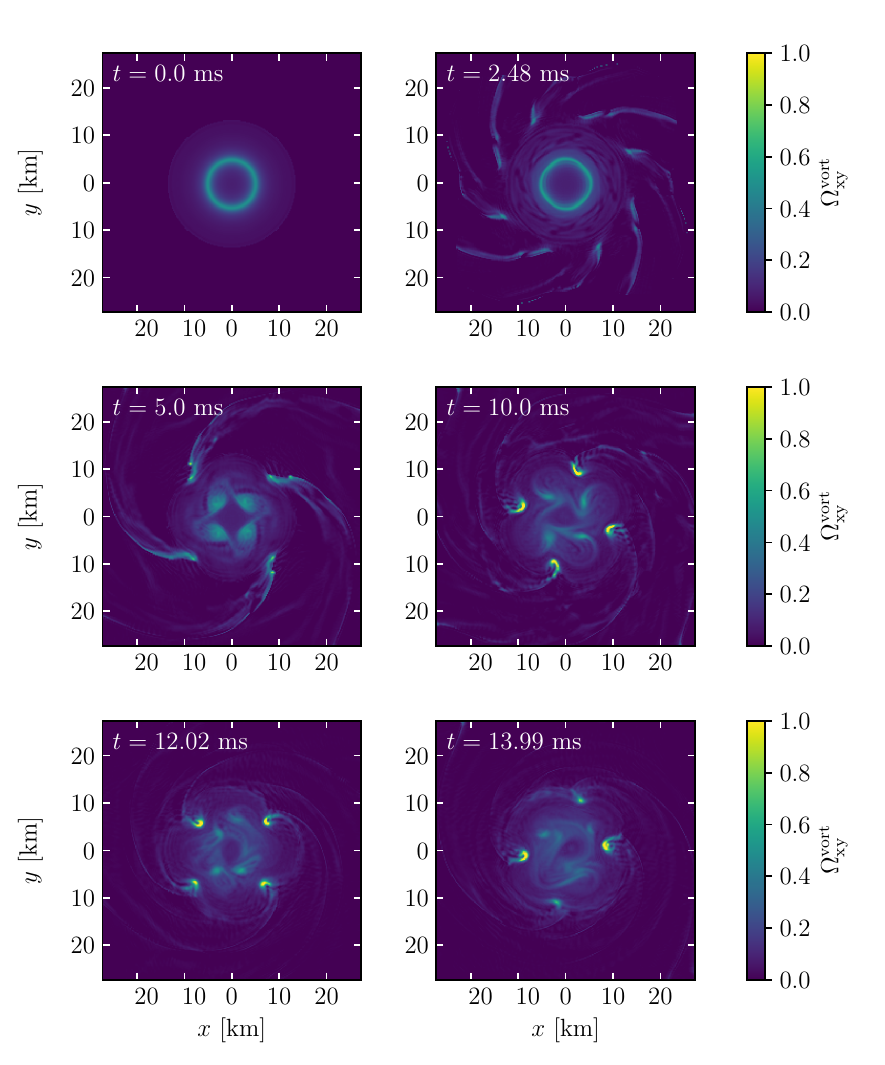}
    \caption{Vorticity snapshots in the equatorial plane for the DD2 reference model. Different panels correspond to different times, matching roughly the times shown in Fig.~\ref{sfho_Vortsnap_plot}.}
    \label{dd2_Vortsnap_plot}
\end{figure*}

The evolution of the vorticity, shown in Fig.~\ref{dd2_Vortsnap_plot}, reveals that 4 primary vortices form in the DD2 stellar configuration, similar to the SFHo reference model (see Fig.~\ref{sfho_Vortsnap_plot}). Compared to the SFHo model, the main vortices exhibit lower vorticity values and we observe less vortex interaction in the DD2 model. The evolution of the vortices leads to some angular momentum transport towards the central parts of the star, as seen for $t\ge 9~\mathrm{ms}$ both in Figs.~\ref{DD2_RotProf_plot} and~\ref{dd2_Vortsnap_plot}. We find that the value of the vorticity in the inner parts of the star is only mildly increased at later times, while the value of the vorticity at the center of the star remains always close to its original value. This aligns with the evolution of the rotation profile, which does feature a small increase in the inner layers of the star at later times, but the angular velocity peak remains off-center throughout the whole evolution. We note that the evolution of the vorticity depends heavily on the dynamics of the stellar configuration (see Sec.~\ref{sec:V}). 
In addition, the observed behaviour does not depend solely on the EOS.
In stellar models described by the DD2 EOS, the angular velocity profile can also evolve to feature a central peak (based on inspection of the additional models discussed in Sec.~\ref{sec:VI}).

\begin{figure}
    \subfloat[]{%
         \includegraphics[width=\columnwidth]{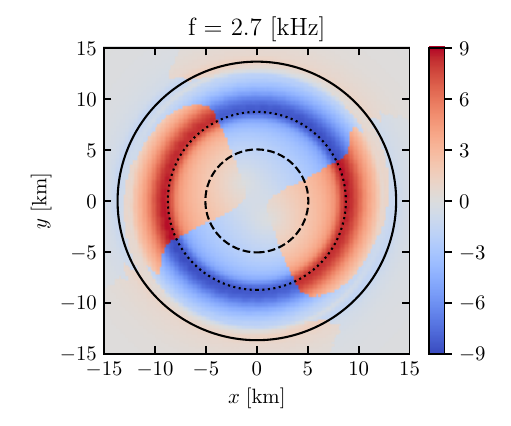}\label{fig:fft_equator_dd2}}
    
    \vspace*{8pt}%
    
    \subfloat[]{%
        \includegraphics[width=\columnwidth]{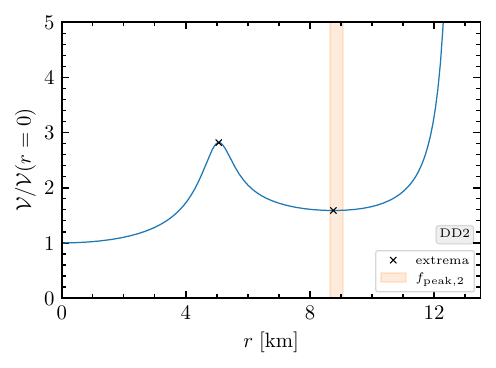}\label{fig:vortensity_dd2}}

    \caption{Panel \protect\subref{fig:fft_equator_dd2}: Same as Fig.~\ref{fig:fft_equator_sfho} for the DD2 reference model and $2.7~\mathrm{kHz}$. Panel \protect\subref{fig:vortensity_dd2}: Same as Fig.~\ref{fig:vortensity_sfho} for the DD2 reference configuration.}
    \label{fig:fft_equator_dd2_total}
\end{figure}

We can identify two corotation points over the first roughly $10~\mathrm{ms}$ for the $f_\mathrm{peak,2}$ mode. At later times, the azimuthally-averaged angular velocity does not admit corotation points. The inner corotation radius is roughly constant and located at $4.9~\mathrm{km}$, while the outer corotation radius varies between $7.5-8.8~\mathrm{km}$. We perform a similar Fourier analysis as in Sec.~\ref{sec:res_sfho} to extract the eigenfunction of the $2.7~\mathrm{kHz}$ mode (see Fig.~\ref{fig:fft_equator_dd2}). We find that the eigenfunction exhibits a characteristic $m=2$ structure. The mode's amplitude drops only for the innermost $\approx 2~\mathrm{km}$, i.e.\ it extends from within the radius corresponding to the local vortensity maximum all the way to the surface. Removing the frequency-independent noise, we find that the peak of the eigenfuction is located very close to the minimum of the reference model's vortensity profile. We depict the radial area where the azimuthally-averaged amplitude is at least $99\%$ of its maximum in Fig.~\ref{fig:vortensity_dd2}. In the case of the DD2 model, the eigenfunction peak is less affected by the surface noise than in the SFHo case, which might allow for a better localization of the peak.

\section{Dynamics for fully developed instabilities}\label{sec:V}

In Sec.~\ref{sec:IV}, we evolve the models for $20~\mathrm{ms}$. This time window suffices to capture the growth timescale of the instability, which however is comparable to the total simulation time. The instability growth rate differs for the two employed EOSs, which also affects the observed general dynamics. For example, the rotation profile at the latest stages of the SFHo and DD2 simulations (see Figs.~\ref{SFHO_RotProf_plot} and \ref{DD2_RotProf_plot}, respectively) admits a different morphology. Notably, the maximum rotation rate is found at the central region for the SFHo configuration, while it remains off-center for the DD2 model.

We perform an additional simulation for each EOS to investigate how the stellar configurations respond to a fully developed instability. To emulate the dynamics after the instability has fully developed, we add a velocity perturbation to the initial data which should strongly excite the quadrupolar fluid oscillation mode. The form of the perturbation is (see e.g.\ \cite{2016EPJA...52...56B})
\begin{equation}
    \delta \upsilon_\theta = 0.4 \sin{\left( \pi \frac{r}{r_\mathrm{surface}(\theta)} \right)} \sin{(\theta)} \cos{(\theta)} \cos{(2 \phi)},
\end{equation}
where $r_\mathrm{surface}(\theta)$ is the coordinate radius as a function of the polar angle. We opt for a rather high perturbation amplitude (i.e.\ $0.4$) to ensure that the mode develops right away. We note that the perturbation is not exact, but captures the angular dependence of the mode (see Figs~\ref{fig:fft_equator_sfho} and \ref{fig:fft_equator_dd2}). Hence, it can excite additional modes as well. We employ a resolution of $\mathrm{dx}=0.25$ in these two additional simulations to reduce the required computational effort. Based on the findings in Appendix~\ref{app:ResStudy}, this resolution suffices to capture the main dynamics. The main deviation from higher resolution results lies in how quickly the instability develops (primarily in the DD2 model), which we already eliminate by introducing a strong initial excitation.

\begin{figure*}
    \subfloat[]{%
        \includegraphics[width=\columnwidth]{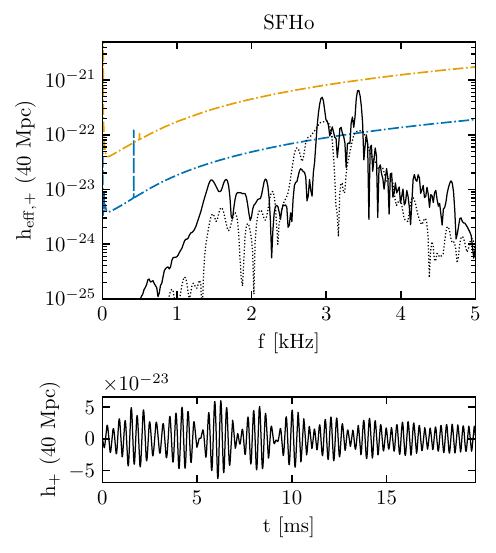}\label{hplusSpectrum_SFHO}}
    \hfill
    \subfloat[]{%
        \includegraphics[width=\columnwidth]{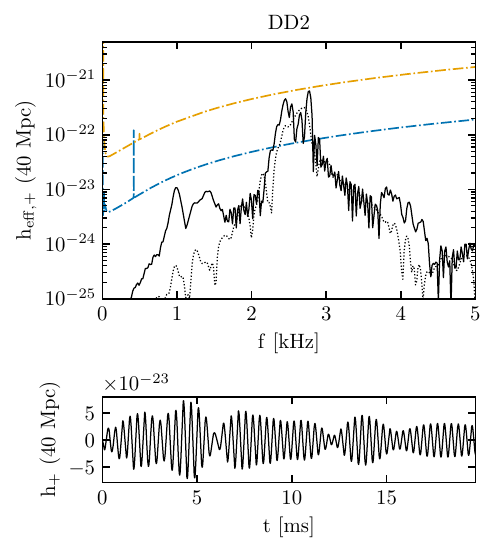}\label{hplusSpectrum_DD2}}
   
    \caption{Plus polarization of the GW strain (bottom figures) and respective GW spectrum (upper figures) for the SFHo (panel \protect\subref{hplusSpectrum_SFHO}) and DD2 (panel \protect\subref{hplusSpectrum_DD2}) models including a velocity perturbation to excite the quadrupolar oscillation. In each GW spectrum plot, the solid line refers to the model which includes the perturbation. The dotted line is the spectrum from the corresponding main simulation without any excitation (i.e. they match Figs.~\ref{SFHO_GWspectrum_plot} and \ref{DD2_GWspectrum_plot} for panels \protect\subref{hplusSpectrum_SFHO} and \protect\subref{hplusSpectrum_DD2}, respectively). The upper (orange) and lower (blue) dash-dotted lines show the design sensitivity of the Advanced LIGO and Einstein Telescope detectors.}\label{hplusSpectrum_Pert}
\end{figure*}

Figure~\ref{hplusSpectrum_Pert} presents the GW signal from each simulation, alongside the respective spectrum. The bottom panels show the GW strain for SFHo (panel~\ref{hplusSpectrum_SFHO}) and DD2 (panel~\ref{hplusSpectrum_DD2}). In both cases, the amplitude of the quadrupole GW component increases as soon as the simulation starts due to the strong excitation of the $l=2$ mode. At late times, the amplitude saturates and preserves a roughly constant amplitude which is expected since our simulations do not include GW emission as a damping mechanism. For both EOSs, the GW strain amplitude exhibits modulation.

The upper panels of Fig.~\ref{hplusSpectrum_Pert} display the GW spectrum for each EOS. In each plot, the solid line presents the spectrum of the simulation which includes the initial perturbation, while the dotted line illustrates the spectrum shown in Figs.~\ref{SFHO_GWspectrum_plot} and \ref{DD2_GWspectrum_plot} for SFHo and DD2 (i.e. without exciting the mode), respectively. Focusing on the SFHo EOS, the two lines are in rather good agreement. A double peak structure is present in the spectrum, regardless of whether the quadrupolar fluid oscillation is excited by hand or not. The positions of the peaks match quite well. We extract frequencies of $2.941~\mathrm{kHz}$ and $3.429~\mathrm{kHz}$ based on the solid line and $3.002~\mathrm{kHz}$, $3.467~\mathrm{kHz}$ for the dotted line (see Sec.~\ref{sec:res_sfho}). In the case of DD2 (i.e. Fig.~\ref{hplusSpectrum_DD2}), the spectrum of the simulation including the perturbation also features a double peak structure. The two peaks lie at frequencies $2.452~\mathrm{kHz}$ and $2.769~\mathrm{kHz}$. This analysis reveals that the main peak at $2.7~\mathrm{kHz}$ found in Sec.~\ref{sec:res_dd2}, may include the contribution from two distinct modes. However, the lower-frequency mode does not have enough time to develop in the main simulation discussed in Sec.~\ref{sec:res_dd2} and form a pronounced distinct feature in the spectrum (see Fig.~\ref{DD2_GWspectrum_plot}).

\begin{figure}
    \includegraphics[width=\columnwidth]{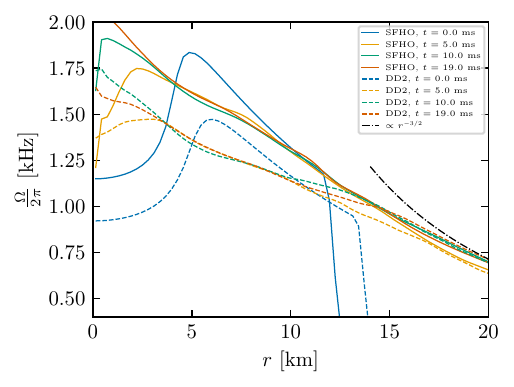}
    \caption{Rotation profile of the SFHo and DD2 models including a velocity perturbation to excite the quadrupolar fluid mode. The solid lines refer to the SFHo model, while the dashed lines illustrate the profile of the DD2 configuration. Different colors are employed to depict different times (see legend). The black dash-dotted line depicts an analytic law which scales as $r^{-3/2}$ for reference.}
    \label{RotProf_Pert_plot}
\end{figure}

Introducing the velocity perturbation leads to a similar evolution of the rotation profile for both EOSs. Figure~\ref{RotProf_Pert_plot} displays the rotation profile at different times. Independently of the EOS, the off-center peak present at $t=0~\mathrm{ms}$ gradually travels inwards in the star. After $10~\mathrm{ms}$, the central region exhibits the highest rotation rate for both configurations. Inspecting the vorticity distribution for both models, we find that the induced quadrupole perturbation leads to the formation of 2 main vortices (not shown). The vortices interact and merge at the center of the star and the accompanying angular momentum transport explains the evolution of the angular velocity profile (see similar discussion in Sec.~\ref{sec:sfho_rotprof_evol}). The angular velocity in the region outside both the stellar models (here defined as $r\ge14~\mathrm{km}$, i.e.\ outside the radius of the initial DD2 configuration) is very similar between the two models and tends to a $\propto r^{-3/2}$ Keplerian rotation law. Furthermore, even though the stellar interior rotates more rapidly for the soft EOS, the orbital frequency for $r\gtrsim 15~\mathrm{km}$ is practically the same for both EOSs (see also e.g.\ \cite{2017PhRvD..96d3004H} for similar behaviour in BNS merger remnants).

\begin{figure}
    \includegraphics[width=\columnwidth]{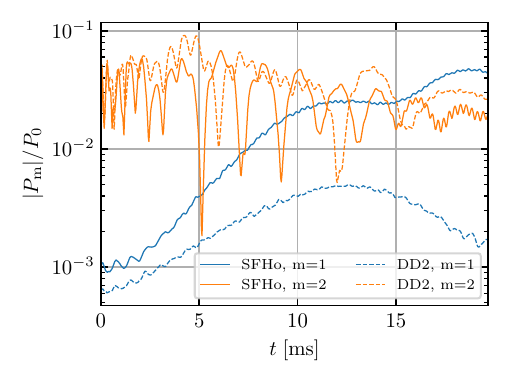}
    \caption{Normalized amplitudes of the $m=1,2$ modes for the SFHo and DD2 configurations including a velocity perturbation in the initial data.}
    \label{ModalDecomp_Pert_plot}
\end{figure}

Finally, we perform a modal decomposition of the density to better understand the dynamics. Figure~\ref{ModalDecomp_Pert_plot} presents the $m=1$ and $m=2$ modes. In agreement with the GW signal, the $m=2$ mode is strongly excited due to the induced perturbation and sustains a high amplitude throughout the whole evolution for both EOSs. Interestingly, the $m=1$ mode develops over time in the SFHo model and at late times becomes more dominant than the $m=2$ mode. Notably, the $m=1$ mode develops and becomes competitive or surpasses the $m=2,3,4$ modes in all the discussed simulations with SFHo (see Fig.~\ref{SFHO_ModalDecomp_plot} and Appendix \ref{app:ResStudy}). In contrast, the $m=1$ lies always below the $m=2$ in the DD2 configuration and does not exhibit a monotonically increasing trend. This result might suggest that soft EOSs are more prone to develop a pronounced $m=1$ mode (see though also \cite{2024MNRAS.527.8812J,2026PhRvD.113b3011G}).

\section{Additional models with different $\beta = T/|W|$ values}\label{sec:VI}

In this section, we examine constant rest-mass sequences of models for both SFHo and DD2, with $M_\mathrm{0}$ matching the value of the respective reference model. We investigate how the value of $\beta$ impacts the results that we report for the reference models in Sec.~\ref{sec:IV}. Our numerical setup details are the same as these employed in Sec.~\ref{sec:IV}.

\subsection{GW signal and the $m=2$ mode}\label{sec:extra_models_gw_signal}
We first focus on the GW signal and the development of the $m=2$ mode. We note that we employ the CFC approximation, which underestimates the GW signal amplitude by some $10\%$ \cite{2005PhRvD..71h4021S,2022EPJA...58...74D}. We do not account for this underestimation in the GW spectra presented in this work. Furthermore, we do not include a GW extraction scheme in our calculations. If the GW damping timescale is comparable to the simulated time (i.e.\ $\approx 20$ ms), it can in principle reduce the height of the various features in the GW spectrum. We employ the same setting in all our simulations, which allows for a fair comparison between different runs. In the following discussion, we compare our spectra to GW detector sensitivity curves, which allows us to highlight the differences between simulations and assess the potential observability.

\begin{figure*}
    \subfloat[]{%
        \includegraphics[width=\columnwidth]{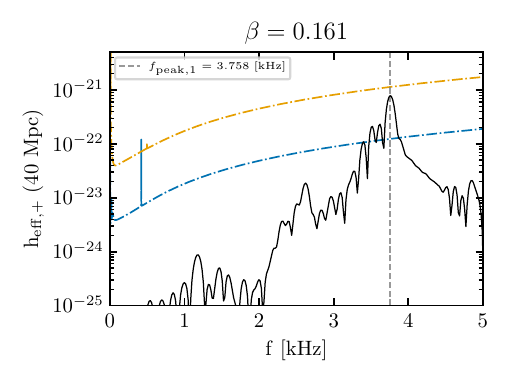}\label{DiffModels_SFHO_1_Log}}
    \hfill
    \subfloat[]{%
        \includegraphics[width=\columnwidth]{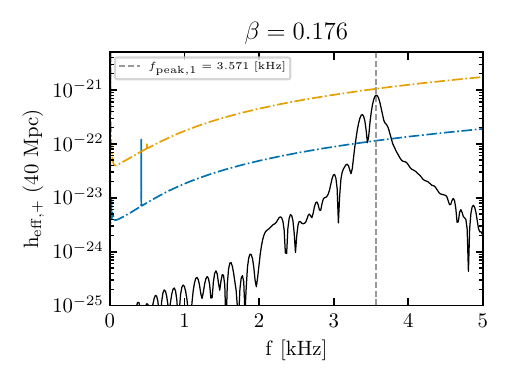}\label{DiffModels_SFHO_2_Log}}

    \vspace*{2pt}%
    
    \subfloat[]{%
         \includegraphics[width=\columnwidth]{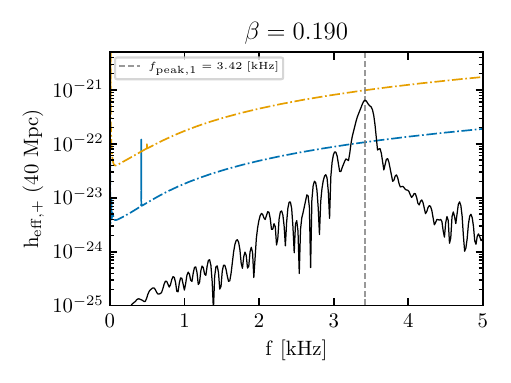}\label{DiffModels_SFHO_3_Log}}
    \hfill
    \subfloat[]{%
        \includegraphics[width=\columnwidth]{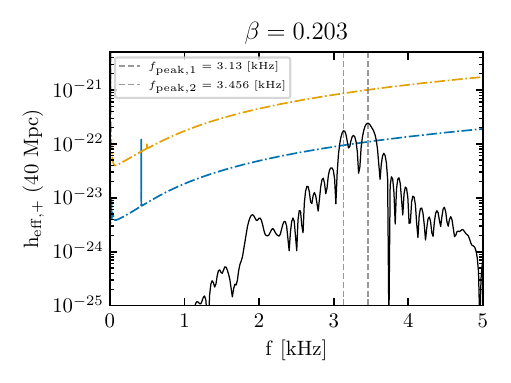}\label{DiffModels_SFHO_4_Log}}

    \vspace*{2pt}%

    \subfloat[]{%
         \includegraphics[width=\columnwidth]{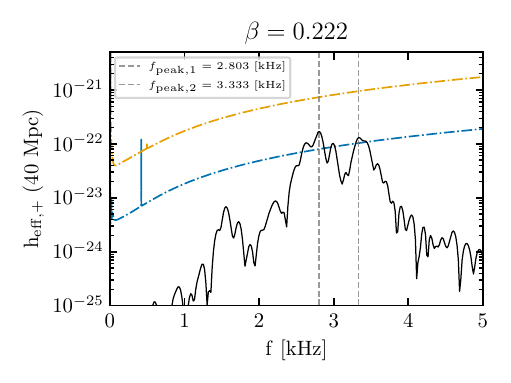}\label{DiffModels_SFHO_5_Log}}
   
    \caption{Each panel refers to a model with different $\beta$ for the SFHo EOS (see panel title). The figures depict the GW spectrum in logarithmic scale. The vertical lines identify the positions of modes (see also Table~\ref{tab:frequencies} and Fig.~\ref{SFHO_GWspectrum_plot}).}\label{DiffModelsSFHO_Log}
\end{figure*}

Figure~\ref{DiffModelsSFHO_Log} shows the GW spectra for the additional models of the SFHo constant rest-mass sequence assuming a distance of $40$~Mpc, similarly to Fig.~\ref{SFHO_GWspectrum_plot} for our reference model. We find that all the SFHo models have at least one peak that lies above the sensitivity of the ET. As the value of $\beta$ increases, by trend the height of the main peak(s) decreases. At the given distance, for low values of $\beta$ the main peak(s) lie well above the ET sensitivity curve, while for the largest values of $\beta=0.203$ and $0.222$ the main peak(s) are only marginally above the sensitivity curve.

For the DD2 EOS, we are able to construct models that cover a broader range of $\beta$ values compared to SFHo (see Table~\ref{tab:SFHO_DD2_M0_const_sequence} and Sec.~\ref{terminal_models}). This leads to spectra that admit a larger set of possible outcomes (see Fig.~\ref{DD2_GWspectrum_plot} and supplementary Fig.~\ref{DiffModelsDD2_Log} in Appendix~\ref{app:Diffbeta}). For models with low values of $\beta=0.064, 0.083$ and $0.105$ the main peak(s) lie well below the ET sensitivity curve. Intermediate $\beta$ models ($\beta=0.132, 0.165, 0.198$) lie clearly above the sensitivity curve. Finally, the peaks in the model with the highest $\beta$ value that we simulate ($\beta=0.222$) lie marginally above the ET curve.

If we define the instability onset based on the height of the features in the spectra, we identify a dependence between the instability window and the EOS. In Fig.~\ref{fig:GWspectra_maxima_simplified}, we present the maxima of the GW spectra as a function of $\beta$. Models with similar $\beta$ behave similarly, regardless of the underlying EOS. However, in the case of DD2 we can construct solutions for lower values of $\beta$, while the SFHo sequence terminates at higher values. Overall, for the soft EOS that we consider, i.e.~SFHo, the instability develops for all values of $\beta$. For lower values, the $m=2$ mode is more strongly excited; see density modal decomposition in Fig.~\ref{SFHO_ModalDecomp_plot} for the reference model and supplementary Fig.~\ref{DiffModelsSFHOModal} in Appendix~\ref{app:Diffbeta} for the additional models. This leads to a quadrupole GW emission which is more favourable for observation. For the stiffer DD2 EOS, we identify that there exists a critical value $\beta_\mathrm{crit}^\mathrm{low}$ below which the $m=2$ mode is only weakly excited (see Fig.~\ref{DD2_ModalDecomp_plot} and supplementary Fig.~\ref{DiffModelsDD2Modal} in Appendix~\ref{app:Diffbeta}) and disfavors GW observations. For the simulated sequence of models, $\beta_\mathrm{crit}^\mathrm{low}$ seems to lie in the range $[0.105, 0.132]$. For both SFHo and DD2, our simulations show that rapidly rotating models approaching the value $\beta=0.222$ undergo a weaker excitation of the $m=2$ mode, leading to a weaker GW emission. We note that the value $\beta=0.222$ is very close to the mass-shedding limit for both sequences ($\Omega_e / \Omega_K \sim 0.83$ for the SFHo sequence and $\sim 0.8$ for DD2; see Table~\ref{tab:SFHO_DD2_M0_const_sequence}).

Overall, the regime where the instability develops seems to have upper/lower bounds with respect to $\beta$. Beyond these bounds, the system transitions into a stable regime. This finding is in agreement with previous studies \cite{2003MNRAS.343..619S, Watts_2005}.

\subsection{Connecting GW spectra with model properties}\label{sec:extra_models_discuss_GW_spectra}
In what follows, we connect the observed behaviour in the GW spectra with various properties of our constant rest-mass sequences, namely: (i) the minima of the normalized Newtonian vortensity, (ii) the depth of the Newtonian vortensity well, (iii) the angular velocity maxima, (iv) the position of each model on the $M-\epsilon_\mathrm{max}$ plot (Figs.~\ref{Mass_density_SFHO} and \ref{Mass_density_DD2}) and (v) the shape of the corotation bands for the two employed EOSs (Figs.~\ref{fig:corot_sfho} and \ref{fig:corot_dd2}).

We note that we identify the local minimum ($\mathcal{V}_\mathrm{min}$) and maximum ($\mathcal{V}_\mathrm{max}$) of the vortensity within each stellar model (away from the stellar center or surface) and define the normalized Netwonian vortensity minimum as $\mathcal{V}_\mathrm{min}/ \mathcal{V}(\mathrm{r=0})$, while the normalized depth of the Newtonian vortensity as
\begin{equation}
        \frac{\Delta \mathcal{V}}{\mathcal{V}_\mathrm{max}} = \frac{\mathcal{V}_\mathrm{max} - \mathcal{V}_\mathrm{min}}{\mathcal{V}_\mathrm{max}} = 1 - \frac{\mathcal{V}_\mathrm{min}}{\mathcal{V}_\mathrm{max}} \label{eq:vort_well_depth}.
\end{equation}
Table~\ref{tab:DeltaV_GWmax} reports the aforementioned values for our dynamically evolved  models, together with the maxima of the GW spectra.

\begin{figure}
    \includegraphics[width=\columnwidth]{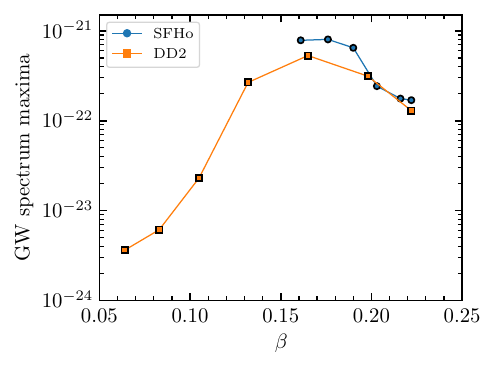}
    \caption{Maxima of the GW spectra calculated at a distance of 40 Mpc versus the rotational parameter $\beta$ for the SFHo (circles) and DD2 (squares) evolved models (marked with a black outline for consistency with Figs.~\ref{fig:newt_vortensity_minima}, \ref{fig:vort_well_depth} and \ref{fig:Omega_maxima}).}
    \label{fig:GWspectra_maxima_simplified}
\end{figure}

\begin{table}
    \centering
    \caption{Newtonian vortensity extrema normalized by the model's central Newtonian vortensity value $\mathcal{V}_c =\mathcal{V} (\mathrm{r=0})$, the respective well depth values and the maxima of the GW spectra calculated at a distance of 40 Mpc for the dynamically evolved models for each EOS.}
    \label{tab:DeltaV_GWmax}
    \begin{tabular}{lccccc}
        \hline
        EOS & $\beta$ & $ \dfrac{\mathcal{V}_\mathrm{min}}{\mathcal{V}_c}  $ & $\dfrac{\mathcal{V}_\mathrm{max}}{\mathcal{V}_c} $ & $\Delta \mathcal{V} / \mathcal{V}_\mathrm{max} $ & $\max \left( h_\mathrm{eff,+}\right)$ \\
        \hline
        SFHo & 0.161 & 1.845 & 8.709 & 0.788 & 7.842e-22 \\
         & 0.176 & 1.822 & 4.043 & 0.549 & 8.028e-22 \\
         & 0.190 & 1.823 & 3.533 & 0.484 & 6.469e-22 \\
         & 0.203 & 1.846 & 3.353 & 0.450 & 2.423e-22 \\
         & 0.216 & 1.913 & 3.291 & 0.419 & 1.762e-22 \\
         & 0.222 & 1.997 & 3.299 & 0.395 & 1.693e-22 \\
        \hline
        DD2 & 0.064 & 1.709 & 2.944 & 0.420 & 3.627e-24 \\
        & 0.083 & 1.671 & 2.882 & 0.420 & 6.115e-24 \\
        & 0.105 & 1.631 & 2.837 & 0.425 & 2.308e-23 \\
        & 0.132 & 1.596 & 2.810 & 0.432 & 2.666e-22 \\
        & 0.165 & 1.571 & 2.797 & 0.438 & 5.305e-22 \\
        & 0.198 & 1.587 & 2.817 & 0.436 & 3.129e-22 \\
        & 0.222 & 1.689 & 2.887 & 0.415 & 1.294e-22 \\
        \hline
\end{tabular}
\end{table}

\begin{figure}
    \includegraphics[width=\columnwidth]{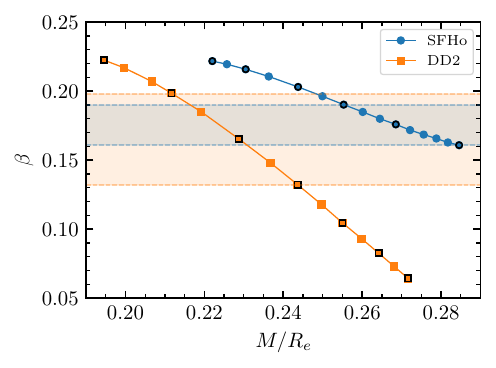}
    \caption{Rotational parameter $\beta$ versus compactness $M / R_e$ for the SFHo (circles) and DD2 (squares) constant rest-mass sequences. Evolved models are marked with a black outline. The shaded regions mark approximately the region of prominent GW emission for each EOS, requesting $\max \left( h_\mathrm{eff,+}\right) \geq 2.5 \times 10^{-22}$ (see Table~\ref{tab:DeltaV_GWmax} and Fig.~\ref{fig:GWspectra_maxima_simplified}).}
    \label{fig:beta_vs_compactness}
\end{figure}

\begin{figure}
    \includegraphics[width=\columnwidth]{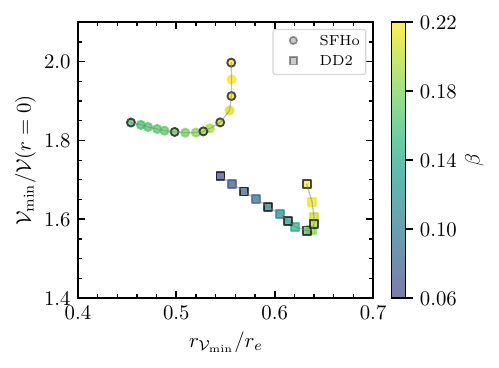}
    \caption{Normalized Newtonian vortensity minima for the SFHo (circles) and DD2 (squares) constant rest-mass sequences versus the location of $\mathcal{V}_\mathrm{min}$ normalized by the respective model's coordinate equatorial radius $r_e$. The marker color indicates different values of the rotational parameter $\beta$ according to the colorbar (see Table~\ref{tab:SFHO_DD2_M0_const_sequence} and Figs.~\ref{fig:corot_sfho} and~\ref{fig:corot_dd2}). A black outline distinguishes the evolved models.}
    \label{fig:newt_vortensity_minima}
\end{figure}

\begin{figure}
    \includegraphics[width=\columnwidth]{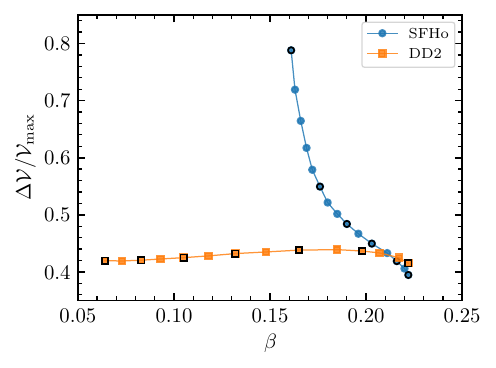}
    \caption{Newtonian vortensity well depth [Eq. \eqref{eq:vort_well_depth}] versus the rotational parameter $\beta$ for the SFHo (circles) and DD2 (squares) constant rest-mass sequences. A black outline distinguishes the evolved models (see Table~\ref{tab:SFHO_DD2_M0_const_sequence}).}
    \label{fig:vort_well_depth}
\end{figure}

\begin{figure}
    \includegraphics[width=\columnwidth]{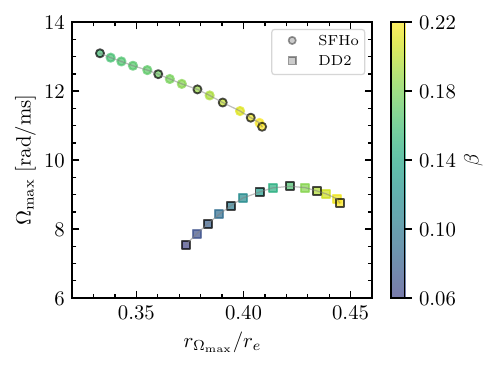}
    \caption{Angular velocity maxima for the SFHo (circles) and DD2 (squares) constant rest-mass sequences versus the location of $\Omega_\mathrm{max}$ normalized by the respective model's coordinate equatorial radius $r_e$. The marker color indicates different values of the rotational parameter $\beta$ according to the colorbar (see Table~\ref{tab:SFHO_DD2_M0_const_sequence} and Figs.~\ref{fig:corot_sfho} and~\ref{fig:corot_dd2}). A black outline distinguishes the evolved models.}
    \label{fig:Omega_maxima}
\end{figure}

In this regard, we make the following observations:
\begin{enumerate}
    \item Figure~\ref{fig:newt_vortensity_minima} shows the minima of the normalized Newtonian vortensity as a function of $r_{\mathcal{V}_\mathrm{min}} / r_e$, i.e. the location of $\mathcal{V}_\mathrm{min}$ normalized by each model's coordinate equatorial radius. Information about the rotational parameter $\beta$ is encoded in the colormap. We note that as rotation increases (moving from left to right i.e. from lower to higher $\beta$) the softer EOS, SFHo, shows a small decrease of the normalized Newtonian vortensity minimum followed by a steep increase for larger values of $\beta$. The minima are pushed towards larger values of the normalized radius as the rotation increases. In the case of the stiffer EOS, DD2, the minima of the normalized Newtonian vortensity assume lower values and appear closer to the surface compared to SFHo. The DD2 sequence exhibits an overall minimum, which is sharper compared to SFHo, potentially also due to the presence of models described by lower $\beta$ (see colorbar). The normalized radial position where the minimum appears with respect to $\beta$ is not monotonic for DD2 and, after an initial increase, starts to decrease for large $\beta$. The minima in Fig.~\ref{fig:newt_vortensity_minima} for SFHo and DD2, appear for $\beta$ equal to $0.176$ and $0.165$, respectively.

    Figure~\ref{fig:GWspectra_maxima_simplified} shows the maxima of our GW spectra as a function of $\beta$.
    Based on the models that we evolve, we find that the height of the main peak in the GW spectrum inversely follows the trend of $\mathcal{V}_\mathrm{min}/ \mathcal{V}(\mathrm{r=0})$ with respect to $\beta$ (cf. Figs.~\ref{fig:GWspectra_maxima_simplified} and \ref{fig:newt_vortensity_minima}). The model with the lowest normalized Newtonian vortensity minimum produces the most pronounced GW spectrum peak for both SFHo and DD2, respectively (see Table~\ref{tab:DeltaV_GWmax}).

    Figure~\ref{fig:beta_vs_compactness} shows the rotational parameter $\beta$ as a function of the compactness $M / R_e$. To demonstrate the EOS behaviour combined with information about the GW emission, across the range of $\beta$ we probe, we plot shaded regions that capture approximately the region of prominent GW emission. The range of the shaded regions is selected by applying the constraint $\max \left( h_\mathrm{eff,+}\right) \geq 2.5 \times 10^{-22}$. This value is an empirical threshold motivated by the maxima of the GW spectra (see Table~\ref{tab:DeltaV_GWmax} and Fig.~\ref{fig:GWspectra_maxima_simplified}), and the corresponding GW spectra for the dynamically evolved models of the SFHo (Fig.~\ref{DiffModelsSFHO_Log}) and DD2 sequences (Fig.~\ref{DiffModelsDD2_Log} in Appendix~\ref{app:Diffbeta}). The $\beta$ range of the shaded region approximating prominent GW emission for SFHo is $[0.161, 0.19]$, while for DD2 it is $[0.132, 0.198]$.
    
    Figure~\ref{fig:vort_well_depth} shows the normalized depth of the Newtonian vortensity $\Delta \mathcal{V} / \mathcal{V}_\mathrm{max}$ with regard to $\beta$.
    The height of the main GW spectrum peak correlates also with $\Delta \mathcal{V} / \mathcal{V}_\mathrm{max}$ (see Figs.~\ref{fig:GWspectra_maxima_simplified} and \ref{fig:vort_well_depth}, also Table~\ref{tab:DeltaV_GWmax}). For a fixed EOS, higher values of the well's depth lead to stronger GW emission. We emphasize that the depth of the Newtonian vortensity well covers a significantly broader range of values for the softer SFHo EOS.
    
    Figure~\ref{fig:Omega_maxima} shows the angular velocity maxima $\Omega_\mathrm{max}$ as a function of $r_{\Omega_{\mathrm{max}}}/r_e$ (i.e. the location of $\Omega_\mathrm{max}$ normalized by each model's coordinate equatorial radius) for the two EOS, with the colormap indicating $\beta$. We find that the height of the GW peak follows a similar trend to $\Omega_\mathrm{max}$ with respect to $\beta$, for both SFHo and DD2. In particular, for DD2 we identify the model with $\beta=0.165$ as having the largest $\Omega_\text{max}$ and the highest maximum in the GW spectrum. This correlation is in agreement with the expectation that models that have a bulk of mass rotating at a higher angular velocity should be efficient GW emitters.
    
    \item The SFHo sequence of models consists of hypermassive configurations. The models of the DD2 sequence are differentially rotating, but they are not hypermassive. In addition, the DD2 sequence crosses and drops below the Kepler sequence in the $M-\epsilon_\mathrm{max}$ plane in Fig.~\ref{Mass_density_DD2} at $\beta\lessapprox0.132$. We observe that all SFHo models develop a pronounced instability, while for DD2 a strong excitation of the $m=2$ mode appears right after the DD2 sequence crosses the respective Kepler sequence.
    
     We also note that, for both EOSs, the corotation region shifts towards lower rotation frequencies as $\beta$ increases, since both edges $\Omega_\mathrm{max}$ and $\Omega_\mathrm{c}$ decrease with increasing $\beta$. For SFHo this applies to the whole $\beta$ range, while for DD2 the trend becomes opposite for $\beta\lesssim0.165$. We note that the change of the behaviour for DD2 coincides with the region where the rotating models cross below the Kepler sequence. Namely, cross from an area which admits only differentially rotating configurations, into the area where both differentially and uniformly rotating configurations can be constructed (see Figs.~\ref{fig:corot_dd2} and \ref{Mass_density_DD2}). Such a transition does not happen for the SFHo sequence of models, where all members of the sequence are hypermassive neutron stars.
\end{enumerate}

\subsection{The $m=\{1, 3, 4\}$ modes}\label{sec:extra_models_m134_modes}
We now turn our attention to the $m=1,3$ and $4$ modes. For both SFHo and DD2, the $m=1$ mode is more strongly excited for models with smaller $\beta$ and less pronounced for higher $\beta$ values. For SFHo the $m=3$ is lower than the $m=2$ for models with smaller $\beta$ (i.e.~$0.161$, $0.176$ and $0.190$) and competitive to $m=2$ for models with higher values of $\beta$ (i.e. $0.203$, $0.216$ and $0.222$). Similarly, the $m=3$ is more dominant than $m=2$ only for the DD2 model with the highest $\beta=0.222$. Furthermore, for both EOSs the two modes become competitive for models with $\beta\approx0.2$ ($\beta=0.203$ for SFHo and $\beta=0.198$ for DD2). The $m=4$ mode is more pronounced at $t=0$ in all simulations due to the Cartesian geometry of the employed grid. At later times, the $m=4$ grows significantly only for the models where the $m=2$ is also strongly excited (all $\beta$ for SFHo and models with $\beta \ge 0.132$ for DD2). We note that in all cases, the $m=4$ mode develops earlier than the $m=2$ mode.

\subsection{GW spectra main peak frequencies}\label{sec:extra_models_frequencies}
For the models simulated in this work, we find that the frequency of the $m=2$ mode decreases as $\beta$ increases (see also \cite{2010CQGra..27k4104C} for the $j$-constant rotation law). Table~\ref{tab:frequencies} reports the frequencies of the main peaks present in the GW spectrum of each simulation (see also Figs.~\ref{SFHO_GWspectrum_plot}, \ref{DD2_GWspectrum_plot}, \ref{DiffModelsSFHO_Log} and \ref{DiffModelsDD2_Log}). We employ the notation $f_\mathrm{peak,i}$ with $i=1,2,3$, where the value of $i$ is $1$ for the lowest frequency peak and increases in case higher frequency peaks are present in the spectrum (i.e.\ numbers the peaks). It is difficult to associate peaks between spectra of models with different values of $\beta$, i.e.\ pick a certain peak in the spectrum of one model and monitor the amplitude of this specific peak as the parameter $\beta$ increases/decreases (as e.g.\ in \cite{2010CQGra..27k4104C})\footnote{For example, we point to model $\beta=0.105$ in Fig.~\ref{DiffModelsDD2_Log}. Associating any of the $3$ identified peaks to those reported for models with $\beta=0.083$ or $\beta=0.132$ is not straightforward.}.

The frequencies of the various peaks tend to increase for SFHo as the rotational parameter $\beta$ decreases. For DD2, we find the highest frequencies for the models with $\beta=0.105$ and $\beta=0.132$. For models with $\beta>0.132$, namely the region where we could construct models for both EOSs, DD2 models behave similarly to SFHo models (i.e.\ frequencies increase as $\beta$ decreases). Similarly, DD2 models with $\beta<0.105$ exhibit frequencies typically below the ones identified for the $\beta=0.105$ case. We emphasize that the corotation bands for SFHo and DD2, shown in Figs.~\ref{fig:corot_sfho} and \ref{fig:corot_dd2}, exhibit an overall similar shape for the common range of $\beta$. With the exception of two cases, the reported frequencies for every model lie within the corotation band of the respective EOS, which is in agreement with the statement that modes with corotation points are unstable \cite{Watts_2005,2006ApJ...651.1068O} and thus produce identifiable features in the GW spectrum. 

\begin{table}
    \centering
    \caption{Extracted frequencies for all the simulated models in this work. The first column lists the EOS, the second column provides the $\beta = T/|W|$ value of the respective model, the third column refers to the frequency of the radial mode, while the remaining columns report the frequencies of the peak(s) in the GW spectrum of the corresponding model (see Figs.~\ref{SFHO_GWspectrum_plot}, \ref{DD2_GWspectrum_plot}, \ref{DiffModelsSFHO_Log} and \ref{DiffModelsDD2_Log}). The frequencies are in kHz.}
    \label{tab:frequencies}
    \begin{tabular}{lccccc}
        \hline
        EOS & Model & $f_0$ & $f_\mathrm{peak,1}$ & $f_\mathrm{peak,2}$ & $f_\mathrm{peak,3}$ \\
         & $\beta$ & [kHz] & [kHz] & [kHz] & [kHz] \\ 
        \hline
        SFHo & 0.161  & 1.125 & 3.758 & - & - \\
         & 0.176  & 1.213 & 3.571 & - & - \\
         & 0.190  & 1.261 & 3.42 & - & - \\
         & 0.203 & 1.27 & 3.13 & 3.456 & - \\
         & 0.216 & 1.265 & 3.002 & 3.467 & - \\
         & 0.222 & 1.256 & 2.803 & 3.333 & - \\
        \hline
        DD2 & 0.064 & 1.565 & 2.101 & 2.318 & - \\
         & 0.083 & 1.597 & 2.284  & 2.488 & - \\
         & 0.105 & 1.591 & 2.402  &  2.731 & 2.944  \\
         & 0.132 & 1.543 & 2.792 & - & - \\
         & 0.165 & 1.46  & 2.605 & 2.761 & - \\
         & 0.198 & 1.345 & 2.453 & 2.697 & - \\
         & 0.222 & 1.242 & 2.306 & 2.487 & 2.808  \\
        \hline
\end{tabular}
\end{table}

We further investigate the two DD2 configurations with $\beta=0.105$ and $0.222$, which admit frequencies outside the corotation band (Fig.~\ref{fig:corot_dd2} and Table~\ref{tab:frequencies}). We aim to understand whether these tertiary frequencies are of a similar character to the corotating ones, or whether they could be explained via a different mechanism, e.g. through a possible mode merger outside the corotation band \cite{2003MNRAS.342.1156W}.
We note that the corotation bands in Figs.~\ref{fig:corot_sfho} and \ref{fig:corot_dd2} correspond to the equilibrium models. 
Throughout the simulation, the rotation profile changes and modes that lie outside the initial corotation band might exhibit corotation points. We find that this is the case for the $\beta=0.222$ model.
Since the $f_\mathrm{peak,3}$ frequency lies marginally outside the initial corotation band (Fig.~\ref{fig:corot_dd2}), as the angular velocity changes, we can identify corotation points for this feature throughout the evolution.

This does not hold for the $\beta=0.105$ configuration. No corotation points develop at any stage for $f_\mathrm{peak,3}$.
For this model, we continue the simulation to roughly $35~\mathrm{ms}$. Based on the spectrogram, we find that this feature develops within the first $\approx9.5~\mathrm{ms}$ to about 10\% of its maximum spectrogram amplitude, followed by a sudden increase reaching its maximum amplitude at $\approx15~\mathrm{ms}$. The feature survives in the GW signal till the end of the simulation. 

To understand this feature's origin we consider a few possibilities. Vortex interaction and pronounced angular momentum transport take place in the $10-15~\mathrm{ms}$ window, which might explain the sudden increase of the mode around this time. 
Furthermore, considering that we identify the feature in the $l=m=2$ GW emission, we investigate possible mode couplings that might cause its appearance. 
However, we cannot identify a combination that can both produce the relevant frequency and is phase-locked to the GW signal. 
In addition, we cannot associate the occurrence of this feature with a mechanism similar to mode merging as in \cite{2003MNRAS.342.1156W}.

\section{Connection between vortensity and unstable modes}\label{sec:VII}

In this work, we evolve a significant number of stellar configurations, modelled by 2 distinct EOSs and employ a rotation law that leads to vortensity profiles with both a local minimum and a local maximum. The combination of these features makes for an attractive playground to test connections reported in the literature between the vortensity profile, corotation points and unstable modes \cite{Watts_2005,2006ApJ...651.1068O,2015MNRAS.446..555P,2010CQGra..27k4104C,2007CoPhC.177..288C,2007CoPhC.177..288C,2017MNRAS.466..600Y,2016PhRvD..94h4032S,2020PhRvD.102d4040X,2020MNRAS.498.5904P}. Unfortunately, extracting the growth timescales of the various modes is rather challenging, since the timescales are comparable to the evolution time of our simulations and not easy to decouple in cases where more than one mode is present (e.g.\ via a fitting process similar to \cite{2010CQGra..27k4104C}, which faces similar issues in extracting the growth timescales). This hinders our ability to make quantitative statements about the growth timescales.

In this section we intend to evaluate to which extent the connections reported in the literature apply to the examined rotation law. We consider (a) corotation is a necessary condition for the onset of an instability \cite{Watts_2005} and (b) local minima of the vortensity profile can amplify the amplitude of unstable modes, while the growth of the unstable mode relates to the distance of the corotation radius from the respective vortensity local minimum \cite{2006ApJ...651.1068O}. We combine these statements, proposed in the literature, to assess their consistency.

\begin{figure}
    \includegraphics[width=\columnwidth]{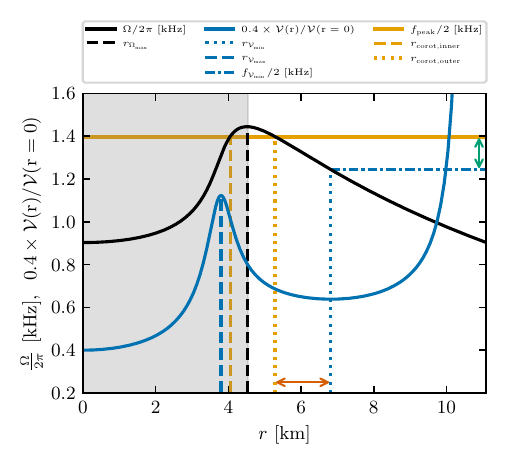}
    \caption{Angular velocity and vortensity (scaled by $0.4$) for the DD2 configuration with $\beta=0.132$. We add the frequency of the unstable mode with an orange solid line. Vertical lines, from left to right, mark the position of the local maximum of the vortensity, the inner corrotation radius of the mode, the maximum of the angular velocity profile, the outer corotation radius of the mode and the local minimum of the vortensity. The blue dash-dotted line indicates the frequency of a mode whose corotation radius lies at the vortensity minimum. Double-sided arrows denote the distance between the respective lines that the arrowheads touch. The gray shaded area illustrates the region to the left of the angular velocity maximum. See the legend and the main text for more details.}
    \label{fig:Sketch}
\end{figure}

To this end, we introduce the following framework based on Fig.~\ref{fig:Sketch}. Figure~\ref{fig:Sketch} depicts the angular velocity (black solid line), as well as the Newtonian vortensity (blue solid line) multiplied by a factor $0.4$ (for better visualization), for the DD2 models with $\beta=0.132$. We present this particular model as an example case, but all stellar configurations in this work share the same characteristics as these shown in Fig.~\ref{fig:Sketch}. We employ horizontal lines to indicate frequencies and vertical lines to mark the radial position where certain features appear.

\begin{enumerate}
    \item For each stellar model which we evolve, we calculate the Newtonian vortensity profile and locate the radius of the local vortensity minimum, $r_{\mathcal{V}_\mathrm{min}}$. (b) suggests that $r_{\mathcal{V}_\mathrm{min}}$ should be an ideal location for the growth of an unstable mode. Combining with (a), this location should be a corotation radius, i.e. the frequency of the unstable mode follows directly from the angular velocity at this point. This allows to compute the frequency of a quadrupole mode at this point as $f_{\mathcal{V}_\mathrm{min}} = \Omega(\mathrm{r=r_{\mathcal{V}_\mathrm{min}}}) / \pi$. We depict the position of the local vortensity minimum with a vertical blue dotted line in Fig.~\ref{fig:Sketch}, while the horizontal blue dash-dotted line corresponds to $f_{\mathcal{V}_\mathrm{min}}/2$.

    \item Due to the off center peak of the adopted rotation law and the fact that $\Omega_c=\Omega_e$ for all our configurations, any mode in the corotation band will have $2$ corotation radii, $r_\mathrm{corot, inner}$ and $r_\mathrm{corot, outer}$. The horizontal orange solid line corresponds to the frequency of the mode (labelled $f_\mathrm{peak}$), while the vertical orange lines mark $r_\mathrm{corot, inner}$ (dashed) and $r_\mathrm{corot, outer}$ (dotted). For all the models in this work, the radius of the local vortensity maximum $r_{\mathcal{V}_\mathrm{max}}$ (vertical blue dashed line) is smaller than the radius of the maximum angular velocity $r_\mathrm{\Omega_\mathrm{max}}$ (vertical black dashed line), which in turn is smaller than the radius of the local vortensity minimum (i.e.\ $r_{\mathcal{V}_\mathrm{max}} < r_\mathrm{\Omega_\mathrm{max}} < r_{\mathcal{V}_\mathrm{min}}$). Since $r_\mathrm{corot, inner} < r_\mathrm{\Omega_\mathrm{max}} < r_\mathrm{corot, outer}$, (b) suggests that the unstable mode around $r_\mathrm{corot, outer}$ should develop over shorter timescales.
    
    \item We want to compare the frequencies extracted from our dynamical simulations (Table~\ref{tab:frequencies}) to the $f_{\mathcal{V}_\mathrm{min}}$ estimate (see Table \ref{tab:estimated_frequencies}) as a measure of the growth timescale following (b). The frequency difference appears as a green double-sided arrow in Fig.~\ref{fig:Sketch}. Previous investigation was on the relation between the growth timescale of a mode and the distance of the associated corotation radius from the location of the local vortensity minimum \cite{2006ApJ...651.1068O}. We depict this distance with a vermillion double-sided arrow in Fig.~\ref{fig:Sketch}\footnote{We remark that our arguments apply also to alternate definitions, e.g. how deep the corotation points lies in the vortensity well.}. For $r > r_\mathrm{\Omega_\mathrm{max}}$ (unshaded area in Fig.~\ref{fig:Sketch}), the rotation law is monotonic and thus larger differences in frequency correspond to greater differences in the distance between $r_\mathrm{corot,outer}$ and $r_{\mathcal{V}_\mathrm{min}}$. Hence, looking at the frequency differences is similar to looking at the distance from the local vortensity minimum\footnote{We note that, this argument also applies in the special case where the mode has a single corotation radius at $r_\mathrm{\Omega_\mathrm{max}}$. Furthermore, based on our evolutions, the rotation profile evolves over time and becomes monotonic. Thus, this argument should hold throughout the evolution as well.}.
    
    \item The difference between the extracted and predicted frequencies relates to the growth timescale of the respective mode, based on (b). In this picture, small frequency differences point to modes that are indeed close to the vortensity minimum and should develop quickly. Large frequency differences should coincide with larger growth timescales. The lack of a mode at the expected frequency may be interpreted as a mode with growth timescale above the evolution time of the respective simulation (i.e.\ we cannot check/verify this case based on our simulation data). We note that an inconsistency between the data and the expected behaviour suggests that both conditions (a) and (b) cannot hold at the same time, while agreement between the evolved models and our simple theoretical modelling hints towards both conditions being valid for the configurations considered and the range of parameters that the models probe.
\end{enumerate}

Table~\ref{tab:estimated_frequencies} presents the frequency $f_{\mathcal{V}_\mathrm{min}}$ for each evolved stellar model. For each configuration, we compare $f_{\mathcal{V}_\mathrm{min}}$ to $f_\mathrm{peak,1}$ and $f_\mathrm{peak,2}$ and report the absolute difference. We exclude $f_\mathrm{peak,3}$ from the current analysis, because it lies outside the corotation band in both cases where such a frequency is present (i.e.\ a corotation point for these frequencies possibly arises at later times, when the rotation profile of the respective configuration changes).

\begin{table}
    \centering
    \caption{For each evolved configuration, the third column reports the frequency of the unstable mode under the assumption that the location of the local vortensity minimum is at a corotation radius. The fourth and fifth columns report the absolute difference between the frequency estimate and the frequencies obtained from the 3D evolutions of the models (see Table~\ref{tab:frequencies}). The final column marks for which model we find a clear inconsistency (denoted by F) between the expectation based on the frequency differences and the respective spectrogram/spectrum.}
    \label{tab:estimated_frequencies}
    \resizebox{\columnwidth}{!}{
    \begin{tabular}{lccccc}
        \hline
        EOS & Model & $f_{\mathcal{V}_\mathrm{min}}$ & $|f_\mathrm{peak,1} - f_{\mathcal{V}_\mathrm{min}}|$ & $|f_\mathrm{peak,2} - f_{\mathcal{V}_\mathrm{min}}|$ & Status \\
         & $\beta$ & [kHz] & [kHz] & [kHz] & \\ 
        \hline
         SFHo & 0.161 & 3.882 & 124.0 & -  & F\\
              & 0.176 & 3.651 & 80.0 & -  &   \\
              & 0.19 & 3.488 & 68.0 & -  &   \\
              & 0.203 & 3.363 & 233.0  & 93.0  & F \\
              & 0.216 & 3.235 & 233.0  & 232.0  & F \\
              & 0.222 & 3.176 & 373.0  & 157.0  &  \\
        \hline
         DD2 & 0.064 & 2.144 & 43.0  & 174.0  &  \\
             & 0.083 & 2.288 & 4.0  & 200.0  & F \\
             & 0.105 & 2.399 & 3.0  & 332.0  & F \\
             & 0.132 & 2.491 & 301.0 & -  &   \\
             & 0.165 & 2.519 & 87.0  & 242.0  &  \\
             & 0.198 & 2.492 & 39.0  & 205.0  &  \\
             & 0.222 & 2.437 & 131.0  & 50.0  &  \\
        \hline
\end{tabular}}
\end{table}

We focus on individual stellar configurations with multiple excited modes and examine their respective spectrogram as a mean to estimate how quickly each mode grows. We find inconsistencies between the frequency differences and (b). In the case of SFHo, we observe that for our reference model $|f_\mathrm{peak,1} - f_{\mathcal{V}_\mathrm{min}}| \approx |f_\mathrm{peak,2} - f_{\mathcal{V}_\mathrm{min}}|$, while the $f_\mathrm{peak,1}$ mode develops more quickly (see Fig.~\ref{SFHO_GWspectrogram_plot}). We find additional cases where the comparative relation between $|f_\mathrm{peak,1} - f_{\mathcal{V}_\mathrm{min}}|$ and $|f_\mathrm{peak,2} - f_{\mathcal{V}_\mathrm{min}}|$ does not align with how quickly the $f_\mathrm{peak,1}$, $f_\mathrm{peak,2}$ modes develop (e.g.\ case $\beta=0.203$ for SFHo and cases $\beta=0.083, 0.105$ for DD2; not shown). 

We also compare the frequency differences that we find with the GW spectrum of the respective stellar configuration as a measure of how strongly the underlying modes are excited. We assume that modes with a short growth timescale should have enough time to develop within the simulation time and lead to pronounced features in the GW spectrum. Considering GW spectra, we do not identify a convincing correlation between the frequency differences and the observability of the various features in the spectra. In the case of SFHo, model $\beta=0.161$ is an exception, since it has a pronounced peak in the spectrum, but the respective frequency deviation is moderately high. In the case of DD2, the inconsistency is more pronounced, since models $\beta=0.083, 0.105$ have unstable modes very close to $f_{\mathcal{V}_\mathrm{min}}$, which do not lead to pronounced peaks in the GW spectrum.

In addition to the local vortensity minimum, we investigate also the phenomenological claim that, for modes within the corotation band, the growth timescale increases towards the band edges and is short for modes close to the center of the band \cite{Watts_2005}. In Figs.~\ref{fig:corot_sfho} and \ref{fig:corot_dd2}, we mark the center of the corotation band with a grey dotted line and the $f_{\mathcal{V}_\mathrm{min}}$ frequencies with green crosses (the green dash-dotted lines connect these frequencies). We notice that the two lines differ, namely the growth times of unstable modes cannot correlate with how deep the mode lies in the corotation band and the vicinity of the corotation point to the local vortensity minimum, at the same time. We inspect the spectrograms for all stellar models which feature multiple modes in the corotation band. In the cases where we can clearly distinguish the two modes in the spectrogram, the lower frequency always develops first. For the evolved models, the lower frequency is always the one deeper in the corotation band (i.e. closer to the dotted grey line). Hence, the prediction that the growth rate of unstable modes is higher for modes closer to the corotation band center appears to hold for the examined data.

Furthermore, we note that with the exception of the SFHo model with the maximum $\beta$, all frequencies populate the upper half of the corotation band. Given that the line related to $f_{\mathcal{V}_\mathrm{min}}$ lies on this side of the band too, it on average provides better estimates for the frequencies of the unstable modes than the line marking the center of the corotation band.

Overall, we find that assuming that the local vortensity minimum is a corotation radius leads to frequency estimates, which are accurate with a maximum relative error\footnote{We define the relative error as $|f_\mathrm{peak,i} - f_{\mathcal{V}_\mathrm{min}}|/f_\mathrm{peak,i} \times 100\%$, where $i=1,2$.} of $13.3\%$ ($12.2\%$) and an average error of $5.6\%$ ($5.4\%$) for the case of SFHo (DD2). Hence, the vortensity minimum may be employed to obtain a reasonable estimate for frequencies of unstable modes based on the equilibrium stellar properties and without the need to perform any additional evolution. Considering that equilibrium models may be viewed as snapshots of the post-merger evolution of the BNS remnant, this approach might be useful to predict how an instability can evolve over time based on instantaneous, azimuthally-averaged vortensity profiles of the merger remnant. For the considered rotation law, we do not find evidence that the distance between the corotation radius of an unstable mode and the position of the vortensity minimum directly informs us about the growth timescale of the respective mode. For the configurations where multiple modes are excited and we can clearly distinguish the modes in the spectrogram, the mode which lies deeper in the corotation band grows more quickly. We remark that our analysis is based on the equilibrium configurations, which we use as initial data for our dynamical simulations, and does not take into account the evolution of the rotation law and the Newtonian vortensity profile. In principle, a more detailed analysis which also locates the region where each mode develops within the star is necessary. Such considerations lie beyond the scope of this work.

\section{Conclusions}
\label{sec:label5}

We construct constant rest-mass sequences of equilibrium models of differentially rotating NSs, adopting a differential rotation law that emulates the angular velocity profile of BNS remnants \cite{Uryu_etal_2017}. We consider two distinct EOSs, the soft SFHo and the stiffer DD2. For each EOS, we evolve a subset of the sequence's models for roughly $20~\mathrm{ms}$ and study their dynamics. We employ a 3D GRHD code, which we introduce in this work. The code solves the GRHD equations, as well as the metric field equations, on a Cartesian grid. For the dynamical spacetime, we employ the CFC approximation. The main simulations discussed in this work employ $3$rd order spatial reconstruction and time integration.

We first discuss two reference models, described by SFHo and DD2, respectively. In both cases, we identify a strong excitation of the quadrupole mode. By extracting the mode pattern speed and comparing to the angular velocity profile of our configurations, we identify corotation radii. This aligns with the expectation that the presence of corotation points within the star can drive low-$T/|W|$ instabilities \cite{Watts_2005}. We test whether the instabilities can be associated with the local minimum of the Newtonian vortensity profile, but only find strong indications that such a connection exists for the DD2 model. We do identify certain differences between the two reference models, related to the underlying EOS. The SFHo model displays a pronounced double peak in the GW spectrum, which is absent for the DD2 configuration, as well as a stronger excitation of the $m=1$ mode. Furthermore, the rotation profile of the SFHo model evolves and, by the end of the simulation, the maximum of the angular velocity appears at the center of the star. In the case of DD2, the rotation profile displays an off-center peak throughout the whole simulation.

In order to further study the differences between the reference models, we perform additional simulations where we introduce a strong initial quadrupolar perturbation. In these setups, the DD2 model also displays a double peak structure in the GW spectrum. Furthermore, the peak of the angular velocity rapidly moves towards the center and the rotation profiles of both the SFHo and DD2 configuration display a monotonic behaviour at late times, which resembels a Keplerian rotation law outside the stars. Interestingly, even though we provide a similar excitation to both reference models, only the SFHo configuration leads to a strong $m=1$ excitation throughout the simulation. This behaviour might hint towards a dependence of the development of the $m=1$ mode on the employed EOS (see also \cite{2024MNRAS.527.8812J,2026PhRvD.113b3011G}).

Considering the full sequence of all evolved models allows to evaluate the properties of the instability with respect to the ratio of the kinetic to potential energy, $\beta$. As a metric of how strongly the instability is excited, we focus on the height of the main features in the GW spectrum of each evolved model. For both EOSs, we find that the instability seems to operate more strongly within a window with respect to $\beta$, in agreement with previous studies in the context of different rotation laws \cite{2003MNRAS.343..619S,Watts_2005}. The upper edge of the instability window seems to occur in the vicinity of $\beta_\mathrm{upper}\approx0.222$ for both EOSs. We note that $\beta = 0.222$ lies close to the mass-shedding limit and are the highest $\beta$ models we construct, for both EOSs. In the case of DD2, we place the lower edge of the window in the range $\beta_\mathrm{lower} \in [0.105,0.132]$, while for SFHo we find $\beta_\mathrm{lower} < 0.161$. We identify a number of correlations between the strength of the peak features in the GW spectrum and properties of the equilibrium models. Specifically, the strength of the GW emission correlates with $\mathcal{V}_\mathrm{min}$ (local minimum value of the Newtonian vortensity), $\Delta \mathcal{V} / \mathcal{V}_\mathrm{max}$ (depth of the vortensity well) and $\Omega_\mathrm{max}$ (maximum angular velocity value). In all cases, the identified mode frequencies fall within the respective corotation band which further supports that models which feature corotation radii are unstable \cite{Watts_2005,2006ApJ...651.1068O}. 

Notably, the constant rest-mass sequences that we construct consist of \textit{hypermassive} models for the softer EOS (SFHo) and differentially rotating but not hypermassive models for the stiffer EOS (DD2). In addition, all SFHo models lie above the respective Kepler sequence for uniform rotation, while the DD2 sequence of configurations crosses the respective Kepler sequence and drops below it for $\beta\lessapprox0.132$. For DD2, the intersection point between the sequence of differentially rotating NSs of constant rest mass and the Kepler sequence coincides well with $\beta_\mathrm{lower}$, i.e.\ differentially rotating models below the Kepler sequence in the $M-\epsilon_\mathrm{max}$ plane do not exhibit strong GW emission due to the low-$T/|W|$ instability. Hence, the imprint of a low-$T/|W|$ instability on the GW spectrum, or the lack of it, might be an observational indication of the position of a BNS remnant relative to the Kepler sequence of uniform rotation in the $M-\epsilon_\mathrm{max}$ space.

We note that in the models that feature prominent GW emission, the amplitude of GW signal is comparable to re-excitations in the GW signal of massive BNS merger remnants in \cite{2022PhRvD.105d3020S}. Thus, the excitation of low-$T/|W|$ instabilities in the remnant might be the underlying reason behind this GW emission. Note that dynamical rotational instabilities are promising for GW detection with future GW observatories \cite{PhysRevD.109.043045}, and thus it is important to determine the parameter range in which they occur. This highlights the importance of our study on how the low-$T/|W|$ instability develops across a range of $\beta$.

Finally, for the examined rotation law, we investigate phenomenological relations proposed in the literature between the growth timescale and (i) the local vortensity minimum \cite{2006ApJ...651.1068O} or (ii) how close the mode lies to the corotation band center/edges \cite{Watts_2005}. Based on the evolved models, we find that the growth timescales of modes correlates better with how deep they lie in the corotation band. Furthermore, under the assumption that the local vortensity minimum corresponds to a corotation radius, we can estimate the unstable mode frequency for every stellar configuration with a maximum error of $13.3\%$. Hence, we obtain a reasonable estimate based on an equilibrium property, without the need to perform the much more expensive 3D simulation.

Differentially rotating NS remnants which occur in BNS mergers feature a complicated temperature profile. In the current study, we construct cold equilibrium models, thus we do not capture potential temperature effects. In the future, we plan to investigate this aspect as well and try to validate the findings of this study based on a more extensive set of EOSs, rotation laws and rest masses.

\begin{acknowledgments}
G.L. acknowledges support by the Klaus Tschira Foundation and by the Deutsche Forschungsgemeinschaft (DFG, German Research Foundation) - MA 4248/3-1. P.I. acknowledges support by ICSC – Centro Nazionale di Ricerca in High Performance Computing, Big Data and Quantum Computing, funded by European Union – NextGenerationEU.  AB acknowledges support by the European Union (ERC Synergy Grant HEAVYMETAL, grant no. 101071865) and the Deutsche Forschungsgemeinschaft (DFG, German Research Foundation) through Project - ID 279384907 - SFB 1245 (subprojects B07). N.~S. acknowledges funding from the H.F.R.I. Project No.~26254. In addition, this publication is part of a project that has received funding from the European Union’s Horizon Europe Research and Innovation Programme under Grant Agreement No 101131928. Virgo is funded, through the European Gravitational Observatory (EGO), by the French Centre National de Recherche Scientifique (CNRS), the Italian Istituto Nazionale di Fisica Nucleare (INFN) and the Dutch Nikhef, with contributions by institutions from Belgium, Germany, Greece, Hungary, Ireland, Japan, Monaco, Poland, Portugal, Spain. KAGRA is supported by Ministry of Education, Culture, Sports, Science and Technology (MEXT), Japan Society for the Promotion of Science (JSPS) in Japan; National Research Foundation (NRF) and Ministry of Science and ICT (MSIT) in Korea; Academia Sinica (AS) and National Science and Technology Council (NSTC) in Taiwan.
\end{acknowledgments}

\appendix

\section{Resolution study}\label{app:ResStudy}
For each reference model discussed in Sec.~\ref{sec:IV}, we perform two additional simulations with $\mathrm{dx}=0.175$ and $\mathrm{dx}=0.25$ to evaluate the impact of the resolution on the results. All other numerical aspects remain identical to the simulations discussed in Sec.~\ref{sec:IV}. We focus the discussion on the GW emission, where the growth of the instability is clearly visible.

\begin{figure*}
    \subfloat[]{%
        \includegraphics[width=\columnwidth]{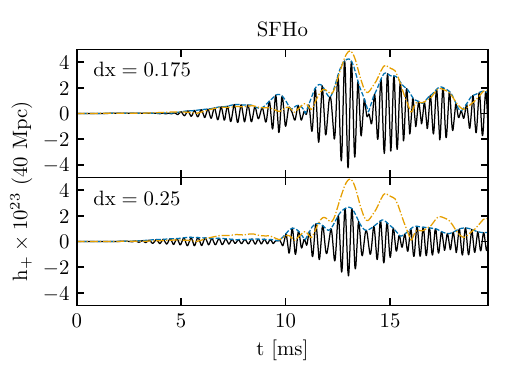}\label{hplus_SFHO_res}}
    \hfill
    \subfloat[]{%
        \includegraphics[width=\columnwidth]{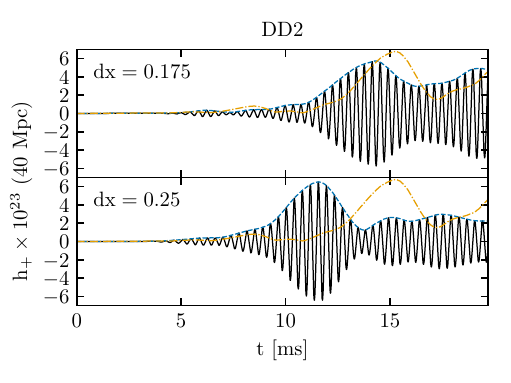}\label{hplus_DD2_res}}
   
    \caption{Plus polarization of the GW strain at a distance of $40~\mathrm{Mpc}$ for the SFHo (panel \protect\subref{hplus_SFHO_res}) and DD2 (panel \protect\subref{hplus_DD2_res}) models at different resolutions. The solid lines illustrate the GW signal for the resolution reported within each panel. The blue dashed line outlines the amplitude of the GW signal in each case. The orange dash-dotted line displays the amplitude of the GW signal in the respective $\mathrm{dx}=0.15$ simulation discussed in Sec.~\ref{sec:res_sfho} (SFHo; panel \protect\subref{hplus_SFHO_res}) and Sec.~\ref{sec:res_dd2} (DD2; panel \protect\subref{hplus_DD2_res}).}\label{hplus_res}
\end{figure*}

Considering the SFHo configuration, the upper panel of Fig.~\ref{hplus_SFHO_res} displays the plus polarization of the GW strain in the $\mathrm{dx}=0.175$ run. The blue dashed line outlines the amplitude of the GW signal. In addition, we include the respective GW amplitude from the $\mathrm{dx}=0.15$ simulation (i.e.\ from Fig.~\ref{SFHO_hplus_plot}) as an orange dash-dotted line. The bottom panel of Fig.~\ref{hplus_SFHO_res} presents the same information for the $\mathrm{dx}=0.25$ run.

We find that the GW amplitude is rather similar in the two higher resolution simulations (i.e.\ $\mathrm{dx}=0.15$ and $\mathrm{dx}=0.175$), reaching similar maximum values and exhibiting an oscillation pattern which matches quite well. Furthermore, the amplitudes of the $m=1,2,3,4$ density modes progress very similarly in these two simulations (not shown). The lower resolution simulation also bears close similarities to the $\mathrm{dx}=0.15$ simulation when examining the modulation of the GW amplitude. The main difference arises in the maximum amplitude of the GW emission, which for the $\mathrm{dx}=0.25$ is lower by a factor of roughly $2$ in the interval $[12,16]~\mathrm{ms}$. Examining the normalized amplitudes of the $m=1,2,3,4$ modes, the $m=1$ does not saturate around $5~\mathrm{ms}$ (as in Fig.~\ref{SFHO_ModalDecomp_plot}), but continues developing until $t=9~\mathrm{ms}$ (not shown). Hence, it becomes the dominant mode around $t=7~\mathrm{ms}$, which might explain why the $m=2$ mode develops less in the $\mathrm{dx}=0.25$ run compared to the $\mathrm{dx}=0.15$ simulation. This might hint that the development of a pronounced $m=1$ mode is up to some extent sensitive to the employed resolution (see also \cite{2016PhRvD..93b4011E,2016PhRvD..94f4011R,2026PhRvD.113b3011G}).

For both the $\mathrm{dx}=0.15$ and $\mathrm{dx}=0.175$ runs, the GW spectrum features a double peak structure as in Fig.~\ref{SFHO_GWspectrum_plot}. The frequencies of both peaks agree within $50~\mathrm{Hz}$ with the main simulation. Furthermore, the rotation profile evolves rather similarly regardless of the resolution. In all cases, the configuration exhibits the maximum angular velocity at the center at times $t>14~\mathrm{ms}$. Hence, we can conclude that our default resolution of $\mathrm{dx}=0.15$ suffices to capture the main dynamics of the system.

Figure~\ref{hplus_DD2_res} is similar to Fig.~\ref{hplus_SFHO_res}, but refers to the DD2 configuration. The GW signal in the $\mathrm{dx}=0.175$ run shows a maximum amplitude similar to that of the $\mathrm{dx}=0.15$ simulation. The amplitude exhibits some modulation, which is also comparable between the two simulations. A minor difference is that the GW strain amplitude grows slightly earlier in the $\mathrm{dx}=0.175$ run. The fast increase of the GW signal amplitude is even more pronounced in the $\mathrm{dx}=0.25$ simulation, where the amplitude reaches its maximum value roughly $4~\mathrm{ms}$ earlier than in the $\mathrm{dx}=0.25$ run.

We further comment on the somehow different growth timescales in the $\mathrm{dx}=0.15$ and the $\mathrm{dx}=0.25$ simulations with DD2. We focus on the first $5~\mathrm{ms}$ to better investigate what drives the early increase of the GW signal amplitude. Based on a Fourier analysis of the GW signal, we extract $2.63~\mathrm{kHz}$ and $2.41~\mathrm{kHz}$ as the dominant frequency in each simulation, respectively. We note that the dominant frequency in the lower resolution simulations matches well with the potential secondary peak seen in Fig.~\ref{DD2_GWspectrum_plot}. Based on the original angular velocity profile (i.e.\ the rotation profile at $t=0~\mathrm{ms}$), these frequencies lead to an inner corotation radius of $5~\mathrm{km}$ and an outer corotation radius of $7.95~\mathrm{km}$ in the $\mathrm{dx}=0.15$ run. The respective corotation radii in the $\mathrm{dx}=0.25$ simulation are $4.6~\mathrm{km}$ and $9.25~\mathrm{km}$. The outer corotation radius of the prevalent mode lies closer to the vortensity minimum in the $\mathrm{dx}=0.25$ run (see location of local minimum in Fig.~\ref{fig:vortensity_dd2}). This observation is in agreement with the expectation that the growth rate of an unstable mode increases when the associated corotation radius lies closer to the minimum of the vortensity profile \cite{2006ApJ...651.1068O}.

We also note that the dominant frequency in the DD2 runs agrees within $25~\mathrm{Hz}$ for the considered resolutions. Examining the amplitudes of the $m=1,2,3,4$ modes, we find that they evolve very similarly in the $\mathrm{dx}=0.15$ and the $\mathrm{dx}=0.175$ simulations. In the $\mathrm{dx}=0.25$ run, we find that the $m=1$ mode constantly increases and at the very end of the simulation its amplitude is comparable with the $m=2$ and $m=4$ modes. This is in agreement with the resolution study of the SFHo model, where the $m=1$ develops in a more pronounced way for the lower resolution. Finally, we comment that the rotation profile evolves similarly for all resolutions. Overall, the employed resolution of $\mathrm{dx}=0.15$ seems to provide an accurate description of the model's dynamics.

\section{Additional figures for models with different $\beta = T/|W|$}\label{app:Diffbeta}

\begin{figure*}
    \subfloat[]{%
        \includegraphics[width=\columnwidth]{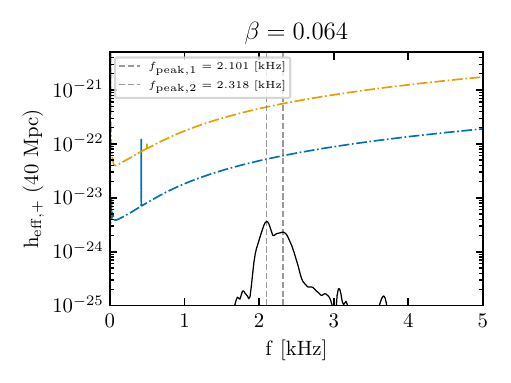}\label{DiffModels_DD2_1_Log}}
    \hfill
    \subfloat[]{%
        \includegraphics[width=\columnwidth]{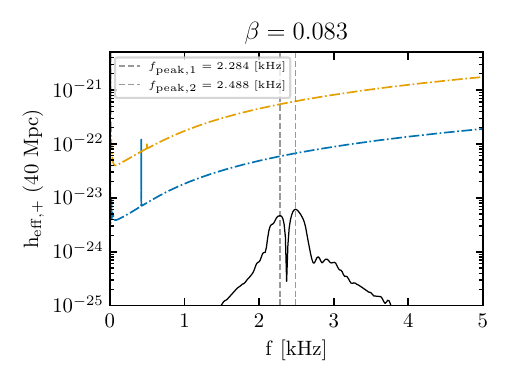}\label{DiffModels_DD2_2_Log}}

    \vspace*{2pt}%
    
    \subfloat[]{%
         \includegraphics[width=\columnwidth]{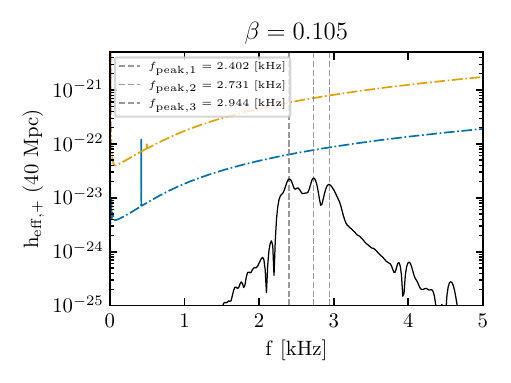}\label{DiffModels_DD2_3_Log}}
    \hfill
    \subfloat[]{%
        \includegraphics[width=\columnwidth]{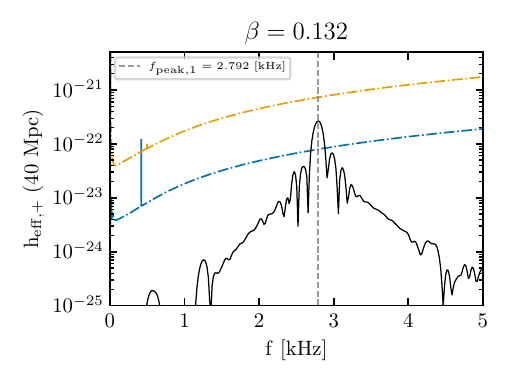}\label{DiffModels_DD2_4_Log}}

    \vspace*{2pt}%

    \subfloat[]{%
         \includegraphics[width=\columnwidth]{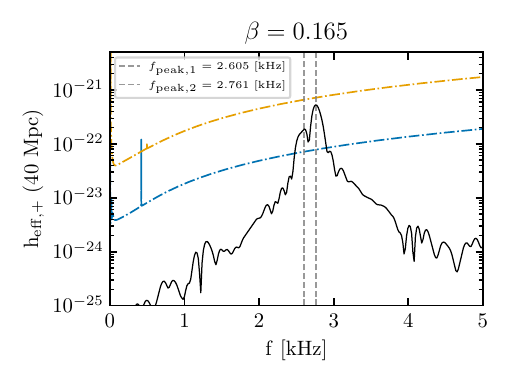}\label{DiffModels_DD2_5_Log}}
    \hfill
    \subfloat[]{%
        \includegraphics[width=\columnwidth]{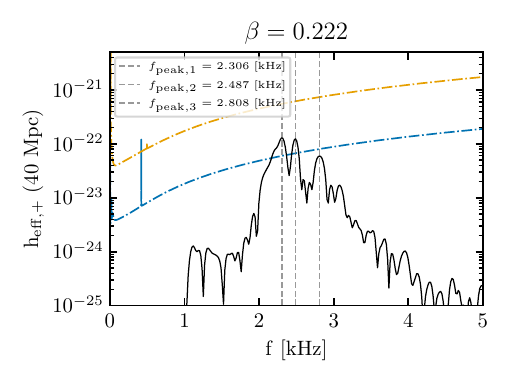}\label{DiffModels_DD2_6_Log}}
   
    \caption{Similar to Fig.~\ref{DiffModelsSFHO_Log} for DD2 (see also Fig.~\ref{DD2_GWspectrum_plot}).}\label{DiffModelsDD2_Log}
\end{figure*}

\begin{figure*}
    \subfloat[]{%
        \includegraphics[width=\columnwidth]{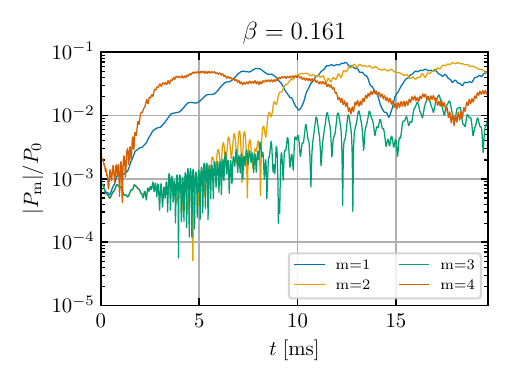}\label{DiffModels_SFHO_1_Modal}}
    \hfill
    \subfloat[]{%
        \includegraphics[width=\columnwidth]{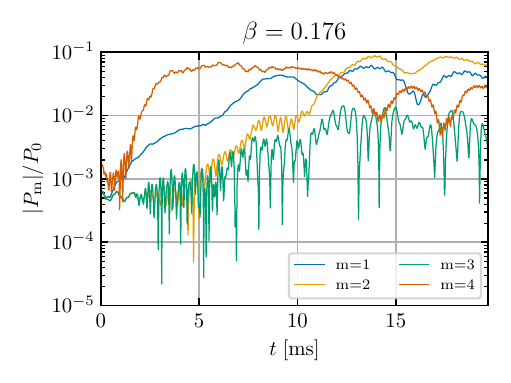}\label{DiffModels_SFHO_2_Modal}}

    \vspace*{2pt}%
    
    \subfloat[]{%
         \includegraphics[width=\columnwidth]{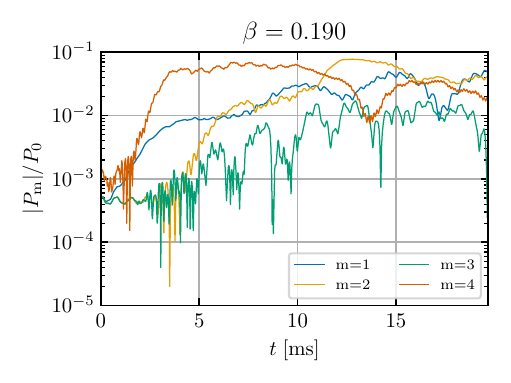}\label{DiffModels_SFHO_3_Modal}}
    \hfill
    \subfloat[]{%
        \includegraphics[width=\columnwidth]{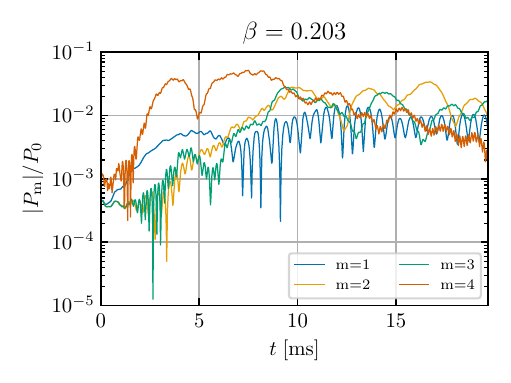}\label{DiffModels_SFHO_4_Modal}}

    \vspace*{2pt}%

    \subfloat[]{%
         \includegraphics[width=\columnwidth]{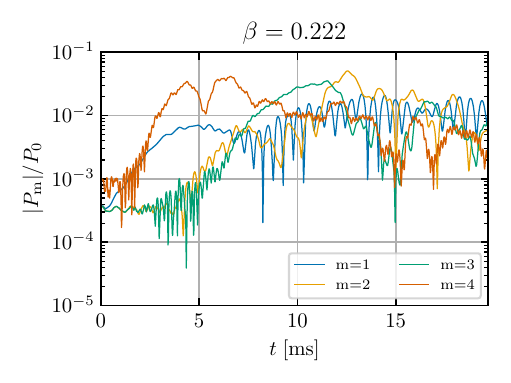}\label{DiffModels_SFHO_5_Modal}}
   
    \caption{Each panel refers to a model with different $\beta$ for the SFHo EOS (see panel title). Each plot shows the modal decomposition for the respective model (see also Fig.~\ref{SFHO_ModalDecomp_plot}). For the respective GW spectra, see Fig.~\ref{DiffModelsSFHO_Log}.}\label{DiffModelsSFHOModal}
\end{figure*}

\begin{figure*}
    \subfloat[]{%
        \includegraphics[width=\columnwidth]{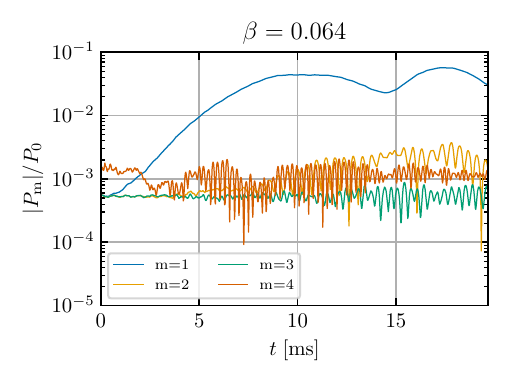}\label{DiffModels_DD2_1_Modal}}
    \hfill
    \subfloat[]{%
        \includegraphics[width=\columnwidth]{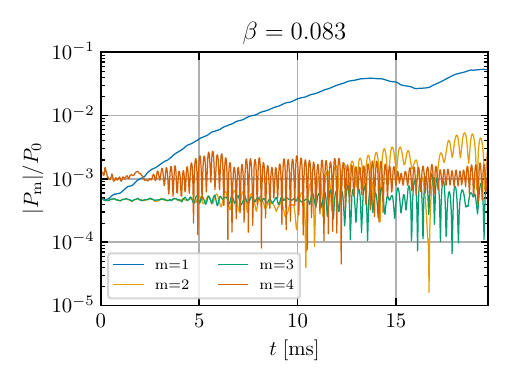}\label{DiffModels_DD2_2_Modal}}

    \vspace*{2pt}%
    
    \subfloat[]{%
         \includegraphics[width=\columnwidth]{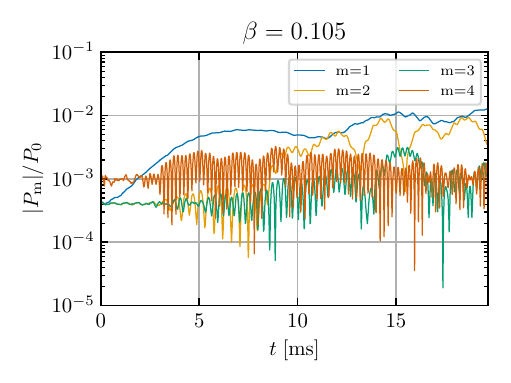}\label{DiffModels_DD2_3_Modal}}
    \hfill
    \subfloat[]{%
        \includegraphics[width=\columnwidth]{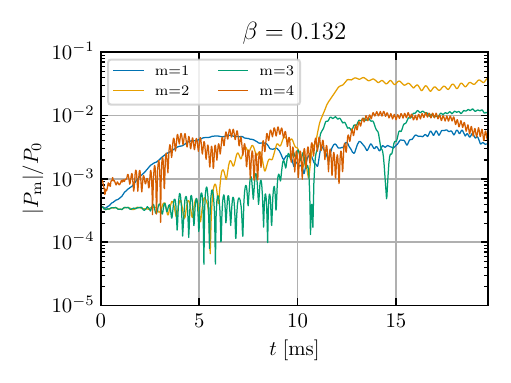}\label{DiffModels_DD2_4_Modal}}

    \vspace*{2pt}%

    \subfloat[]{%
         \includegraphics[width=\columnwidth]{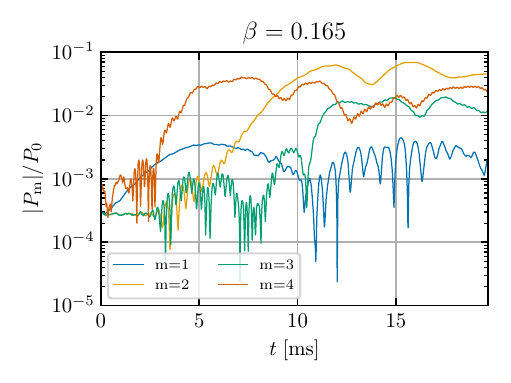}\label{DiffModels_DD2_5_Modal}}
    \hfill
    \subfloat[]{%
        \includegraphics[width=\columnwidth]{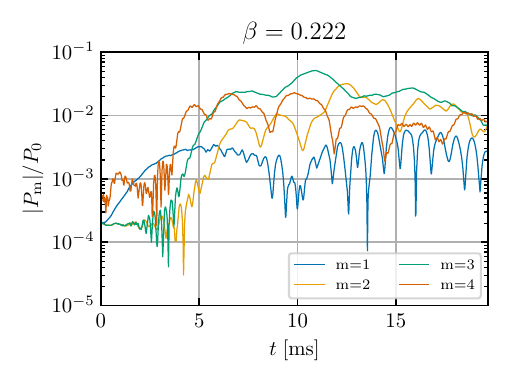}\label{DiffModels_DD2_6_Modal}}
   
    \caption{Similar to Fig.~\ref{DiffModelsSFHOModal} for DD2 (see also Fig.~\ref{DD2_ModalDecomp_plot}).  For the respective GW spectra, see Fig.~\ref{DiffModelsDD2_Log}.}\label{DiffModelsDD2Modal}
\end{figure*}

We provide supplementary figures for all the models discussed in Sec.~\ref{sec:VI}. Figure~\ref{DiffModelsDD2_Log} presents the GW spectra for the DD2 models in a similar manner to Fig.~\ref{DiffModelsSFHO_Log} (which refers to the SFHo models). Figures~\ref{DiffModelsSFHOModal} and \ref{DiffModelsDD2Modal} show the respective modal decomposition plots for the $m=1, 2, 3, 4$ modes. For a discussion of the plots, see Sec.~\ref{sec:VI}.

\section{Numerical methods of three-dimensional relativistic hydrodynamics code}\label{app:Codeimplementation}
In this section, we describe the numerical schemes employed in our code. Considering that our implementation employs well-established schemes, which can be considered textbook knowledge (see e.g.\ \cite{2008LRR....11....7F,rezzolla2013relativistic,shibata2015numerical,baumgarte_shapiro_2010}), we refrain from an overly extensive presentation. Instead, we provide references to works where the schemes were introduced and where detailed discussions can be found. We note that we solve both the hydrodynamical, as well as the metric equations on a uniform Cartesian grid (see also Sec.~\ref{sec:NumMethMetric}).

\subsection{Hydrodynamical evolution}
For the solution of the hydrodynamical equations, we employ commonly used high-resolution shock-capturing schemes. For the reconstruction of cell states to the interfaces, we implement total variation diminishing (TVD) slope-limited reconstruction (see e.g.\ \cite{1983JCoPh..49..357H,1984SJNA...21....1H,toro_riemann_2009}). We consider a number of different forms for the slope limiter, including the monotonized central \cite{1977JCoPh..23..276V}, minmod \cite{1979JCoPh..32..101V} and superbee \cite{1985ams..conf..163R} limiters. In addition, we implement the piecewise parabolic method (PPM) \cite{1984JCoPh..54..174C}, which is our standard choice for the simulations discussed in this work. At the interfaces, we employ the HLLE Riemann solver \cite{10.2307/2030019,1988SJNA...25..294E} for the computation of the fluxes.

The code can carry out the time update based on TVD Runge--Kutta (RK) methods of different order \cite{1988JCoPh..77..439S,1998MaCom..67...73G}. These include a $1^\mathrm{st}$ order (i.e.\ forward Euler), $2^\mathrm{nd}$ order (also known as Euler-Heun) and $3^\mathrm{rd}$ order RK integrator (see $2^\mathrm{nd}$ and $3^\mathrm{rd}$ optimal RK methods in \cite{1998MaCom..67...73G}). Our standard choice is the $3^\mathrm{rd}$ order RK time integrator. We apply the Courant–Friedrichs–Lewy (CFL) condition \cite{1928MatAn.100...32C,1967IBMJ...11..215C} and compute the appropriate time step for each cell. The time integration uses a global time step, which corresponds to the minimum among all the individual cell time steps.

GRHD implementations require a numerical scheme to recover the primitive variables from the conserved variables. Depending on the form of the EOS (see Sec.~\ref{sec:EOS}), different recovery methods might be more appropriate. Hence, we implement a series of different approaches (see e.g.\ \cite{rezzolla2013relativistic} for more details):
\begin{enumerate}
    \item In the case of the polytropic EOS, $p$ and $\epsilon$ (i.e.\ also $h$) follow analytically from $\rho$. Therefore, to recover all the primitive variables, it suffices to solve a non-linear equation of the form
    \begin{equation}
        \rho \hat{W}_\mathrm{L} - D = 0,
    \end{equation}
    where
    \begin{equation}
        \hat{W}_\mathrm{L} = \sqrt{1+\frac{S^2}{(Dh)^2}}.
    \end{equation}
    Here, $S^2=\gamma^{ij} S_i S_j$, while $h$ depends on $\rho$ via Eqs.~\eqref{eq:EOSpolyP}-\eqref{eq:EOSpolyE}. We employ a Newton-Raphson method as the root-solving method.

    \item For the more generic case, where the EOS is not a function of $\rho$ alone, we rephrase the rest-mass density and specific internal energy as
    \begin{eqnarray}
        \rho     &=& \frac{D\sqrt{Q^2-S^2}}{Q}, \label{eq:Recov2rho} \\
        \epsilon &=& \left(\sqrt{Q^2-S^2} - \frac{pQ}{\sqrt{Q^2-S^2}} - D\right) \Bigg/ D, \label{eq:Recov2e}
    \end{eqnarray}
    where $Q=\tau+p+D$. The recovery problem then reduces to identifying the value of $p$ which satisfies
    \begin{equation}
        p - \hat{p}\left[\rho\left(\bm{\hat{U}},p\right),\epsilon\left(\bm{\hat{U}},p\right)\right] = 0,
    \end{equation}
    where $\bm{\hat{U}}=\bm{U}/\sqrt{\gamma}$ are the undensitised conserved variables and $\hat{p}$ denotes the value that the EOS predicts for the given density and specific internal energy (via Eqs.~\eqref{eq:Recov2rho}-\eqref{eq:Recov2e}). For the solution of the non-linear equation, we employ a Newton-Raphson method using the value of $p$ from the previous time step in the respective cell as our starting guess.

    \item The recovery scheme which is presented in detail in Appendix C of \cite{2013PhRvD..88f4009G}. This approach avoids the use of a root-solving method which relies on derivatives, because these might be noisy for the case of tabulated EOSs. Instead, we use a Regula-Falsi method (see \cite{2013PhRvD..88f4009G} for more details).
\end{enumerate}

As is typical in GRHD applications, we employ a low density artificial atmosphere to represent vacuum regions. In regions where no matter is present, we set the rest-mass density to $\rho_\mathrm{atm}$ and the velocities to zero. The remaining hydrodynamical quantities follow from the EOS. The value of $\rho_\mathrm{atm}$ is a multiple of the initial maximum rest-mass density within the numerical domain $\rho_\mathrm{max}(t=0)$. This parameter can be freely chosen and for the simulations discussed in Secs.~\ref{sec:IV} and \ref{sec:V} is set to $10^{-8}$ (i.e. $\rho_\mathrm{atm} = 10^{-8}\times \rho_\mathrm{max}(t=0)$). Furthermore, we define a threshold density $\rho_\mathrm{thr}$. If, throughout the evolution, the density of any cell drops below $\rho_\mathrm{thr}$, the cell is reset to atmospheric values. In the simulations presented in Secs.~\ref{sec:IV}, \ref{sec:V} and \ref{sec:VI}, $\rho_\mathrm{thr} = 2 \times \rho_\mathrm{atm}$. In general, the value of $\rho_\mathrm{thr}$ is also a free parameter.

Finally, we impose outflow boundary conditions in our runs. Namely, we remove matter that reaches the boundaries of the hydrodynamical grid from the simulation.

\subsection{Metric equations}\label{sec:NumMethMetric}
We solve the metric field Eqs.~\eqref{eq:DEcfcpsi},\eqref{eq:DEcfcalppsi},\eqref{DEcfc:Beta} and \eqref{eq:DEcfcchi} iteratively with a multigrid solver (see e.g.\ \cite{10.5555/357695}). To obtain boundary conditions for Eq.~\eqref{eq:DEcfcpsi}, we write the formal solution to the Poisson equation, which reads
\begin{equation}\label{Eq:MetricExp}
\Delta\psi = S_\psi \Rightarrow \psi(\bm{r}) = -\frac{1}{4\pi} \int \frac{S_\psi(\bm{r}^\prime)}{|\bm{r}-\bm{r}^\prime|} d^3\bm{r}^\prime,
\end{equation}
where $\bm{r}$ and $\bm{r}^\prime$ are coordinate vectors and $S_\psi = - 2\pi\psi^5 E - \frac{1}{8}\psi^5K_{ij}K^{ij}$ denotes the source term of Eq.~\eqref{eq:DEcfcpsi}. We then approximate the solution at the metric grid boundaries based on expanding $S_\psi$ in Eq.~\eqref{Eq:MetricExp} up to quadrupole order in the first term (i.e.\ $- 2\pi\psi^5 E$) and monopole order in the non-compact term (i.e.\ $- \frac{1}{8}\psi^5K_{ij}K^{ij}$). The same procedure applies for the boundary conditions of Eq.~\eqref{eq:DEcfcalppsi}. In the case of Eqs.~\eqref{DEcfc:Beta} and \eqref{eq:DEcfcchi}, we follow the approach described in \cite{1998PhRvD..57.7299B} and impose fall-off boundary conditions. The original implementation of the metric solver originates from \cite{2002PhRvD..65j3005O,2007A&A...467..395O}, where more details can be found.

A relatively common practice in simulations which employ the conformal flatness condition is to solve the metric field equations only every few time steps and approximate the metric based on an extrapolation scheme in the intermediate time steps (see e.g.\ \cite{2002A&A...388..917D,2011A&A...528A.101B,2020CQGra..37n5015C,2021MNRAS.508.2279C,2024MNRAS.528.1906L}). This approach is justified in cases where the metric fields do not change significantly over the course of a time step and aims to reduce the computational effort. In our application, we can freely set how frequently the metric field equations are solved. In this work, we opt to update the metric in the first substep of the RK integrator. We call the metric solver in the first $10$ time steps. We then explicitly solve the field equations every $5$ time steps. In the intermediate time steps, we extrapolate the metric via parabolic extrapolation using the last three metric solutions as input. Overall, we find that these settings produce accurate results compared to solving the metric field equations in every time step. The accuracy and possible limitations of this approach have been investigated for a number of different physical scenarios, such as evolutions of TOV stars \cite{2024MNRAS.528.1906L}, rapidly rotating NSs \cite{2020CQGra..37n5015C}, BNS mergers \cite{2024MNRAS.528.1906L} and core collapse \cite{2002A&A...388..917D}.

In our standard setup, the metric and hydro grids are identical. Namely, the hydro cell centers match the metric grid points and, thus, the hydro and metric resolutions are the same. However, in realistic scenarios involving NSs, the hydrodynamical variables vary considerably more than the metric fields. For additional flexibility in how we allocate computational resources, we implement in the code the functionality to completely decouple the metric and hydro grids. In this case, we solve the metric and GRHD equations on two distinct uniform Cartesian grids, which can have different resolutions and sizes. We perform the necessary interpolations between the two grids via a $3^\mathrm{rd}$ order Lagrange polynomial. In special cases where the position of a metric grid point coincides with a hydro cell center \footnote{For example, the hydro grid resolution might be higher by a factor of $2$ compared to the metric grid resolution by placing one metric grid point at every second hydro cell center.}, Lagrange interpolation ensures that the exact value of the respective field is passed from one grid to the other at these positions. For the results discussed in Secs.~\ref{sec:IV} and \ref{sec:V} the metric and hydro grids are identical.

We employ the quadrupole formula to extract the GW amplitude (see e.g.\ \cite{1990MNRAS.242..289B})
\begin{equation}
    h_{ij} = \frac{2}{r} P_{ijkl}(\bm{\eta}) \frac{d^2 I_{kl}}{dt^2},
\end{equation}
where $r$ is the distance to the source, $\bm{\eta}$ denotes the unit vector from the source to the observer, $P_{ijkl}(\bm{\eta}) = (\delta_{ik}-\eta_i \eta_k) (\delta_{jl}-\eta_j \eta_l) - \frac{1}{2} (\delta_{ij}-\eta_i \eta_j) (\delta_{kl}-\eta_k \eta_l)$ is the transverse-traceless projection tensor and $I_{ij}$ is the quadrupole moment (see \cite{1990MNRAS.242..289B,2007A&A...467..395O} for a definition).

\section{GRHD Code tests}\label{app:Codetests}
In the following sections, we present a series of benchmark problems which test the validity of our implementation. Appendix~\ref{app:Shocktube} demonstrates the performance of our application in a one-dimensional shock tube test, while Appendix~\ref{app:TOV} presents evolutions of initial data referring to static stars in the Cowling approximation (Appendix~\ref{app:TOV_Cowling}), as well as models with a dynamical spacetime (Appendix~\ref{app:TOV_DynSp}).

\subsection{Relativistic shock tube}\label{app:Shocktube}
We set up a one-dimensional special relativistic shock tube, which is commonly employed to test relativistic hydrodynamics codes (see e.g.\ \cite{2003LRR.....6....7M,2015LRCA....1....3M}). We cover the interval $[0,1]$ with $N$ cells with a fixed grid spacing. The initial conditions consist of the left state ($x\le0.5$) with $(\rho_L, p_L) = (10, 40/3)$ and the right state ($x>0.5$) $(\rho_R, p_R) = (1, 10^{-6})$, while the velocity is initially zero everywhere. We assume an ideal gas equation of state with $\Gamma=5/3$. We employ a 3rd-order Runge--Kutta scheme with a CFL factor of $0.3$, for the time integration, and the 3rd-order PPM, for the spatial reconstruction. We assume a Minkowski spacetime and evolve the system up to $t=0.4$.

\begin{figure}
    \subfloat[]{%
         \includegraphics[width=\columnwidth]{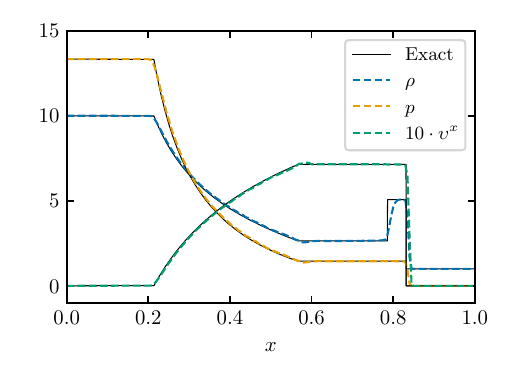}\label{Shocktube_fields}}

    \vspace*{8pt}%
    
    \subfloat[]{%
        \includegraphics[width=\columnwidth]{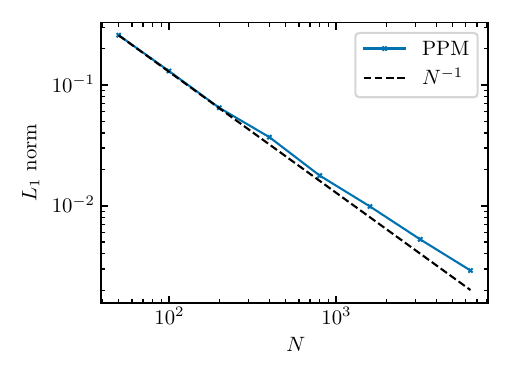}\label{Shocktube_L1_rho}}
    
    \caption{Panel \protect\subref{Shocktube_fields} shows the rest-mass density $\rho$, pressure $p$ and velocity $\upsilon^x$ at $t=0.4$ for the shock tube test and a resolution of $N=200$. The solid lines refer to the analytic result. Panel \protect\subref{Shocktube_L1_rho} presents the L1 norm of the density field as a function of the number of grid points for the shock tube test. The dashed line indicates first-order convergence.}
    \label{Shocktube_plot}
\end{figure}

Figure~\ref{Shocktube_fields} shows our numerical results for a resolution $N=200$ (grid spacing $\Delta x = 0.005$). The solid line refers to the analytic solution, which we compute using the code from the supplemental material of \cite{2003LRR.....6....7M}. For all the physical fields, there is good agreement between our numerical evolution and the analytic solution. In addition, we perform simulations for a number of different resolutions. Figure~\ref{Shocktube_L1_rho} presents the $L_1$ norm of the error between the analytic solution for the rest-mass density and the respective result from our numerical evolutions. In the current test, we find that the $L_1$ norm converges at first-order, which is the expected convergence rate in the presence of shocks.

\subsection{Tolman-Oppenheimer-Volkoff star}\label{app:TOV}
We construct a static 1.4~$M_\odot$ star by solving the TOV equations and assuming the polytropic EOS with $\Gamma=2$ and $K=100$. The central rest-mass density of the star is $\rho_c=1.28\times 10^{-3}$. We map the TOV data to our three-dimensional code and perform calculations at different resolutions. The purpose of this test is to determine whether the evolved three-dimensional star remains close to the original equilibrium solution obtained from the TOV system of equations. In addition, truncation errors excite the radial mode (e.g.\ \cite{2002PhRvD..65h4024F}). Monitoring the radial pulsation frequencies and comparing to independent calculations is a well-known test for hydrodynamical applications.

As a first test, we validate our GRHD implementation by evolving the stellar model in the Cowling approximation (Section~\ref{app:TOV_Cowling}). We then perform simulations that include a dynamical spacetime (Appendix~\ref{app:TOV_DynSp}), which serves as a test to the GRHD implementation, the metric solver, as well as their coupling. In all cases, we perform the time integration with a 3rd-order Runge--Kutta scheme with a CFL factor of $0.3$ and the spatial reconstruction to the cell interfaces with the 3rd-order PPM. The atmosphere density is set to $\rho_\mathrm{atm} = 10^{-8}\times \rho_\mathrm{max}(t=0)$ and any cell with a density below $2\times\rho_\mathrm{atm}$ is reset to atmospheric values. We perform calculations with both adiabatic evolutions, where we explicitly assume that the specific internal energy follows from the polytropic EOS, as well as evolutions with the ideal gas EOS. 

\subsubsection{Cowling approximation}\label{app:TOV_Cowling}
\begin{figure}
    \subfloat[]{%
         \includegraphics[width=\columnwidth]{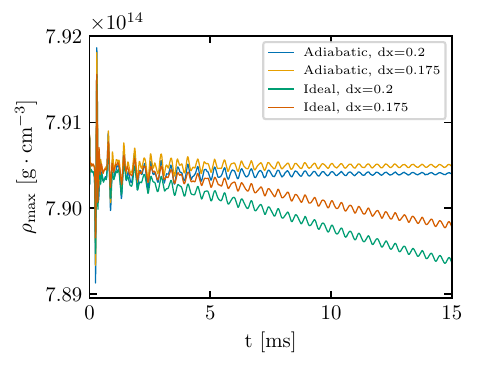}\label{ct_tov_Cowling_rho}}

    \vspace*{8pt}%
    
    \subfloat[]{%
        \includegraphics[width=\columnwidth]{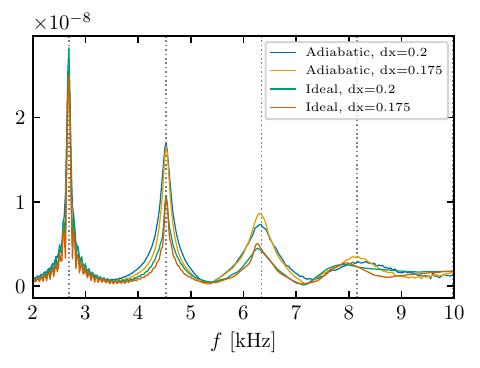}\label{ct_tov_Cowling_fft}}
    
    \caption{Panel \protect\subref{ct_tov_Cowling_rho}: Maximum rest-mass density evolution of a $1.4~M_\odot$ polytropic star with $\Gamma=2$ and $K=100$ in the Cowling approximation. Both adiabatic and ideal gas evolutions are shown at two different resolutions (see legend). Panel \protect\subref{ct_tov_Cowling_fft}: Fourier transformation of the maximum rest-mass density for the various simulations shown in panel \protect\subref{ct_tov_Cowling_rho}. The vertical dashed lines correspond to the frequencies from an independent result (see main text).}
    \label{ct_tov_Cowling}
\end{figure}

Figure~\ref{ct_tov_Cowling_rho} displays the evolution of the maximum density over roughly $15$~ms. For both the adiabatic and ideal gas evolutions, we perform simulations at two different resolutions, $\mathrm{dx}=0.175$ and $\mathrm{dx}=0.2$. In all the simulations, we find that the star remains close to the initial equilibrium solution for the time considered. Focusing on the adiabatic evolutions, the maximum density features practically no secular drift. The radial pulsation, as seen in the oscillation of the maximum density, exhibits some damping over time due to numerical viscosity, but can be clearly identified at the end of both simulations. Increasing the resolution leads to less damping of the oscillation at the late stages of the simulations. Adopting an ideal gas EOS leads to a very minor drift of the maximum density towards lower densities, which remains less than $0.2\%$ for both resolutions. The density drift is less pronounced in the higher resolution simulation. Compared to the adiabatic evolutions, the maximum density oscillation when assumming an ideal gas EOS is less damped over time. We note that the secular drift trends are in agreement with previous studies (see e.g.\ \cite{2002PhRvD..65h4024F}).

We perform a Fourier transform of the maximum density oscillation for all four simulations. Figure~\ref{ct_tov_Cowling_fft} shows the power spectrum of the maximum density oscillation. The dashed vertical lines refer to the frequencies of the fundamental radial mode, as well as the first three overtones, as they are reported in \cite{2002PhRvD..65h4024F}. We find excellent agreement with the frequencies from the independent code. Regarding the fundamental mode, we extract $2.683$~kHz (adiabatic, $\mathrm{dx}=0.2$), $2.683$~kHz (adiabatic, $\mathrm{dx}=0.175$), $2.684$~kHz (ideal gas, $\mathrm{dx}=0.2$), $2.685$~kHz (ideal gas, $\mathrm{dx}=0.175$), which perfectly align with the value of $2.696$~kHz in \cite{2002PhRvD..65h4024F}. For the higher overtones, the relative differences between our higher resolution simulations and the values reported in \cite{2002PhRvD..65h4024F} are $-0.04\%$, $-0.58\%$, $0.17\%$ (adiabatic, $\mathrm{dx}=0.175$) and $-0.04\%$, $-1.28\%$, $-2.55\%$ (ideal gas, $\mathrm{dx}=0.175$). The better agreement of the adiabatic evolution in the second and third overtones is expected, since in \cite{2002PhRvD..65h4024F} they extract the frequencies based on adiabatic evolutions.

\subsubsection{Dynamical spacetime}\label{app:TOV_DynSp}
We evolve the same stellar configuration with a dynamical spacetime. In this test, we employ the functionality to decouple the hydro and metric grids (see Sec.~\ref{sec:NumMethMetric}). We set up two uniform Cartesian grids that have the same resolution $\mathrm{dx}$, but different sizes. The hydro grid consists of $127$ cells per direction ($127^3$ cells in total). The cell centers have coordinates $(x_\mathrm{hydro},y_\mathrm{hydro},z_\mathrm{hydro})$, where $x_\mathrm{hydro}(i)=(i-64)\times\mathrm{dx}$ with $i$ an integer in $[1,127]$ (similarly for $y_\mathrm{hydro}$ and $z_\mathrm{hydro}$). The metric grid consists of $257$ grid points per direction (i.e.\ a total of $257^3$ grid points). The metric grid points have coordinates $(x_\mathrm{metric},y_\mathrm{metric},z_\mathrm{metric})$, where $x_\mathrm{metric}(i)=(i-129)\times\mathrm{dx}$ with $i$ an integer in $[1,257]$ (same goes for $y_\mathrm{metric}$ and $z_\mathrm{metric}$). In this arrangement, the central part of the metric grid is identical to the hydro grid, but the metric grid extends beyond the hydro grid. For the metric grid points outside the hydro grid, we monitor that no matter reaches the hydro grid boundaries and provide a vacuum hydro state as input to the metric solver. This setup allows us to reduce the computational effort to solve the GRHD equations by a factor of roughly $8$ (compared to using a hydro grid identical to the metric grid). We note that the metric grid box has a side of $5.5$ and $6.3$ times the stellar radius in isotropic coordinates ($r_s$) for the simulations with resolution $\mathrm{dx}=0.175$ and $\mathrm{dx}=0.2$, respectively. Thus, the metric boundary conditions are computed at a distance of at least $2.75 \times r_s$ from the stellar center in the runs with $\mathrm{dx}=0.175$ and a distance of $3.15\times r_s$ in the calculations with $\mathrm{dx}=0.2$.

\begin{figure}
    \subfloat[]{%
         \includegraphics[width=\columnwidth]{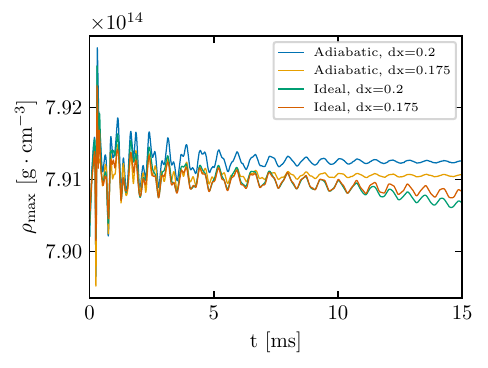}\label{ct_tov_DynSp_rho}}

    \vspace*{8pt}%
    
    \subfloat[]{%
        \includegraphics[width=\columnwidth]{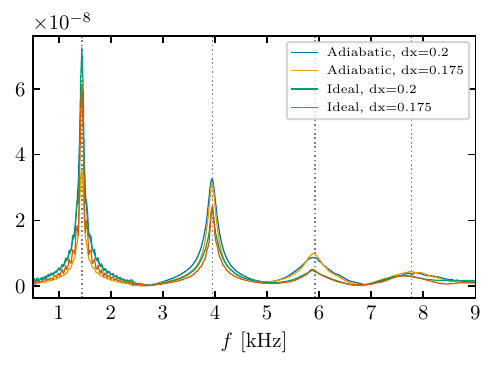}\label{ct_tov_DynSp_fft}}
    
    \caption{Panel \protect\subref{ct_tov_DynSp_rho}: Evolution of the maximum rest-mass density of a $1.4~M_\odot$ star modelled by the polytropic EOS ($\Gamma=2$, $K=100$), including a dynamical spacetime. For information on the resolution and whether the internal specific energy is dynamically evolved or follows analytically from the polytropic EOS, see the legend. Panel \protect\subref{ct_tov_DynSp_fft}: Fourier spectrum of the maximum density oscillations for all simulations displayed in panel \protect\subref{ct_tov_DynSp_rho}. The vertical dashed lines refer to the radial mode frequencies computed with perturbation theory (see \cite{2002PhRvD..65h4024F}).}
    \label{ct_tov_DynSp}
\end{figure}

We present the time evolution of the rest-mass density for all the simulations in Fig.~\ref{ct_tov_DynSp_rho}. We find that in all cases the star oscillates stably around the initial TOV solution. Adopting the ideal gas EOS leads to a slightly more pronounced secular drift of the rest-mass density, which however is very minor and less than $0.1\%$. Similar to the respective tests in the Cowling approximation, the oscillation is more strongly damped in the adiabatic evolutions. In the evolutions with the ideal gas EOS, the oscillation amplitude is well-preserved until the end of the simulation.

We monitor the radial pulsation frequencies and present the power spectrum of the maximum density oscillation in Fig.~\ref{ct_tov_DynSp_fft}. The vertical dashed lines refer to a perturbative calculation of the respective frequencies (see \cite{2002PhRvD..65h4024F}). We measure the fundamental frequency to be $1.439$~kHz (adiabatic, $\mathrm{dx}=0.2$), $1.44$~kHz (adiabatic, $\mathrm{dx}=0.175$), $1.444$~kHz (ideal gas, $\mathrm{dx}=0.2$), $1.442$~kHz (ideal gas, $\mathrm{dx}=0.175$), which is in perfect agreement with the perturbative frequency ($1.442$~kHz in \cite{2002PhRvD..65h4024F}). In addition, we can identify three higher frequency overtones, which also very well agree with the perturbative results. For the adiabatic evolution with $\mathrm{dx}=0.175$, the relative differences between the overtones and the calculation in perturbation theory are $-0.02\%$, $-0.49\%$ and $-0.58\%$. The respective relative differences for the ideal gas simulation with $\mathrm{dx}=0.175$ are $-0.02\%$, $-0.86\%$ and $-1.72\%$.

We note that, in addition to the simulations discussed in this section, we also evolved a 1.35$M_\odot$ star modelled by the DD2 EOS on a dynamical spacetime. The evolution takes into account the full temperature and composition dependence of the DD2 model. The purpose of this test was to fully examine and validate the setups that we employ in our study. At a resolution of $\mathrm{dx}=0.2$, we find that the fundamental mode frequency agrees with the value reported in the literature based on a perturbative calculation \cite{2023Galax..11...60S}.

\bibliography{references}

\end{document}